\RequirePackage{lineno} 

\documentclass[twocolumn,showpacs,preprintnumbers,amsmath,amssymb,nofootinbib]{revtex4}

\usepackage{graphicx}
\usepackage{dcolumn}
\usepackage{bm}
 
\usepackage{epstopdf}

\def\bea{\begin{eqnarray}}
\def\eea{\end{eqnarray}}

\def\pp{\mbox{$p$-$p$}}

\def\pa{\mbox{$p$-A}}
\def\da{\mbox{$d$-A}}
\def\auau{\mbox{Au-Au}}

\def\pbpb{\mbox{Pb-Pb}}
\def\ppb{\mbox{$p$-Pb}}
\def\dau{\mbox{$d$-Au}}
\def\pn{\mbox{$p$-N}}
\def\aa{\mbox{A-A}}
\def\nn{\mbox{N-N}}

\def\ee{\mbox{$e^+$-$e^-$}}

\def\pt{$p_t$}

\def\yt{$y_t$}

\def\nch{$n_{ch}$}
\def\mmpt{$\bar p_t$}

\begin{document} 

\setlength{\pdfpagewidth}{8.5in}
\setlength{\pdfpageheight}{11in}

\setpagewiselinenumbers
\modulolinenumbers[5]

* \addtolength{\footnotesep}{-10mm}\

\preprint{version 2.4}

\title{A two-component model for identified-hadron $\bf p_t$ spectra from 5 TeV p-Pb collisions
}

\author{Thomas A.\ Trainor}\affiliation{CENPA 354290, University of Washington, Seattle, WA 98195}


\date{\today}

\begin{abstract}
In preparation for the heavy ion program at the relativistic heavy ion collider (RHIC) $d$-Au collisions were designated as a control experiment for possible discovery of a quark-gluon plasma (QGP) in more-central Au-Au collisions, and contrasting results from the two systems seemed to support such a discovery. In contrast, recent results ($p_t$-spectrum and angular-correlation features) from $p$-Pb collisions at the large hadron collider (LHC) have been interpreted to support claims of hydrodynamic flows and QGP formation even in small collision systems. The present study addresses such claims via a two-component (soft + hard) model (TCM) of identified-hadron (PID) $p_t$ spectra from 5 TeV $p$-Pb collisions. $p$-Pb centrality is adopted from a previous study of ensemble-mean $\bar p_t$ data from the same system. $p$-Pb $p_t$ spectra for pions, kaons, protons and Lambdas are described by the TCM within their point-to-point uncertainties.  Invariance of the TCM hard component vs $p$-Pb centrality indicates that jet formation remains unchanged in $p$-Pb collisions relative to $p$-$p$ collisions, and radial-flow contributions to \pt\ spectra are negligible. These $p$-Pb TCM results have implications for interpretation of similar data features from A-A collisions in terms of QGP formation.
\end{abstract}


\maketitle

 \section{Introduction}
   
 It is conventionally asserted that conditions arising in more-central high-energy nucleus-nucleus (\aa) collisions (high temperatures and densities) are sufficient to achieve deconfinement of colored quarks and gluons from nucleons to form a quark-gluon plasma (QGP)~\cite{perfect}. But interpretation of certain data manifestations from \aa\ collisions to confirm QGP formation relies on establishment of control experiments involving low-density \pp\ and \pa\ collision systems where a QGP is unlikely:   ``The interpretation of heavy-ion results depends crucially on the comparison with results from smaller collision systems such as proton-proton (pp) or proton-nucleus (pA)''~\cite{aliceppbpid}. Initial results from the relativistic heavy ion collider (RHIC) seemed to confirm those expectations by the presence (in \auau) vs absence (in \dau) of jet quenching in spectra and angular correlations~\cite{daufinalstate}.
   
 However, claims have emerged, based mainly on data from the large hadron collider (LHC), that evidence for ``collectivity'' (conventionally interpreted to mean hydrodynamic flows) has been observed in \pa\ data~\cite{ppbridge}, and some even conclude that collectivity is evident in \pp\ collisions~\cite{ppcms}, thus negating the intended role of $p$-A or $d$-A and \pp\ collisions as experimental controls. The revised conclusion has emerged that QGP and flows must be universal phenomena for all high-energy nuclear collisions~\cite{dusling}.  However, it has been recognized that such interpretations, and the implication that there may be no density threshold for QGP formation, pose a central problem for interpretation of high-energy particle data~\cite{thoughts}.
  
 The new experimental evidence is derived from two-particle angular correlations and from \pt\ spectra. 2D angular correlations on $(\eta,\phi)$ from high-charge-multiplicity \nch\ \pp\ collisions exhibit a same-side (on azimuth) single ``ridge'' (maximum at $\phi = 0$) extending over a large pseudorapidity $\eta$ interval~\cite{ppcms}. A so-called ``double ridge'' structure (with maxima at $\phi = 0,~\pi$) has also been observed in \ppb\ data~\cite{ppbridge}. The latter is formally the same quadrupole structure $\cos(2\phi)$ associated with elliptic flow parameter $v_2$ in \aa\ collisions. The ridge structures have been interpreted by some as flow manifestations~\cite{dusling}.
  
 Evidence from \pt\ spectra relates to apparent indications of radial flow in the form of ``flattening'' or ``hardening'' of spectra (increase of slope parameter $T$) with increasing \nch\ or centrality. The changes are more pronounced for higher-mass hadrons. Fits to spectra with a blast-wave model return parameters $T_{kin}$ and $\bar \beta_t$, the latter interpreted as a measure of radial flow~\cite{blastwave}. The general trends with \nch\ and hadron mass are similar to those encountered in \aa\ collisions where they are interpreted to indicate radial flow increasing with \aa\ centrality~\cite{kolb}.
 
However, such a reversal of the control function of \pp\ and \pa\ systems is questionable. It is reasonable to maintain that if a phenomenon is observed in low-density \pp\ or \pa\ collisions it is unlikely to demonstrate formation of a dense medium in \aa\ collisions and should not be interpreted as such. Alternative analysis methods applied to small-system data have lead to interpretations based on minimum-bias (MB) jets rather than flows, casting doubt on flow interpretations inferred from the same data~\cite{mbdijets}. Abandonment of the control function of \pp\ and \pa\  or $d$-A data should therefore be reconsidered.

High-energy nuclear data exhibit a  basic property:  certain {\em composite} structures require a two-component (soft + hard) model (TCM) of hadron production as demonstrated initially for \pp\ collisions~\cite{ppprd,ppquad}. The TCM has since been applied successfully to \auau~\cite{hardspec,anomalous} and \ppb~\cite{tommpt,tomglauber} collisions. As a composite  production model the TCM is apparently required by spectrum~\cite{alicetomspec} and correlation~\cite{ppquad} data for all A-B collision systems.

In a previous study~\cite{tommpt} a TCM for spectra and ensemble-mean \pt\ or \mmpt\ data from 5 TeV \ppb\ collisions was formulated based on certain assumptions: (a) hadron production near midrapidity proceeds via two distinct mechanisms (i.e., the TCM), (b)  one mechanism (hard component) represents MB jets and (c)  jet production is unmodified in \pa\ collisions relative to isolated \pp\ collisions. The \ppb\ TCM describes \mmpt~\cite{tommpt} and spectrum data (present study) within their published uncertainties and requires only minor modification of the \pp\ TCM. 

In the present study the \ppb\ TCM is extended to identified-hadron (PID) \pt\ spectra to provide the most-differential possible test of the TCM and its conclusions for \pp\ and \pa\ collision systems: almost all hadron production arises from the two mechanisms represented by the TCM -- longitudinal projectile-nucleon dissociation (soft) and transverse MB dijet production (hard) -- as manifested in yields, spectra and two-particle correlations. Differential \pt\ spectra for four species of identified hadrons from seven centrality classes of 5 TeV \ppb\ collisions are analyzed. The TCM description of data is exhaustive and allows no room for flow interpretations.

This article is arranged as follows:
Section~\ref{alicedata}  presents PID \pt\ spectra from 5 TeV \ppb\ collisions.
Section~\ref{alternative} compares alternative data descriptions.
Section~\ref{spectrumtcm} describes a \pt\ spectrum TCM for composite A-B collisions.
Section~\ref{ppbgeom}  derives a TCM for \mmpt\ data and centrality parameters for \ppb\ collisions.
Section~\ref{pidspec}  presents a TCM for PID \pt\ spectra from 5 TeV \ppb\ collisions.
Section~\ref{pidmptt}  describes a TCM for PID \mmpt\ data.
Section~\ref{pidratios}  discusses PID spectrum ratios derived from TCM spectra in Sec.~\ref{pidspec}.
Section~\ref{sys} reviews systematic uncertainties.
Sections~\ref{disc} and~\ref{summ} present discussion and summary. Appendix~\ref{ppmptapp} describes a TCM for \mmpt\ data from \pp\ collisions.

\section{5 $\bf TeV$ $\bf p$-$\bf Pb$ PID Spectrum data} \label{alicedata}

The identified-hadron spectrum data obtained from Ref.~\cite{aliceppbpid} for the present analysis were produced by the ALICE collaboration at the LHC.  The event sample for charged hadrons is 12.5 million non-single-diffractive (NSD) collisions and for neutral hadrons 25 million NSD collisions. Collision events are divided into seven charge-multiplicity \nch\ or \ppb\ centrality classes. Corresponding estimated centrality parameters from Ref.~\cite{aliceppbpid} are shown in Table \ref{ppbparams1}. A more detailed analysis of \ppb\ centrality including estimates of systematic biases for different methods is reported in Ref.~\cite{aliceglauber}.
Hadron species include charged pions $\pi^\pm$, charged kaons $K^\pm$, K-zero shorts $K^0_S$, protons $p,~\bar p$ and Lambdas $\Lambda,~\bar \Lambda$. Spectra for charged vs neutral kaons and particles vs antiparticles are reported to be statistically equivalent.

\subsection{$\bf p$-Pb PID  spectrum data} \label{piddataa}

\begin{figure}[h]
	\includegraphics[width=1.65in]{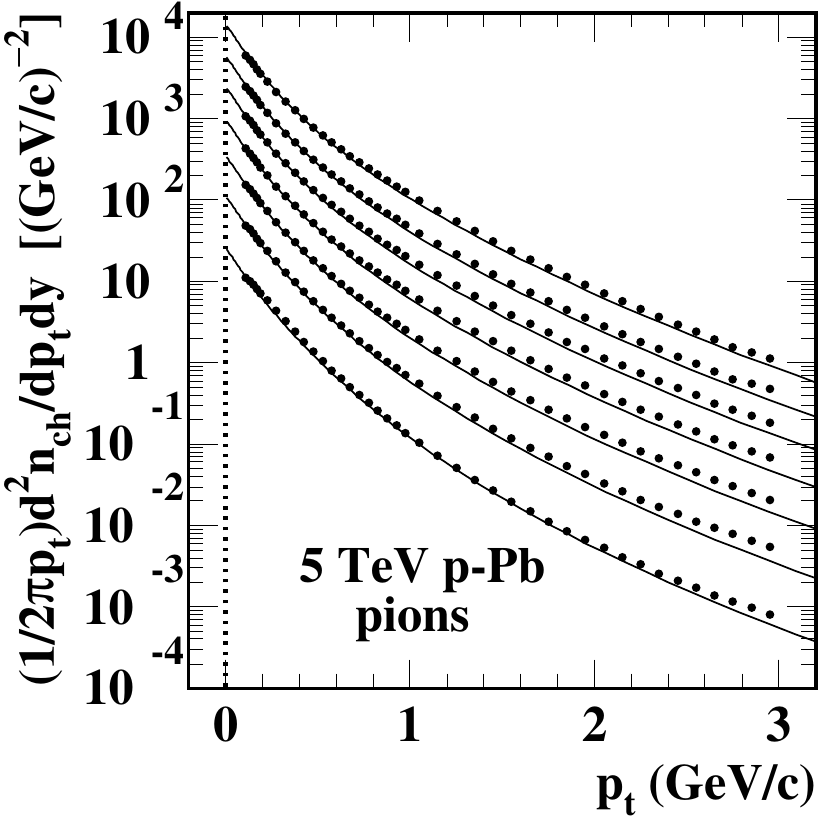}
	\includegraphics[width=1.65in]{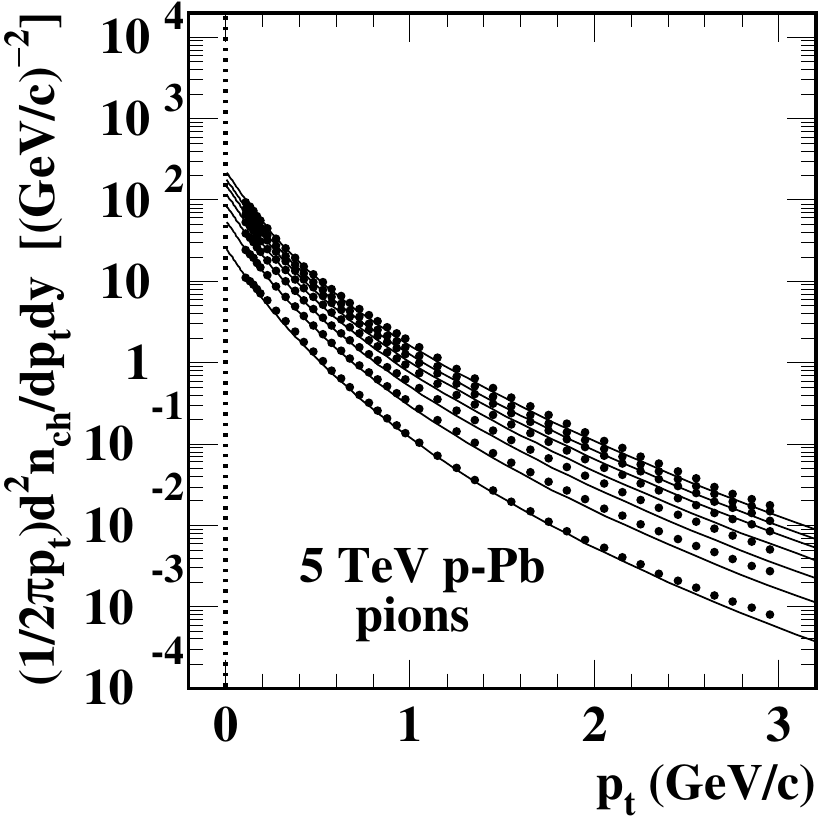}
\put(-142,105) {\bf (a)}
\put(-23,105) {\bf (b)}\\
	\includegraphics[width=1.65in]{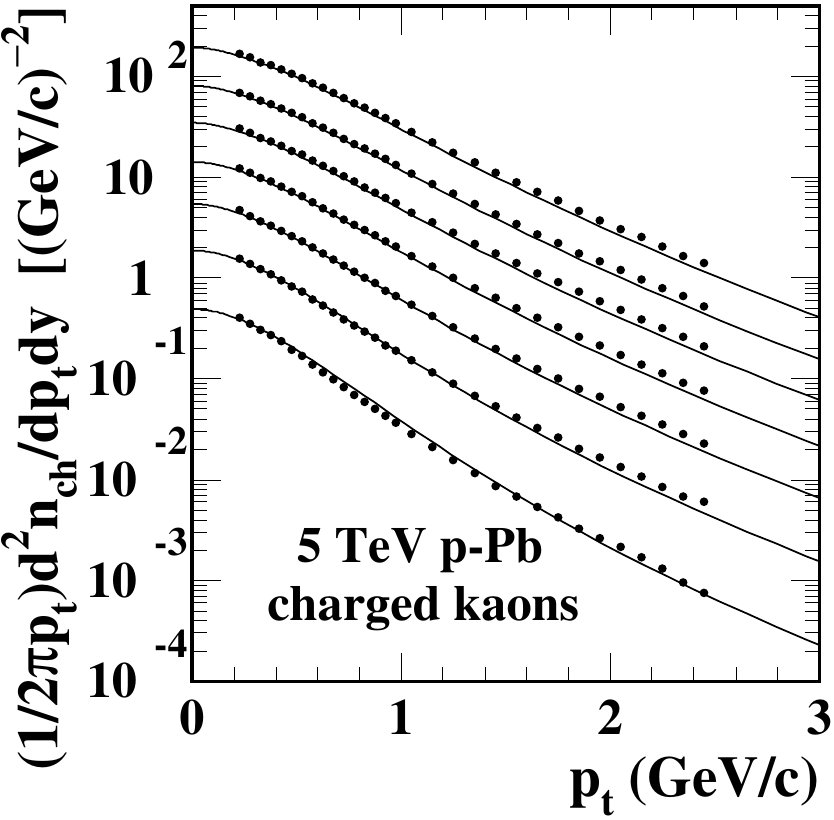}
\includegraphics[width=1.65in]{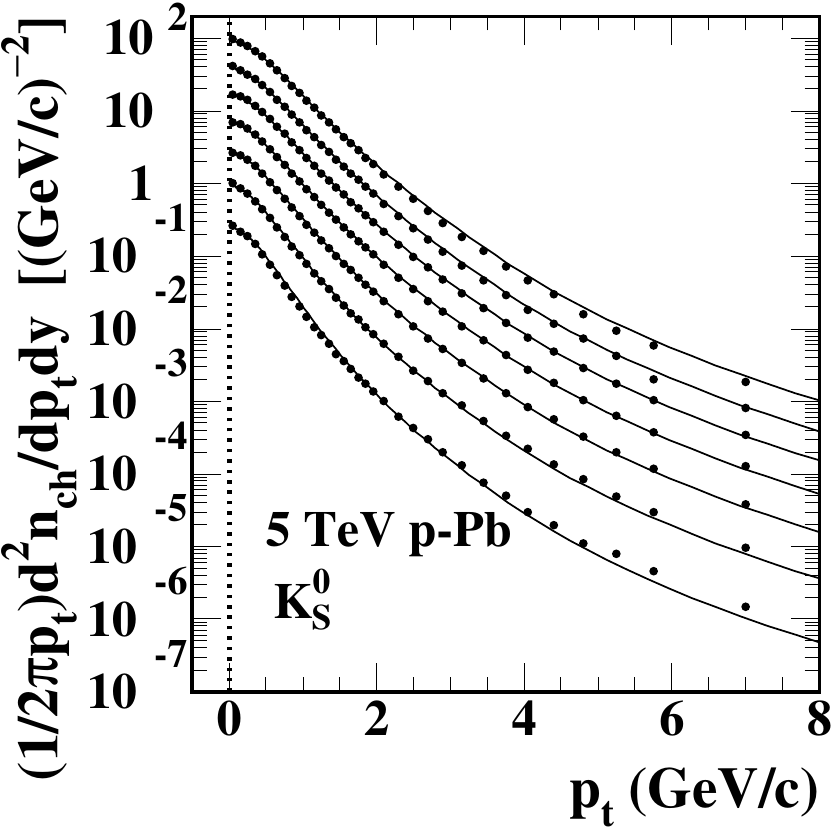}
\put(-142,105) {\bf (c)}
\put(-23,105) {\bf (d)}\\
	\includegraphics[width=1.65in]{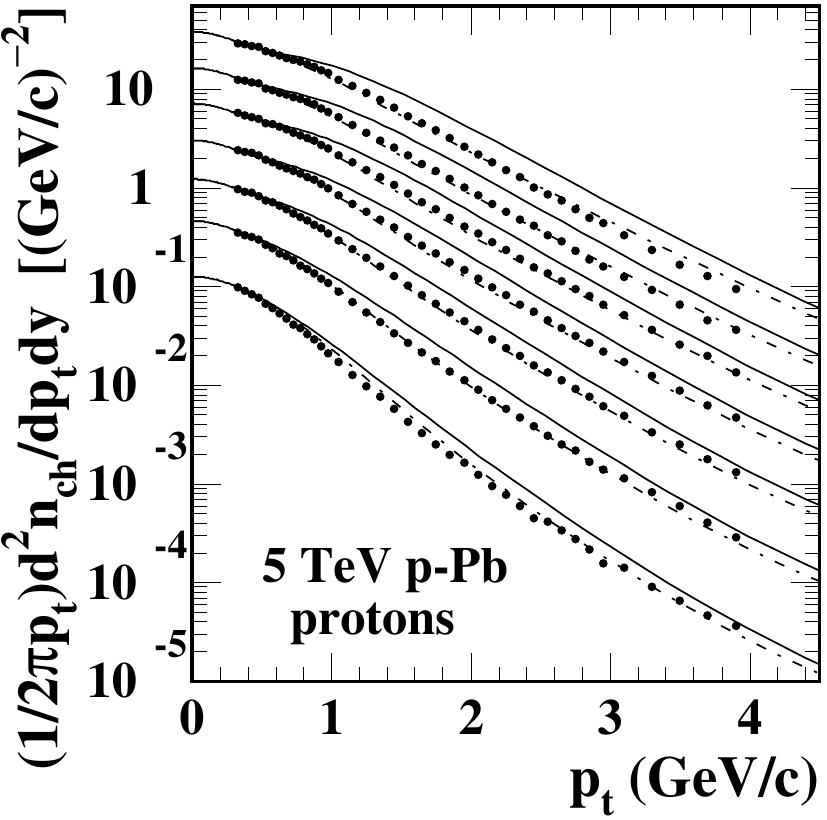}
\includegraphics[width=1.65in]{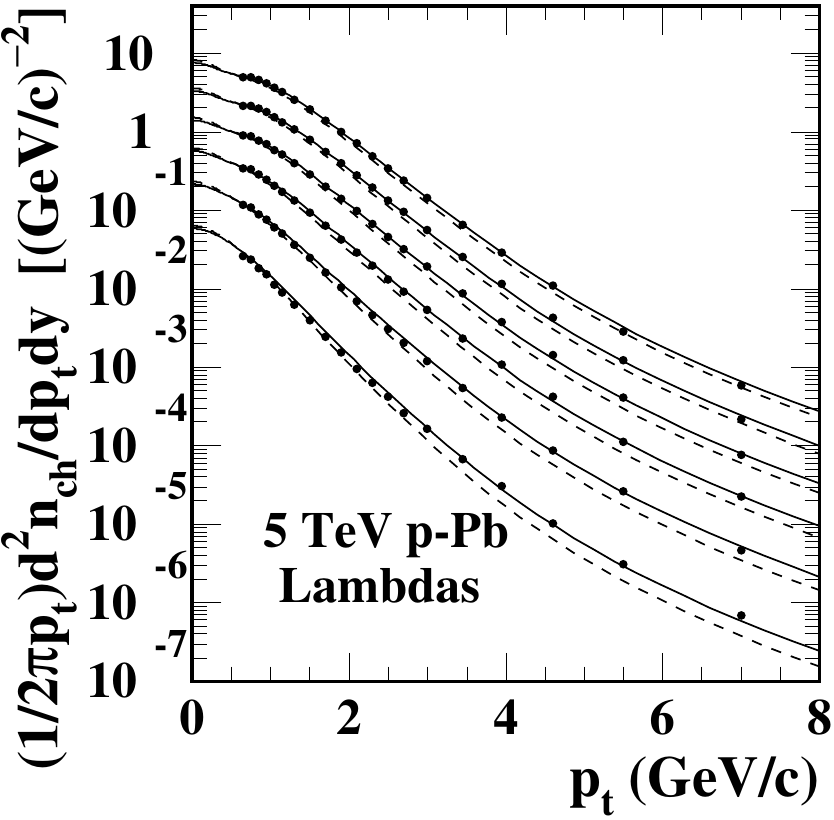}
\put(-142,105) {\bf (e)}
\put(-23,105) {\bf (f)}\\
	\caption{\label{piddata}
\pt\ spectra for identified hadrons from 5 TeV \ppb~\cite{aliceppbpid}:
(a) pions,
(b) pions without multiplicative factors,
(c) charged kaons,
(d) neutral kaons,
(e) protons,
(f) Lambdas.
Solid curves represent the PID spectrum TCM from Sec.~\ref{pidspec}. The dashed curves in (f) repeat the proton solid curves in  (e). The dash-dotted curves in  (e) are described just below Fig.~\ref{protons}.
} 
\end{figure}

Figure~\ref{piddata} shows PID spectrum data from Ref.~\cite{aliceppbpid} (points) in a conventional  semilog plotting format vs linear hadron \pt. The curves are TCM parametrizations derived in Sec.~\ref{pidspec}. The spectra for panels (a) and (c-f) have been scaled by powers of 2 according to $2^{n-1}$ where $n \in [1,7]$ is the centrality class index and $n = 7$ is most central. Panel (b) shows pion spectra with no such scaling, the variation due solely to the different \ppb\ centrality classes. The elevation of pion data (points) in (a) and (b) above TCM (solid curves) at higher \pt\ corresponds to the hard-component excess for pions in Fig.~\ref{pions} (right). It is notable that the maximum spectrum variation corresponds to the most peripheral \nch\ classes $n \in [1,3]$ where the actual \ppb\ centrality {\em varies most slowly} (see Table~\ref{rppbdata}). The chosen plot format limits visual access to {\em differential} spectrum features, especially as they vary with hadron species and \ppb\ centrality and especially at lower \pt\ where most jet fragments appear~\cite{ppprd,fragevo}. Compare with corresponding figures in Sec.~\ref{piddiff}.

The baryon data in panels (e) and (f) are expected to correspond closely [compare solid (Lambda) and dashed (proton) TCM curves in panel (f)]. The differences there arise from the mass difference (soft component) and the hard-component centroid difference (see Table~\ref{pidparams}). However, the proton data in panel (e) fall well below the proton TCM expectation [solid curves, same as dashed curves in panel (f)]. To characterize the discrepancy precisely the proton TCM was modified (dash-dotted curves) to include an additional suppression factor $\approx 0.6$ that accommodates the proton data, as described in Sec.~\ref{piddiff}.

Reference~\cite{aliceppbpid} interprets the spectrum data as follows: The spectra become ``harder as the multiplicity increases,'' where ``harder'' corresponds to decreasing spectrum slope (``flattening'') at lower \pt\ and is said to be similar to that observed in \aa\ collisions~\cite{9,10}. 
It is suggested that commonalities between \ppb\ data and those from \pbpb\ collisions imply the presence of collective flow in \ppb\ collisions: ``In heavy-ion [\aa] collisions, the flattening of transverse momentum distribution and its mass ordering find their {\em natural explanation in the collective radial expansion of the system} [emphasis added].''

Several parametrizations of PID spectra were explored according to Ref.~\cite{aliceppbpid}, of which the so-called blast-wave (BW) model is said to ``give the best description of the data over the full $p_T$ range.'' The BW model assumes ``a locally-thermalized medium, expanding collectively with a common velocity field....'' It is acknowledged that ``the actual values of the [BW] fit parameters depend substantially on the fit range [on  \pt].'' The chosen fit ranges ``...have been defined according to the available data at low \pt\ and {\em based on the agreement with the data} at high \pt'' [emphasis added]. It is concluded from BW fits to \ppb\ spectrum data that results are ``consistent with the presence of radial flow in \ppb\ collisions.'' It is further noted that ``a larger [inferred] radial velocity in \ppb\  [vs \pbpb] collisions has been suggested as a consequence of {\em stronger radial gradients}'' [emphasis added]. 

However, Ref.~\cite{aliceppbpid} includes the following disclaimer in its Sec.~4: "Other processes not related to hydrodynamic collectivity could also be responsible for the observed results" and goes on to cite BW model fits to \pp\ spectra from the PYTHIA Monte Carlo with and without color reconnection (CR). The fit results with CR are observed to be similar to those from \ppb\ and \pbpb\ spectra. It is concluded that ``This generator study shows that other final state mechanisms, such as color reconnection, can mimic the effects of radial flow.''

Reference~\cite{aliceppbpid} concludes that \ppb\ PID spectra ``represent a crucial set of constraints for the modeling of proton-lead collisions at the LHC. The transverse momentum distributions show a clear evolution with multiplicity, similar to the pattern observed in high-energy pp and heavy-ion collisions, where in the latter case {\em the effect is usually attributed to collective radial expansion} [emphasis added].   Models incorporating final state effects give a better description of the data.''

\subsection{$\bf p$-Pb Glauber-model centrality parameters}

Table~\ref{ppbparams1} shows centrality parameters for 5 TeV \ppb\ collisions from Table~2 of Ref.~\cite{aliceglauber} nominally corresponding to the spectra in Fig.~\ref{piddata}. The $\bar \rho_0 = n_{ch} / \Delta \eta$ charge densities are measured quantities inferred from Fig.~16 of that reference, whereas the centrality parameters are inferred from a Glauber model Monte Carlo~\cite{aliceglauber}. Specifically, charge density distributions in Fig.~16 of Ref.~\cite{aliceppbpid} were averaged over $|\eta_\text{lab}| < 0.5$. The results in the fifth column agree with those in Table 1 of Ref.~\cite{aliceppbpid} within the data uncertainties presented in that table. 

\begin{table}[h]
	\caption{V0A Glauber parameters for 5 TeV \ppb\ collisions are from Table 2 and $\bar \rho_0 = n_{ch}/ \Delta \eta$ densities are derived from Fig.~16, both in Ref.~\cite{aliceglauber}.  The event sample is $\approx$ NSD. Comparable charge densities are reported in Table 1 of Ref.~\cite{aliceppbpid}.
	}
	\label{ppbparams1}
	\begin{center}
		\begin{tabular}{|c|c|c|c|c|} \hline
			centrality (\%) & $b$ (fm) & $N_{part}$ & $N_{bin}$ & $n_{ch} / \Delta \eta$  \\ \hline
			0 - 5   & 3.12 & 15.7  & 14.7 &  44.6  \\ \hline
			~\,5 - 10  & 3.50 &  14.0 & 13.0 & 35.9  \\ \hline
			10 - 20  & 3.85 &  12.7 &11.7  & 30.0    \\ \hline
			20 - 40 & 4.54 &  10.4 & 9.36 & 23.0    \\ \hline
			40 - 60   & 5.57 & 7.42  & 6.42 & 15.8   \\ \hline
			60 - 80  & 6.63 & 4.81  & 3.81 & 9.7  \\ \hline
			~\,80 - 100  & 7.51 &  2.94 &1.94  & 4.2   \\ \hline
		\end{tabular}
	\end{center}
\end{table}

To interpret the PID \pt\ spectra from Ref.~\cite{aliceppbpid} properly centralities and geometry parameters should be estimated as accurately as possible. Reference~\cite{aliceglauber} examines several centrality estimation methods identified by detector designations (e.g.\ V0A, V0C, ZNA) and emphasizes estimates based on V0A, a large-$\eta$ detector on the Pb-going side. A detailed analysis of centrality biases is included in that study. The spectrum data from Ref.~\cite{aliceppbpid} that form the basis for the present analysis are based on the V0A method. However, the V0A centrality determination (corresponding to primed quantities in Table~\ref{rppbdata}) is not utilized in the present study. Instead, an independent \ppb\ centrality determination based on TCM descriptions of \ppb\ \mmpt\ data as reported in Ref.~\cite{tomglauber}  (corresponding to unprimed quantities in Table~\ref{rppbdata}) is utilized. The large differences between V0A and TCM geometries exceed the systematic biases estimated in Ref.~\cite{aliceglauber} (see Sec.~\ref{geom}).

\section{Alternative interpretations} \label{alternative}

The conclusions in this paper contradict popular interpretations of collision data, especially that QGP may be formed in possibly all collision systems as demonstrated by experimental and theoretical evidence for flows and jet quenching in a dense QCD medium. This section reviews evidence seen as supporting the flow/QGP paradigm and evidence that appears to contradict such conclusions in order to provide a balanced context for the present study.


\subsection{Collectivity in A-A collision systems}

The terms ``collectivity'' and ``collective phenomena'' are commonly understood to represent data features from high-energy nuclear collisions interpreted as manifestations of transverse expansion in the form of hydrodynamic flow(s). Perceived evidence for strong expansion of a bulk medium in heavy-ion (\aa) collisions has been based on PID \pt\ spectra, azimuth correlations, Bose-Einstein correlations and high-\pt\ suppression (jet quenching). The evidence as of 2004 and summarizing the first years of RHIC operation was reported in so-called white papers~\cite{whitebrahms,whitestar,whitephob,whitephen}.
The data as analyzed were interpreted to indicate that a QGP is formed and undergoes nearly-ideal hydrodynamic expansion (a nearly perfect liquid)~\cite{perfect}. 

Evidence for transverse flows is derived mainly from ``blast-wave'' fits to \pt\ spectra to infer radial flow~\cite{blastwave,starblast} and from Fourier-series fits to azimuth angular correlations to infer elliptic flow ($v_2$)~\cite{2004} and ``higher harmonics'' ($v_3$, etc.)~\cite{aliceflows}. Aside from blast-wave model fits indications in \pt\ spectra of ``hardening'' (increased slope parameter) with collision centrality and with hadron mass are seen as indicators of radial flow. Azimuthal asymmetries (i.e.\ any azimuth correlation structure) are interpreted to indicate azimuthal modulations of transverse flow, with elliptic flow ($v_2$) being most prominent. Data features and conjectured physical phenomena are often treated as synonymous. The higher harmonics, as modulations of transverse flow, are seen as arising from fluctuations in the initial-state (IS) collision geometry~\cite{alver,luzum}.

An important element in such arguments is a correlation feature denoted by ``the ridge,'' a peak at the origin on azimuth difference extending symmetrically over large $\eta$ difference intervals (``long-range'' correlations). The ridge in \auau\ collisions has been associated with jet structure resulting from certain \pt\ cuts in which the   2D jet peak apparently develops tails extending over substantial $\eta$ intervals~\cite{ridge1,ridge2,ridge3}. The ridge in that case is attributed to bulk matter as opposed to jets based on observed large baryon/meson ratios. Note that if no \pt\ cuts are applied the 2D jet peak itself is observed to broaden substantially on $\eta$ for more-central \auau\ collisions~\cite{anomalous}.

In Ref.~\cite{alver} the concept of triangularity and triangular flow is introduced, in which conjectured fluctuations of initial-state geometry (triangularity) are transformed to observed triangular flow via hydrodynamic evolution of a bulk medium. The concept is then generalized to ``higher harmonics'' in which all ``long-range'' (on $\eta$) azimuth structure (i.e.\ the ridge) is a flow manifestation represented by a Fourier series with amplitudes $v_n$~\cite{luzum}. 

Interpretation of $v_n$ data as representing flow of a dense medium in \aa\ is conventionally justified by data comparisons  with viscous-hydro theory descriptions invoking a small viscosity ($\eta/s$ ratio) (e.g.\ Ref.~\cite{gale5}). Based on the apparent success of many such comparisons with an assortment of $v_n$ data, formation of a low-viscosity QGP in \aa\ collisions is considered to be broadly accepted~\cite{dusling}.

\subsection{Collectivity in small collision systems}

In light of the sequence of developments responding to correlation data from RHIC \auau\ collisions the first observation of a ``ridge'' in 7 TeV \pp\ collisions at the LHC~\cite{ppridge} was very surprising. Given interpretation of the ridge feature in \aa\ collisions as indicating collective flow of a dense bulk medium the \pp\ result suggested the possibility of collectivity in the smallest collision system for some imposed conditions (high charge multiplicity, certain \pt\ cuts). A follow-up study of \ppb\ collisions revealed a similar ridge structure with larger amplitudes, comparable to those observed in \aa\ collisions~\cite{ppbridge}.

More generally, \pp\ and \pa\ \pt\ spectra have trends similar to those found in \aa\ collisions, including ``hardening'' (increased slope parameter) increasing with \nch\ or centrality and with hadron mass (relevant to radial flow). Equivalently, ensemble-mean \pt\ also increases with centrality or \nch\ faster for more-massive hadrons. $v_2$ and $v_3$ data for \ppb\ are similar to those for \pbpb, and mass ordering observed in PID $v_2(p_t)$ data for \aa\ collisions is also observed for \ppb\ data~\cite{ppbmassorder}. The $v_2(p_t)$ mass ordering is interpreted as further evidence for radial flow.

Just as for \aa\ collision systems description of relevant data by hydro models is seen as confirming a QGP/flow interpretation of \pa\ and even \pp\ data. In Ref.~\cite{dusling}, commenting on novel collective phenomena in small collision systems, the situation is summarized by the statement ``...it is possible to describe all characteristic features measured in p-p and p/d/$^3$He-A collisions with models based on the collective response to an initial state geometry. In particular hydrodynamic models can reproduce the azimuthal anisotropies  of charged hadrons $v_n$, the mass splitting of the mean transverse momentum and $v_2$ for identified particles and the HBT radii.'' However, the report warns that ``Depending on the assumptions made about how the initial shape of  the system is generated, {\em final results can vary dramatically} [emphasis added].'' 


\subsection{Responses to claims for collectivity}

The evidence and arguments summarized above overlook a number of issues that present a strong challenge to the flow/QGP paradigm. In general, much of the information carried by basic particle data tends to be suppressed or ignored, by data selection and by specific choices of analysis methods and variables. Focus is maintained on certain collision centralities and not others. Spectrum models are applied to selected \pt\ intervals and not others. 2D angular correlations are projected onto 1D azimuth for certain key analyses, thereby discarding much information.  \pt\ cuts are applied as biases to angular correlations based on {\em a priori} assumptions that may be invalid. Some characterizations of data features are qualitative rather then quantitative, such as ``hardening'' of spectra and ``mass ordering'' for PID $v_2(p_t)$ data. The ill-defined term ``ridge'' is applied to more than one data feature leading to confusion. Extensive quantities are invoked (e.g.\ $N_{track}$ based on some arbitrary $\eta$ acceptance) when intensive counterparts (e.g.\ a mean charge density) would facilitate clearer comparisons among A-B systems.
 
More specifically,  a growing body of negative evidence is ignored. In \aa\ collisions the centrality trend for jet modification (``quenching'') as revealed by modeling of 2D angular correlations~\cite{anomalous} corresponds to the trend for \pt\ spectra~\cite{hardspec} but is very different from the trend for $v_2$ data also inferred by modeling of 2D angular correlations~\cite{nonjetquad}. A close correspondence should be expected if both trends are related to a common flowing dense medium. In fact, the $v_2$ trend on centrality is generally inconsistent with hydro expectations. The blast-wave model conventionally used to infer radial flow from spectra is an inferior spectrum model applied to limited \pt\ intervals, in some cases determined solely by the fit quality~\cite{aliceppbpid}. The contrast with an accurate and generalized spectrum model with no \pt\ limits is noted in Sec.~\ref{pidspec}.

The use of Fourier series alone to model 1D azimuth correlations is strongly rejected by Bayesian analysis of model quality~\cite{bayes}. For 200 GeV \auau\ collisions a model consisting of a narrow Gaussian with $\cos(\phi)$ and $\cos(2\phi)$ terms only is strongly preferred on the basis of {\em how little information is acquired by a model} upon encountering new data. For example, with increasing data acquisition the Fourier model requires additional terms to maintain a low $\chi^2$ whereas the model with a Gaussian does not. In effect, ``higher harmonics'' in the Fourier series are substituted for a fixed two-parameter Gaussian. In more-central \aa\ collisions, and for certain \pt\ cuts, the 1D Gaussian on azimuth includes projected contributions from the same-side ``ridge''  that has not been effectively ruled out as an aspect of jet modification.

The ``mass ordering'' of PID $v_2(p_t)$ data is said to confirm the presence of transverse expansion of a dense medium, but differential analysis of such data provides much more information. The same data plotted on transverse rapidity $y_t$ (with proper mass for each hadron species) reveal a common zero intercept corresponding quantitatively to a particle-source boost distribution~\cite{quadspec,njquad}. But the effective boost distribution is consistent with a single value, not the broad distribution expected from Hubble expansion of a bulk medium. Further analysis isolates a ``quadrupole spectrum'' associated with the $\cos(2\phi)$ correlation feature that is quite different from what is inferred for the great majority of hadrons.

In \pp\ and \pa\ collisions misuse of the term ``ridge'' has caused confusion since it is applied to at least two different phenomena. The ridge observed in 7 TeV \pp\ collisions~\cite{ppridge} is actually one lobe of a quadrupole $\cos(2\phi)$ correlation. The same-side lobe is easily visible as a ridge because the curvature of that lobe is opposite in sign to the background, whereas superposition of the second lobe at $\pi$ as an ``away-side'' ridge increases the magnitude of the like-sign curvature there which is usually overlooked. In Ref.~\cite{dusling} the \pp\ ridge is described as ``seen'' only for high event multiplicities, is ``not present'' in minimum-bias events and is not well understood. However, the systematics (e.g.\ \nch\ dependence) of the nonjet quadrupole for 200 GeV \pp\ collisions have been accurately determined via 2D model fits~\cite{ppquad} and differ strongly from any reasonable hydro expectation. The same trend is followed from lowest to highest \pp\ \nch\ corresponding to a {\em thousand-fold increase} in the associated number of correlated pairs. The factorized \pp\ quadrupole trend is formally equivalent to the trend for \auau\ collisions~\cite{nature}.

The nonjet quadrupole has also been observed in 5 TeV \ppb\ collisions~\cite{ppbridge} where the quadrupole $\cos(2\phi)$ feature has been isolated via a background subtraction but is described as a ``double ridge.'' It is notable that the ``ridge'' first observed in more-central \auau\ collisions as a centrality-dependent elongation of the same-side 2D jet peak~\cite{anomalous} projecting to a narrow 1D Gaussian on azimuth is distinct from the nonjet quadrupole $\cos(2\phi)$ feature that has been described as a ``ridge'' in \pp\ collisions or ``double ridge'' in \ppb\ collisions but is correctly identified as a cylindrical quadrupole structure in \aa\ data.

\ppb\ PID \pt\ spectra are described as exhibiting ``hardening'' with increasing \nch\ or centrality and with hadron mass~\cite{aliceppbpid}. The \ppb\ trend is seen as matching a similar trend in PID spectra from \aa\ collisions that is interpreted as an indicator for ``collectivity'' in the form of radial flow. A similar argument is applied to ensemble-mean \pt\ data~\cite{alicempt} which simply reflect measured \pt\ spectra. Those specific conjectures motivated the present \ppb\ spectrum study: Hadron mass- and centrality-dependent ``hardening'' and blast-wave model fits to PID spectra are interpreted to demonstrate radial flow upon which the ridge and associated higher harmonics are assumed to be modulations. If the same data features can be identified with confidence as arising from a nonflow mechanism (e.g.\ jets)  then the flow conjecture is unlikely.

Hydrodynamic modeling of \aa\ collisions is reviewed in Ref.~\cite{galehydro}. The agreement between data and theory is apparently very good as illustrated in Ref.~\cite{galehydro2}. The hybrid model utilized in the latter reference consists of components IP-Glasma + Music where IP-Glasma~\cite{glasmafluc} models initial conditions and MUSIC~\cite{music} models viscous hydro evolution. The hydro model utilized for a recent analysis reported in Ref.~\cite{nature0}, comparisons with $v_n\{\text{EP}\}$ data from small asymmetric $x$-Au collisions, is based on the SONIC hydro model. According to Refs.~\cite{naglesonic,naglesonic2} SONIC combines  Monte-Carlo Glauber initial conditions with a 2+1 viscous hydrodynamics evolution and hadronic-cascade afterburner. As noted, Ref.~\cite{dusling} warns that such models are typically very sensitive to initial conditions.

Although hydro models may appear to describe selected data quite precisely there are several major issues:

(a) Sensitivity to initial conditions (IC): If a model is very sensitive to some of its parameters then for parameter values varying across some {\em a priori} reasonable intervals the model may disagree strongly with data. To achieve good agreement with data the parameters must then be confined to a small volume within the parameter space that corresponds to the data. But that is simply a data-fitting procedure. As noted in connection with Bayesian model evaluation~\cite{bayes}, a model that acquires much information from new data is disfavored compared to a model that acquires little (i.e.\ a {\em predictive} model). 

(b) Validity of some IC estimators: Some estimators may be questioned, especially for modeling small asymmetric collision systems. For example, a conventional Glauber Monte Carlo applied to 5 TeV \ppb\ collisions produces strongly-biased estimates for $N_{part}$, $N_{bin}$ and collision centrality~\cite{tomglauber} (primed numbers in Table~\ref{rppbdata}). For \ppb\ collisions assigned to 0-5\% centrality via Monte Carlo Glauber the mean charge density is 45~\cite{aliceglauber} whereas high-statistics ensemble-mean \pt\ (\mmpt) data for the same collision system extend out to charge density 115~\cite{alicetommpt}. The number of participants estimated by the Glauber MC for more-central collisions is roughly 3 times larger  than what is consistent with a TCM description of the \mmpt\ data (unprimed numbers in Table~\ref{rppbdata}). Note that Ref.~\cite{aliceglauber} includes a detailed study of possible biases from several estimation methods, but the difference between Glauber and TCM estimates substantially exceeds such biases.

If the IP-Glasma estimator is used the IC geometry depends strongly on the projectile proton transverse structure, especially its fluctuations. Strong fluctuations of the IC geometry are considered essential to generate ``higher harmonics'' and the ridge structure(s). However, spectrum and correlation data from \pp\ collisions and \mmpt\ data from \ppb\ collisions imply that transverse geometry is not relevant for nucleon-nucleon collisions. Any \nn\ collision appears to achieve full overlap, and simultaneous multiple collisions are excluded. Those principles emerge from the \pp\ rate of dijet production~\cite{ppprd,ppquad} and from analysis of \ppb\ geometry~\cite{tommpt,tomglauber,tomexclude}.

(c) Superiority of alternative data descriptions: Figure~4 of Ref.~\cite{galehydro2} (ALICE data) or Fig.~20 of Ref.~\cite{dusling} (ATLAS data) show IP-Glasma + MUSIC calculations compared to $v_n\{2\}(b)$ data from 2.76 TeV \pbpb\ collisions. Again the agreement appears to be very good. However, the ALICE $v_n$ data have been previously described within data uncertainties by the combination of a nonjet quadrupole (inconsistent with hydro) and multipoles that are Fourier components of the same-side 2D jet peak as demonstrated in Figs.~17 (200 GeV \auau) and 18 (2.76 TeV \pbpb) of Ref.~\cite{harmonics}. $v_2\{2\}$ is known to have a strong jet contribution despite a cut on $\eta$ acceptance intended to exclude ``nonflow'' imposed by the ALICE analysis. And $v_3\{2\}$ and higher coefficients are consistent with the jet peak alone as the source, i.e.\ are Fourier components of a narrow Gaussian. The full centrality dependence of ``higher harmonics'' is exactly as expected from jets, including jet modification (quenching) in more-central collisions as reported in Ref.~\cite{anomalous}.

\subsection{Conclusions}

The material in this section includes only a sampling of an abundance of apparent evidence both for and against the flow/QGP paradigm. Nevertheless, some critical issues raised above remain unresolved after nearly ten years. As noted, recent claims for flow/QGP appearing in small asymmetric collision systems and possibly even in \pp\ collisions have been recognized as presenting a major puzzle for the nuclear physics community. Any evidence that might resolve the puzzle should be welcomed.

The combination of data features interpreted to indicate transverse flow(s) and jet modification (quenching) in \aa\ collisions have been accepted as demonstrating formation of a dense flowing medium or QGP. It is reasonable to apply the same criteria to any collision system in which QGP formation is claimed. The present study reports arguably the most accurate PID spectrum analysis to date, applied in this case to \ppb\ data. The intent is to test for the presence simultaneously of both radial flow and jet modification to the statistical limits of available spectrum data. In the event of a null result claims of QGP in small systems should be strongly questioned.

\section{$\bf p$-$\bf Pb$ Spectrum TCM} \label{spectrumtcm}

The TCM for \pp\ and \ppb\ collisions utilized in this study is the product of phenomenological analysis of data from a variety of collision systems and data formats~\cite{ppprd,ppquad,alicetomspec,tommpt}. As such it does not represent imposition of {\em a priori} physical models but does assume approximate {\em linear superposition} of \pn\ collisions within \ppb\ collisions consistent with no significant jet modification. Physical interpretations of TCM soft and hard components have been derived {\em a posteriori} by comparing inferred TCM characteristics with other relevant measurements~\cite{hardspec,fragevo}, in particular measured MB jet characteristics~\cite{eeprd,jetspec2}. Development of the TCM contrasts with data models based on {\em a priori} physical assumptions such as PYTHIA~\cite{pythia} and the BW model~\cite{blastwave}.  It is notable that the TCM does not result from fits to individual spectra (or other data formats), which would require many parameter values. The few TCM parameters are required to have simple $\log(\sqrt{s})$ trends on collision energy and simple extrapolations from \pp\ trends.

\subsection{Spectrum TCM for unidentified hadrons} \label{unidspec}

The \pt\ or \yt\ spectrum TCM is by definition the sum of soft and hard components with details inferred from data (e.g.\ Ref.~\cite{ppprd}). For \pp\ collisions
\bea  \label{rhotcm}
\bar \rho_{0}(y_t;n_{ch}) &\approx& \bar \rho_{s}(n_{ch}) \hat S_{0}(y_t) + \bar \rho_{h}(n_{ch}) \hat H_{0}(y_t),
\eea
where \nch\ is an event-class index, and factorization of the dependences on \yt\ and \nch\ is a central feature of the spectrum TCM inferred from 200 GeV \pp\ spectrum data in Ref.~\cite{ppprd}. The motivation for transverse rapidity $y_{ti} \equiv \ln[(p_t + m_{ti})/m_i]$ (applied to hadron species $i$) is described in Sec.~\ref{tcmmodel}. The \yt\ integral of Eq.~(\ref{rhotcm}) is $\bar \rho_0 = n_{ch} / \Delta \eta = \bar \rho_s + \bar \rho_h$, a sum of soft and hard charge densities. $\hat S_{0}(y_t)$ and $\hat H_{0}(y_t)$ are unit-normal model functions approximately independent of \nch, and the centrally-important relation $\bar \rho_{h} \approx \alpha \bar \rho_{s}^2$ with $\alpha \approx O(0.01)$ is inferred from \pp\ spectrum data~\cite{ppprd,ppquad,alicetomspec}. 

For composite A-B collisions the spectrum TCM is generalized to
\bea \label{ppspectcm}
\bar \rho_{0}(y_t;n_{ch}) 
&\approx& \frac{N_{part}}{2}  \bar \rho_{sNN} \hat S_{0}(y_t) + N_{bin}  \bar \rho_{hNN} \hat H_{0}(y_t),~~
\eea
which includes a further factorization of charge densities $\bar \rho_x = n_x / \Delta \eta$ into A-B Glauber geometry parameters $N_{part}$ (number of nucleon participants N) and $N_{bin}$ (\nn\ binary collisions) and mean charge densities $\bar \rho_{xNN}$  per \nn\ pair averaged over all \nn\ interactions within the A-B system.  For A-B collisions $\bar \rho_{s} = [N_{part}(n_s)/2]\bar \rho_{sNN}(n_s)$ is a factorized soft-component density and $\bar \rho_h(n_s) = N_{bin}(n_s) \bar \rho_{hNN}(n_s)$ is a factorized hard-component density.   

Integrating Eq.~(\ref{ppspectcm}) over  \yt\ the mean charge density is
\bea \label{nchppb}
\bar \rho_0 &=& \frac{N_{part}}{2} \bar \rho_{sNN}(n_s) + N_{bin} \bar \rho_{hNN}(n_s)
\\ \nonumber
\frac{\bar \rho_0}{\bar \rho_{s}} &=& \frac{n_{ch}}{n_s} ~=~ 1 + x(n_s)\nu(n_s),
\eea
where the hard/soft ratio is $x(n_s) \equiv \bar \rho_{hNN}/\bar \rho_{sNN}$ and the mean number of binary collisions per participant pair is $\nu(n_s) \equiv 2 N_{bin} / N_{part}$. If the \pt\ acceptance is limited by a low-\pt\ cutoff
\bea \label{prime}
\frac{\bar \rho_0'}{\bar \rho_{s}} &=& \frac{n_{ch}'}{n_{s}} ~=~ \xi + x(n_s)\nu(n_s),
\eea
where $\xi \leq 1$ is the fraction of $\hat S_0(p_t)$ admitted by a low-\pt\ acceptance cut $p_{t,cut}$, and primes indicate corresponding uncorrected (biased) quantities. It is assumed that a typical $p_{t,cut}$ is below the effective $\hat H_0(y_t)$ lower limit.

To obtain details of model functions and other aspects of the TCM the measured hadron spectra are normalized by charge-density soft component $\bar \rho_s$. Normalized spectra then have the form
\bea  \label{norm}
\frac{\bar \rho_{0}(y_t;n_{ch})}{\bar \rho_{s}}  
&=&  \hat S_{0}(y_t) +   x(n_s) \, \nu(n_s) \hat H_{0}(y_t),
\eea
where $n_s$ is the soft component of event-class index \nch\ integrated within some $\eta$ acceptance $\Delta \eta$. For A-B collisions $x(n_s)$ is generally inferred from data.  For \pp\ collisions $x(n_s) \equiv \bar \rho_{h}/\bar \rho_{s} \approx \alpha \bar \rho_{s}$ is inferred with $\alpha \approx O(0.01)$ over a broad range of \pp\ collision energies~\cite{alicetomspec}.  For \pa\ collisions $x(n_s)  \approx \alpha \bar \rho_{sNN}$ is assumed by analogy with \pp\ collisions, and other \ppb\ TCM elements  are in turn defined in terms of $x(n_s)$. For unidentified hadrons the normalization factor in Eq.~(\ref{norm}) is 
\bea \label{overrhos}
\frac{1}{\bar \rho_s} &=& \frac{1}{\bar \rho_{sNN} N_{part}/2}= \frac{1 + x(n_s) \nu(n_s)}{\bar \rho_0(n_s)}.
\eea

\subsection{Spectrum TCM model functions} \label{tcmmodel}

Given normalized spectrum data as in Eq.~(\ref{norm}) and the trend $x(n_s) \sim n_s \sim n_{ch}$ the spectrum soft component is defined as the asymptotic limit of normalized data spectra as \nch\ goes to zero. Hard components of data spectra are then defined as complementary to soft components.

The data soft component for a specific hadron species $i$ is typically well described by a L\'evy distribution on $m_{ti}  = \sqrt{p_t^2 + m_i^2}$. The unit-integral soft-component model is 
\bea \label{s00}
\hat S_{0i}(m_{ti}) &=& \frac{A}{[1 + (m_{ti} - m_i) / n T]^n},
\eea
where $m_{ti}$ is the transverse mass-energy for hadrons $i$ of mass $m_i$, $n$ is the L\'evy exponent, $T$ is the slope parameter and coefficient $A$ is determined by the unit-integral condition. Reference parameter values for unidentified hadrons from 5 TeV \pp\ collisions reported in Ref.~\cite{alicetomspec} are $(T,n) \approx (145~ \text{MeV},8.3)$. Model parameters $(T,n)$ for each species of identified hadrons as in Table~\ref{pidparams} are determined from \ppb\ spectrum data as described below.
 
The unit-integral hard-component model  is a Gaussian on $y_{t\pi} \equiv \ln((p_t + m_{t\pi})/m_\pi)$ (as explained below) with exponential (on $y_t$) or power-law (on $p_t$) tail for larger \yt\
\bea \label{h00}
\hat H_{0}(y_t) &\approx & A \exp\left\{ - \frac{(y_t - \bar y_t)^2}{2 \sigma^2_{y_t}}\right\}~~~\text{near mode $\bar y_t$}
\\ \nonumber
&\propto &  \exp(- q y_t)~~~\text{for larger $y_t$ -- the tail},
\eea
where the transition from Gaussian to exponential on \yt\ is determined by slope matching~\cite{fragevo}. The $\hat H_0$ tail density on \pt\ varies approximately as power law $1/p_t^{q + 2}$. Coefficient $A$ is determined by the unit-integral condition. Model parameters $(\bar y_t,\sigma_{y_t},q)$ for identified hadrons as in Table~\ref{pidparams} are also derived from \ppb\ spectrum data.

All spectra are plotted vs pion rapidity $y_{t\pi}$ with pion mass assumed. The motivation is comparison of spectrum hard components assumed to arise from a common underlying jet spectrum on \pt, in which case $y_{t\pi}$ serves simply as a logarithmic measure of hadron \pt\ with well-defined zero. $\hat S_0(m_{ti})$ in Eq.~(\ref{s00}) is converted to $\hat S_0(y_{t\pi})$ via the Jacobian factor $m_{t\pi} p_t / y_{t\pi}$, and $\hat H_0(y_{t})$ in Eq.~(\ref{h00}) is always defined on $y_{t\pi}$ as noted. For unidentified hadrons a pion mass is assumed. In general, plotting spectra on a logarithmic rapidity variable permits superior access to important low-\pt\ structure where the {\em majority of jet fragments appear}. In what follows, hadron species index $i$ may be suppressed for simplicity.

\section{$\bf p$-$\bf Pb$ Mean-$\bf p_t$ TCM} \label{ppbgeom}

Appendix~\ref{ppmptapp} describes a TCM for \mmpt\ data from \pp\ collisions which provides a context for \ppb\ \mmpt\ analysis. With the dominant role of MB jets established for \pp\ (\pn, \nn) collisions  and elements of the \pp\ \mmpt\ TCM introduced the \ppb\ \mmpt\ TCM is presented here in more detail. \ppb\ \mmpt\ data can in turn be used to infer \ppb\ centrality parameters with improved accuracy~\cite{tommpt}.

\subsection{TCM for p-Pb Mean-pt vs $\bf n_{ch}$} \label{}

Given the TCM for \pa\ or A-B \pt\ spectra $\bar \rho_0(p_t)$ as described in the previous section the associated TCM for \mmpt\ vs \nch\ data is simply determined~\cite{tommpt}. The ensemble-mean {\em total} \pt\ for unidentified hadrons integrated over all \yt\ and within some angular acceptance $\Delta \eta$ is
\bea \label{ptint}
\bar P_t &=& \Delta \eta \int_0^\infty dp_t\, p_t^2\, \bar \rho_0(p_t)
~=~ \bar P_{ts} + \bar P_{th}
\\ \nonumber
&=& \frac{N_{part}}{2}  n_{sNN}(n_s) \bar p_{tsNN} + N_{bin}  n_{hNN}(n_s) \bar p_{thNN},
\eea 
where $\bar p_{tsNN}$ and $\bar p_{thNN}$ are determined by model functions $\hat S_0(y_t)$ and $\hat H_0(y_t)$.
Data indicate that $\bar p_{tsNN} \rightarrow \bar p_{ts} \approx 0.40$ GeV/c is a universal quantity (for unidentified hadrons) corresponding to spectrum slope parameter $T \approx 145$ MeV~\cite{alicetomspec}. 
A mean-\pt\ expression based on the TCM (with $n_s = n_{sNN} N_{part} / 2$) has the simple form
\bea \label{pampttcm1}
\frac{\bar P_t}{n_{s}} &=& \bar p_{ts} + x(n_s)\nu(n_s) \, \bar p_{thNN}(n_s).
\eea
In general, $\bar p_{thNN}(n_s)$ may depend on the imposed multiplicity condition $\bar n_{ch}$~\cite{alicetomspec}. However, for this analysis it is assumed that $\bar p_{thNN}(n_s) \rightarrow \bar p_{th0}$ fixed.
If the \pt\ integral in Eq.~(\ref{ptint}) does not extend down to zero because of limited \pt\ acceptance (e.g.\ termination at some $p_{t,cut}$) the expression is modified. The corresponding TCM for {\em uncorrected} conventional ratio $\bar p_t'$ with $p_{t,cut} > 0$ is
\bea \label{pampttcm}
\frac{\bar P_t'}{n_{ch}'} &\equiv&  \bar p_t' ~\approx~ \frac{\bar p_{ts} + x(n_s) \nu(n_s) \, \bar p_{th0}}{\xi + x(n_s)\, \nu(n_s)},
\eea
where $\xi$ is the fraction of the \pt\ spectrum soft component included by acceptance cut $p_{t,cut}$, and that cut does not affect the hard component. The lower limit for $\bar p_t'$ is $\bar p_{ts}' \equiv \bar p_{ts} / \xi$ with $\xi \approx 0.75$ for a $p_{t,cut} \approx 0.15$ GeV/c~\cite{tommpt}.

\subsection{Centrality parameter $\bf x(n_s)$ model} \label{xnsmodel}

Formulation of a TCM for \ppb\ data requires accurate determination of centrality parameter $N_{part} = N_{bin} + 1$  which in turn requires an expression for $x(n_s)$ from which other model parameters may be derived. An expression for $x(n_s)$ can be established by generalizing from $x(n_s) \approx \alpha \bar \rho_s$ for \pp\ collisions. The relation $x(n_s) \approx \alpha \bar \rho_{sNN}(n_s)$ then defines $\bar \rho_{sNN}(n_s)$ and  $N_{part}(n_s)/2 = \alpha \bar \rho_s / x(n_s)$ from which $\nu(n_s) = 2 N_{bin}(n_s) / N_{part}(n_s)$ follows. Parameter $n_s$ is the independent variable for the model.

In an analysis of   \mmpt\ vs \nch\ data from 5 TeV \ppb\ collisions a simple algebraic expression for $x(n_s)$, as an extrapolation of the \pp\ $\alpha \bar \rho_{s}$ trend, is found to describe \mmpt\ data  accurately~\cite{tommpt}. For \ppb\ data the evolution of factors $x(n_s)\, \nu(n_s)$ from strictly \pp--like to alternative behavior is observed near a transition point  $\bar \rho_{s0}$,%
\footnote{The transition arises from competition between two probability distribution, not between physical mechanisms. See Sec.~\ref{geomestimate}.}
but $\bar p_{thNN}(n_s) \rightarrow \bar p_{th0}$ is assumed to maintain a fixed \pp\ (\pn) value in the \pa\ system (i.e.\ no jet modification per supporting evidence in Sec.~\ref{pidspec}). The derivation follows.

\begin{figure}[h]
	\includegraphics[width=1.65in]{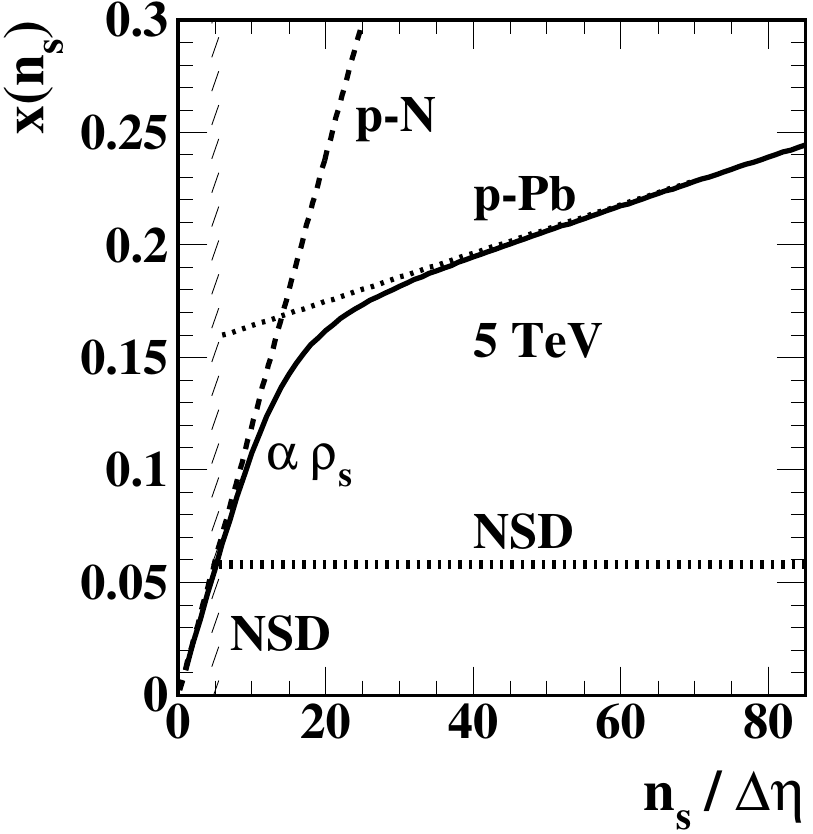}
	\includegraphics[width=1.65in]{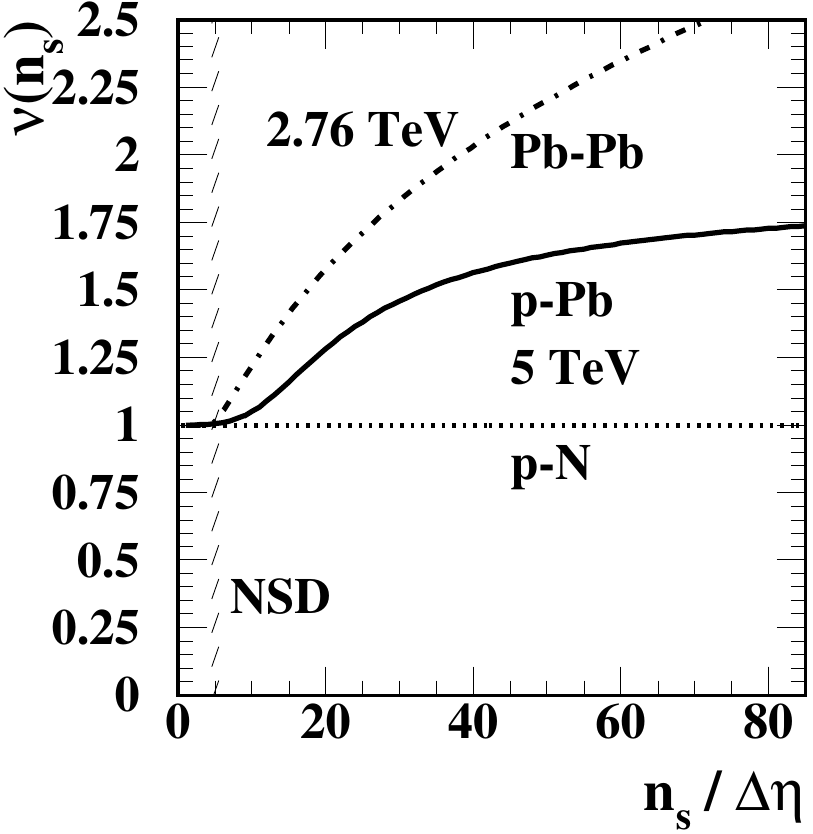}
	\caption{\label{paforms}
		Left: Evolution of TCM hard/soft  ratio parameter $x(n_s)$ with mean soft charge density $\bar \rho_s = n_s / \Delta \eta$ following a linear \pn\ (\pp) trend (dashed) for lower multiplicities and a trend with ten-fold reduced slope for higher multiplicities (dotted) to describe \ppb\ \mmpt\ data.
		Right: Mean participant path length $\nu \equiv 2N_{bin} / N_{part}$ vs $\bar \rho_s$ (solid) as determined by the $x(n_s)$ trend in the left panel (see text). A $\nu$ trend for \pbpb\ collisions (dash-dotted) is included for comparison.
	} 
\end{figure}

Figure~\ref{paforms} (left) shows a model for $x(n_s)$ expressed as
\bea \label{xmodel}
x(n_s) &=& \frac{\alpha}{\left\{[1/ \bar \rho_s]^{n_1} + [1/f(n_s)]^{n_1}\right\}^{1/n_1}},
\eea
where $f(n_s) = \bar \rho_{s0} + m_0(\bar \rho_s - \bar \rho_{s0})$.
Below a transition point at $\bar \rho_{s0}$, $x(n_s) \approx \alpha \bar \rho_s$ as for \pp\ collisions (dashed line). Above the transition $x(n_s)$ still increases linearly but with  reduced slope controlled by parameter $m_0 < 1$ (dotted line). Exponent $n_1$ controls the transition width. The horizontal dotted line and vertical hatched band estimate values of $x(n_s)$ and $\bar \rho_s$ for NSD \pp\ collisions.

Figure~\ref{paforms} (right) shows $\nu \equiv 2N_{bin} / N_{part}$ for \ppb\ data (solid curve) based on $N_{part}(n_s)/2 = \alpha \bar \rho_s / x(n_s)$ and $N_{bin} = N_{part} - 1$ with $x(n_s)$ as described in the left panel (solid curve). 
The dash-dotted curve indicates a $\nu  \sim (N_{part}/2)^{1/3}$ trend for \pbpb\ collisions for comparison, consistent with the eikonal approximation assumed for the \aa\ Glauber model. For \pbpb\ collisions $\nu \in [1,8]$ whereas for \ppb\ $\nu \in [1,2]$. The resulting \mmpt\ TCM is compared with \ppb\ \mmpt\ data in Fig.~\ref{padata}.

\subsection{$\bf \bar p_t$ TCM for $\bf p$-$\bf Pb$ collisions  vs data} \label{ppb}

\begin{figure}[h]
	\includegraphics[width=1.65in]{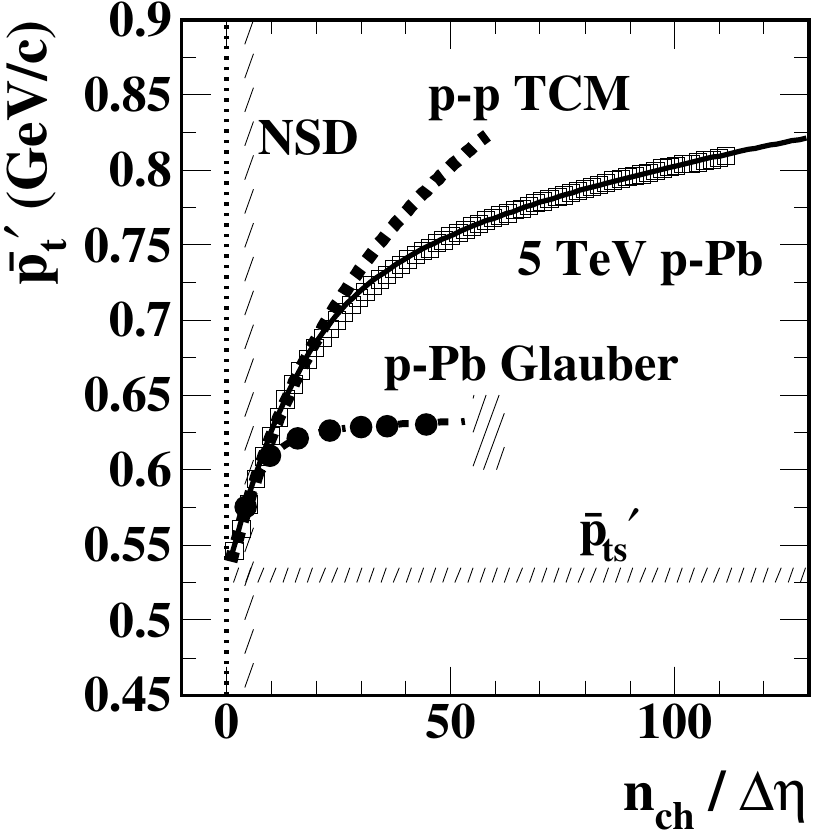}
	\includegraphics[width=1.65in]{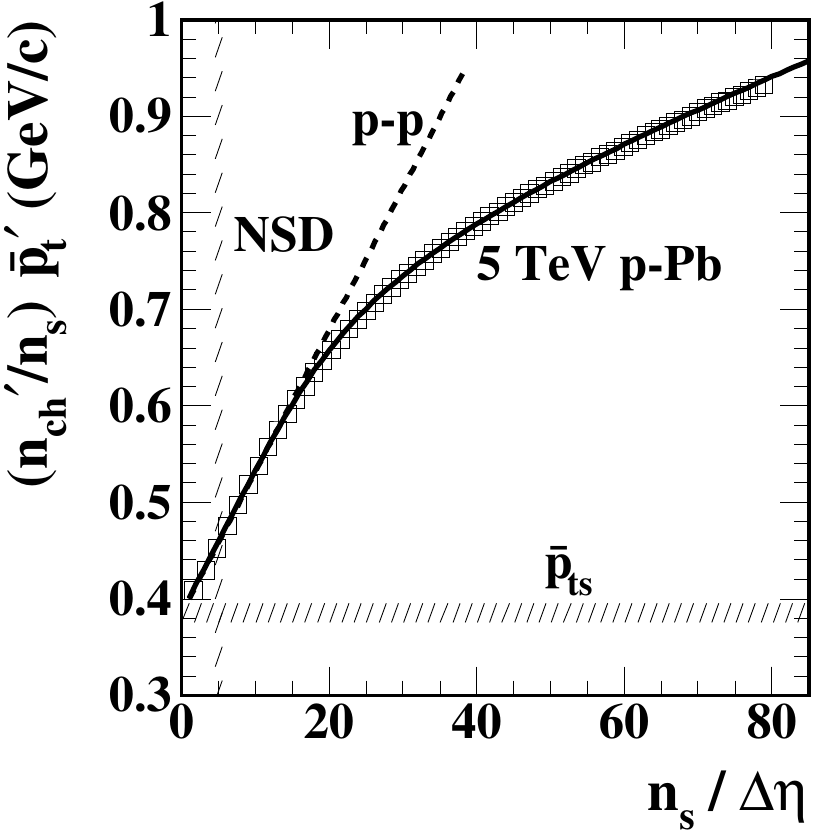}
	\caption{\label{padata}
		Left: Uncorrected ensemble-mean \mmpt\ data from 5 TeV \ppb\ collisions (open squares) vs corrected charge density $\bar \rho_0 = n_{ch} / \Delta \eta$ from Ref.~\cite{alicempt}. Solid and dashed curves are TCM data descriptions from Ref.~\cite{tommpt}. Solid dots are derived from a Glauber analysis~\cite{aliceglauber}.
		Right: Curves and data in the left panel transformed by factor $\bar \rho_0'/\bar \rho_s = n_{ch}' / n_s$ from Eq.~(\ref{prime}).
	} 
\end{figure}

Figure~\ref{padata} (left) shows uncorrected $\bar p_t'$ data for 106 million 5 TeV \ppb\ collisions vs corrected \nch\ (open boxes) from Ref.~\cite{alicempt}. The dashed curve is the TCM for 5 TeV \pp\ collisions given by Eq.~(\ref{ppmpttcm}) with $\alpha = 0.0113$, $\bar p_{ts} \approx 0.4$ GeV/c, $\bar p_{th0} = 1.3$ GeV/c and $\xi = 0.73$~\cite{tommpt}. The solid curve through points is the TCM described by Eqs.~(\ref{pampttcm}) and (\ref{xmodel})  with parameters  $\alpha = 0.0113$ and $\bar p_{th0} = 1.3$ GeV/c held fixed as for 5 TeV \pp\ collisions (assuming no jet modification).  Parameters $\bar \rho_{s0} \approx 3 \bar \rho_{sNSD} \approx 15$ and $m_0 \approx 0.10$  are adjusted to accommodate the \ppb\ data. Exponent $n_1=5$ affects the TCM  only near $\bar \rho_{s0}$.
Solid dots and dash-dotted curve represent a \mmpt\ trend implied by a Glauber analysis of \ppb\ centrality~\cite{aliceglauber}, with $N_{part}$ and $N_{bin}$  taken from Table~\ref{ppbparams1} and with $x = 0.06$, $\bar p_{ts} = 0.4$ GeV/c  and $\bar p_{th0} = 1.3$ GeV/c fixed at their \pp\ values.
 
Figure~\ref{padata} (right) shows data in the left panel converted to $(n_{ch}' / n_s)\, \bar p_t' \approx \bar P_t / n_{s}$ by factor $\xi + x(n_s) \nu (n_s)$ as in Eq.~(\ref{prime}). The dashed line is the TCM for 5 TeV \pp\ collisions defined by Eq.~(\ref{niceeq}). The solid curve is the \ppb\ TCM defined by Eq.~(\ref{pampttcm1}) corresponding also to the solid curve in the left panel. Transforming data from left to right panels requires an estimate of $n_s$ for the $\bar p_t'$ data to evaluate the required conversion factor $\xi + x(n_s) \nu(n_s)$. The map $n_s \rightarrow n_{ch}$ for the TCM from Eq.~(\ref{nchppb}) (second line) is inverted  via linear interpolation to provide the map $n_{ch} \rightarrow n_s$ for data. 
Figure~\ref{padata} demonstrates via \mmpt\ data that the \ppb\ TCM for unidentified-hadron  \pt\ spectra and their centrality evolution is quite accurate.

\subsection{p-Pb geometry inferred from non-PID data} \label{geom}

With the TCM relations derived in the previous subsection it is possible to associate with any measured \ppb\ charge density $\bar \rho_0 = n_{ch} / \Delta \eta$ a complete set of TCM spectrum and geometry parameters. For each value of $\bar \rho_0$ the corresponding $\bar \rho_s$ or $n_s$ is obtained by inverting Eq.~(\ref{nchppb}) (second line). $x(n_s)$ is obtained from Eq.~(\ref{xmodel}) with parameters $N_{part}(n_s)$ and $\nu(n_s)$ as described below Fig.~\ref{paforms}.

Table~\ref{rppbdata} presents geometry parameters for 5 TeV \ppb\ collisions inferred from a Glauber-model analysis in Ref.~\cite{aliceglauber} (primed values) compared to TCM values from the analysis as in Ref.~\cite{tomglauber} and as described above  (unprimed values). The large differences between Glauber and TCM values are explained in Ref.~\cite{tomglauber}. The primed quantities (also see Table~\ref{ppbparams1}) based on a Glauber Monte Carlo study result from assumptions inconsistent with \mmpt\ data. Large differences in fractional cross section $\sigma / \sigma_0$ and binary-collision number $N_{bin}$ are especially notable. Charge densities $\bar \rho_0$ are derived directly from data and correctly characterize the seven centrality classes, but the centralities reported in Ref.~\cite{aliceglauber} may be questioned.

\begin{table}[h]
	\caption{Nominal (primed~\cite{aliceglauber}) and  TCM (unprimed~\cite{tomglauber}) fractional cross sections and  Glauber parameters, midrapidity charge density $\bar \rho_0$, \nn\ soft component $\bar \rho_{sNN}$ and TCM hard/soft ratio $x(n_s)$ used for 5 TeV \ppb\ PID spectrum data.
	}
	\label{rppbdata}
	\begin{center}
		\begin{tabular}{|c|c|c|c|c|c|c|c|c|} \hline
			$\sigma' / \sigma_0$ &   $\sigma / \sigma_0$    & $N_{bin}'$ & $N_{bin}$ & $\nu'$ & $\nu$ & $\bar \rho_0$ & $\bar \rho_{sNN}$ & $x(n_s)$ \\ \hline
			0.025         & 0.15 & 14.7  & 3.20 & 1.87  & 1.52 & 44.6 & 16.6  & 0.188 \\ \hline
			0.075  & 0.24 &  13.0   & 2.59  &  1.86 & 1.43 & 35.9 &15.9  & 0.180 \\ \hline
			0.15  & 0.37 &  11.7 & 2.16 & 1.84 &  1.37 & 30.0  & 15.2  & 0.172 \\ \hline
			0.30 & 0.58 &  9.4 & 1.70 & 1.80  & 1.26  & 23.0  & 14.1  & 0.159  \\ \hline
			0.50   &0.80  & 6.42  & 1.31 & 1.73  & 1.13 & 15.8 &   12.1 & 0.137  \\ \hline
			0.70  & 0.95 & 3.81  & 1.07 & 1.58  & 1.03  & 9.7  &  8.7 & 0.098 \\ \hline
			0.90 & 0.99 &  1.94 & 1.00 & 1.32  & 1.00  &  4.4  & 4.2 &0.047  \\ \hline
		\end{tabular}
	\end{center}
\end{table}

Note that aside from the experimentally-determined $\bar \rho_0$ values all TCM (unprimed) parameters in Table~\ref{rppbdata} are determined by three numbers: $\alpha$, $\bar \rho_{s0}$ and $m_0$. $\alpha$ is defined for all \pp\ collision energies by Eq.~(15) of Ref.~\cite{alicetomspec} based on measured jet properties and is not adjusted for individual collision systems or hadron species. $\bar \rho_{s0}$ and $m_0$ are inferred by comparing \pp\ and \ppb\ \mmpt\ data as in Sec.~\ref{xnsmodel}. That parameter combination represents a transition from individual peripheral \pn\ collisions at lower \nch\ with $N_{part} = 2$ to increase of $N_{part}$ above 2 for higher multiplicities, which in turn depends on the relation between the \pp\ probability distribution on \nch\ and the \ppb\ cross-section distribution on $N_{part}$ (Sec.~\ref{geomestimate}).

The TCM (unprimed) geometry parameters in Table~\ref{rppbdata}, derived from \ppb\ \pt\ spectrum and \mmpt\ data for unidentified hadrons, are assumed to be valid for each identified-hadron species and are used unchanged to process PID spectrum data below. However, certain additions to the spectrum TCM of Sec.~(\ref{unidspec}) are required to accommodate PID data as described next.

\section{$\bf p$-$\bf Pb$ PID spectrum TCM} \label{pidspec}

 PID spectrum data from Sec.~\ref{alicedata} are  analyzed to develop a TCM for each hadron species. Spectrum hard components for the individual hadron species are isolated and compared to \ee\ fragmentation functions for identified hadrons to support inference of jet-related origins.

\subsection{Spectrum TCM for identified hadrons}

To establish a TCM for \ppb\ PID \pt\ spectra it is assumed that (a) \nn\ parameters $\alpha$, $\bar \rho_{sNN}$ and $\bar \rho_{hNN}$ have been inferred from unidentified-hadron data and (b) geometry parameters $N_{part}(n_s)$, $N_{bin}(n_s)$ are a common property (i.e.\ centrality) of \ppb\ collisions independent of detected hadron species.  The required TCM parameters inferred from previous ensemble-mean \mmpt\ data analysis of the same centrality classes for 5 TeV \ppb\ collisions~\cite{tommpt} are presented in Table~\ref{rppbdata}.

Given the \ppb\ spectrum TCM for unidentified-hadron spectra in Eq.~(\ref{ppspectcm}) a corresponding TCM for identified hadrons can be generated by assuming that each hadron species $i$ comprises certain {\em fractions} of soft and hard TCM components denoted by $z_{si}$ and $z_{hi}$  (both $\leq 1$). The PID spectrum TCM can then be written as
\bea \label{pidspectcm}
\bar \rho_{0i}(y_t)
&\approx& \frac{N_{part}}{2} z_{si} \bar \rho_{sNN} \hat S_{0i}(y_t) + N_{bin}  z_{hi} \bar \rho_{hNN} \hat H_{0i}(y_t) 
\nonumber \\
\frac{\bar \rho_{0i}(y_t)}{ \bar \rho_{si}} &=&  \hat S_{0i}(y_t) +  (z_{hi}/z_{si})x(n_s)\nu(n_s) \hat H_{0i}(y_t),
\eea
where unit-integral model functions $\hat S_{0i}(y_t)$ and $\hat H_{0i}(y_t)$ may depend on hadron species $i$.
For identified hadrons of species $i$ the normalization factor $1/ \bar \rho_{si}$ in the second line follows the form of Eq.~(\ref{overrhos}) but can be re-expressed in terms of $1/ \bar \rho_{s}$ for unidentified hadrons already inferred
\bea \label{rhosi}
\frac{1}{\bar \rho_{si}} &=& \frac{1 + (z_{hi}/z_{si}) x(n_s) \nu(n_s)}{\bar \rho_{0i} \equiv z_{0i} \bar \rho_0}
\\ \nonumber
&\approx& \left\{\frac{1 + (z_{hi}/z_{si}) x(n_s) \nu(n_s)}{1 + x(n_s) \nu(n_s)} \right\}  \frac{1}{z_{0i}} \cdot \frac{1}{\bar \rho_{s}}.
\eea
For each hadron species $i$ ratio $z_{hi} / z_{si}$ is first adjusted to achieve coincidence of all seven normalized spectra as $y_t \rightarrow 0$. Parameter $z_{0i}$ is then adjusted to match those rescaled spectra to unit-normal $\hat S_{0i}(y_t)$, also as $y_t \rightarrow 0$. 

Unit-normal model functions $\hat S_{0i}(y_t)$ and $\hat H_{0i}(y_t)$ must be determined for each hadron species; however a close relation to unidentified-hadron models is expected. As noted in Sec.~\ref{tcmmodel} $\hat S_{0i}(y_t)$ is first defined on proper $m_{ti}$ for a given hadron species $i$ and then transformed to $y_{t\pi}$. $\hat H_{0i}(y_t)$ is defined on $y_{t\pi}$ in all cases.

\subsection{$\bf p$-$\bf Pb$ differential PID spectrum data} \label{piddiff}

In the figures below, PID \pt\ spectra from Sec.~\ref{alicedata} are replotted in {\em left panels} in the normalized form of Eq.~(\ref{pidspectcm}) (second line) and compared to TCM soft components $\hat S_{0i}(y_t)$ (bold dotted curves). For each hadron species published spectra are plotted on $y_{t\pi}$ as a logarithmic representation of \pt\ with well-defined zero since the underlying jet (parton) \pt\ spectrum determines the spectrum hard component~\cite{fragevo}. Data spectra on \pt\ are transformed to densities on $y_{t\pi}$ via Jacobian factor $m_{t\pi} p_t / y_{t\pi}$ where $m_{t\pi}^2 = p_t^2 + m_\pi^2$ and $y_{t\pi} = \ln[(m_{t\pi} + p_t)/m_\pi]$. The plot boundaries for full spectra (left) and for spectrum hard components (right) are maintained consistent among the several hadron species to facilitate comparisons. The exception is for kaon data where the left panels are extended down to zero to accommodate the $K^0_S$ data.

\begin{figure}[h]
	\includegraphics[width=3.3in]{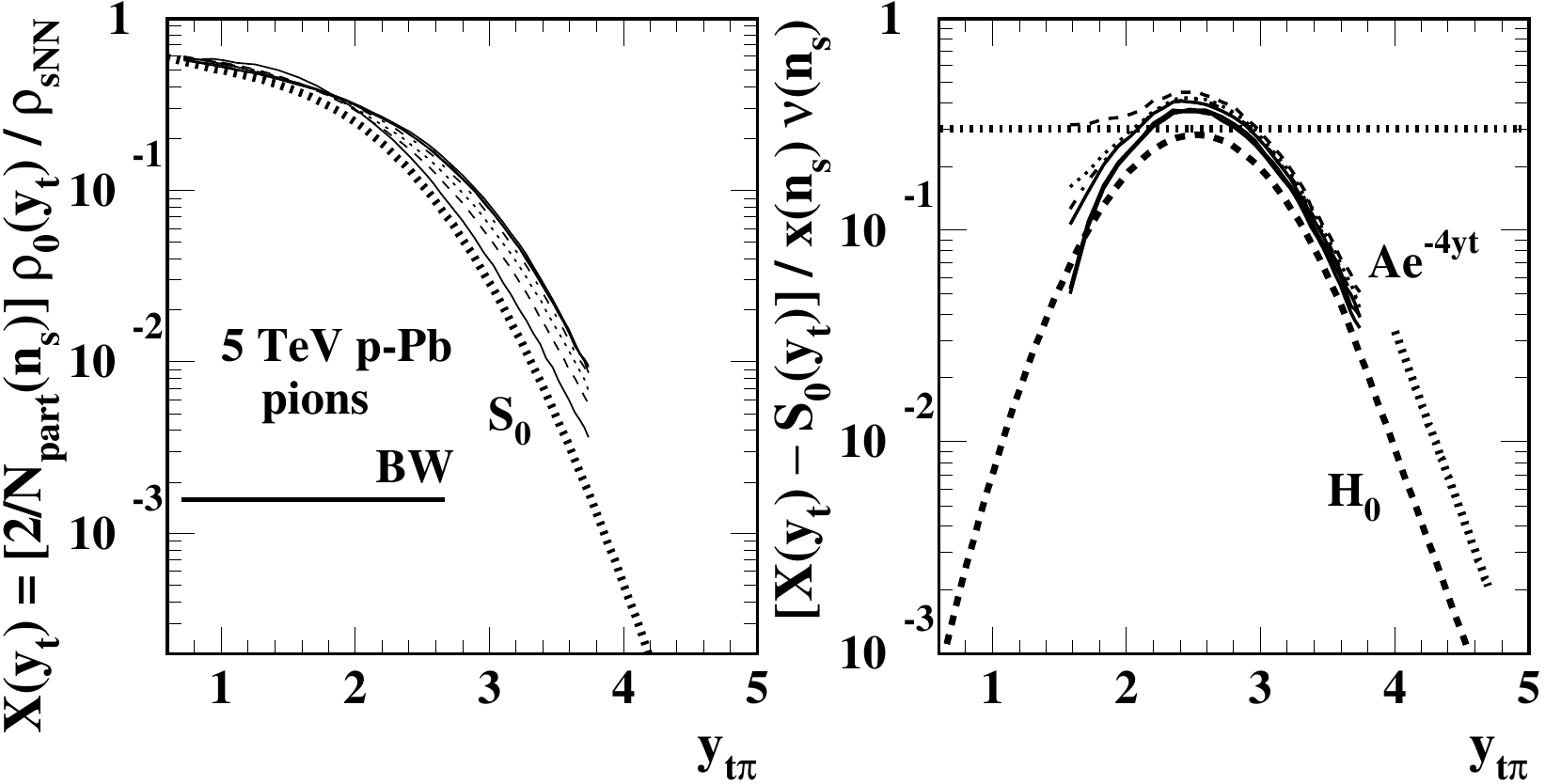}
	\caption{\label{pions}
		Left: Identified-pion spectra for 5 TeV \ppb\ collisions from Ref.~\cite{aliceppbpid} transformed to \yt\ with Jacobian $m_t p_t / y_t$ and normalized by TCM quantities in Table~\ref{rppbdata} (7 thinner curves of several styles). $\hat S_0(y_t)$ (bold dotted) is the soft-component model.
Right: Difference $X(y_t) - \hat S_0(y_t)$ normalized by $x(n_s) \nu(n_s) \approx \alpha \bar \rho_{sNN} \nu(n_s)$ using TCM values from Table~\ref{rppbdata} reported in Ref.~\cite{tommpt} (6 thinner curves, most-peripheral curve is omitted). The bold dashed curve is hard-component model $\hat H_0(y_t)$ with exponential tail. 
		}  
\end{figure}

Figure~\ref{pions} (left) shows identified-pion spectra from Fig.~\ref{piddata} (a). The published spectra have been multiplied by $2\pi$ to be consistent with the $\eta$ densities used in this study. The spectra are then normalized by soft-component density $\bar \rho_{si}$ as defined in Eq.~(\ref{rhosi}) with TCM parameter values reported in Tables~\ref{rppbdata}, \ref{pidparams} and \ref{otherparams}. The normalized spectra $X(y_t)$ can then be compared with spectrum soft-component model $\hat S_0(y_{t\pi})$ shown as the bold dotted curve: a L\'evy distribution defined on $m_{t\pi}$ with parameters $T = 145$ MeV and $n = 8.5$ transformed to $y_{t\pi}$ that also describes unidentified hadrons from 5 TeV \pp\ collisions as reported in Ref.~\cite{alicetomspec}. The solid line labeled BW marks the \pt\ ($y_{t\pi}$) interval over which a BW model fit was imposed as reported in Ref.~\cite{aliceppbpid}.

Figure~\ref{pions} (right) shows difference $X(y_t) - \hat S_0(y_t)$ normalized by $(z_{hi}/z_{si})x(n_s) \nu(n_s)$ using TCM values as reported in Tables~\ref{rppbdata} and \ref{pidparams}. The result should be directly comparable to the \pp\ spectrum hard-component model in the form $\hat H_0(y_t)$ per Eq.~(\ref{pidspectcm}). The bold dashed curve is $\hat H_0(y_t)$ with model parameters  $(\bar y_t,\sigma_{y_t},q)$ for pions as in Table~\ref{pidparams}. The dotted line is a reference to verify that the model is properly normalized.
Any deviations from $\hat H_0(y_t)$ in the right panels are the ``fit'' residuals for the model, but the model is highly constrained with only a few adjustable parameters. There is substantial uncertainty in the  data hard component for the first \nch\ class, so those data are omitted to improve access to the other centrality classes. It is evident that the pion hard components are systematically above the model, by $\approx 40$\%.

\begin{figure}[h]
	\includegraphics[width=3.3in]{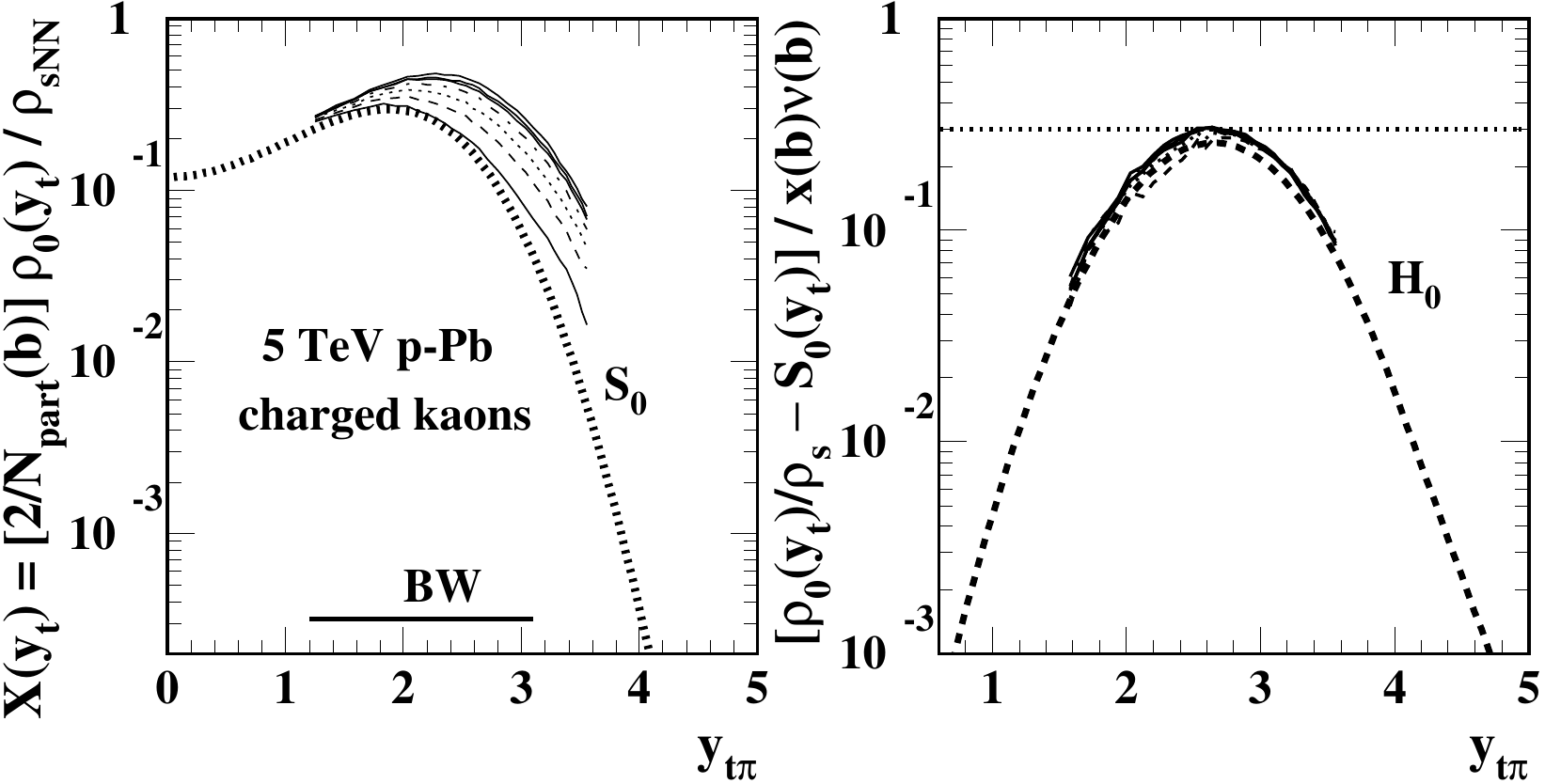}
\put(-142,105) {\bf (a)}
\put(-23,105) {\bf (b)}\\
	\includegraphics[width=3.3in]{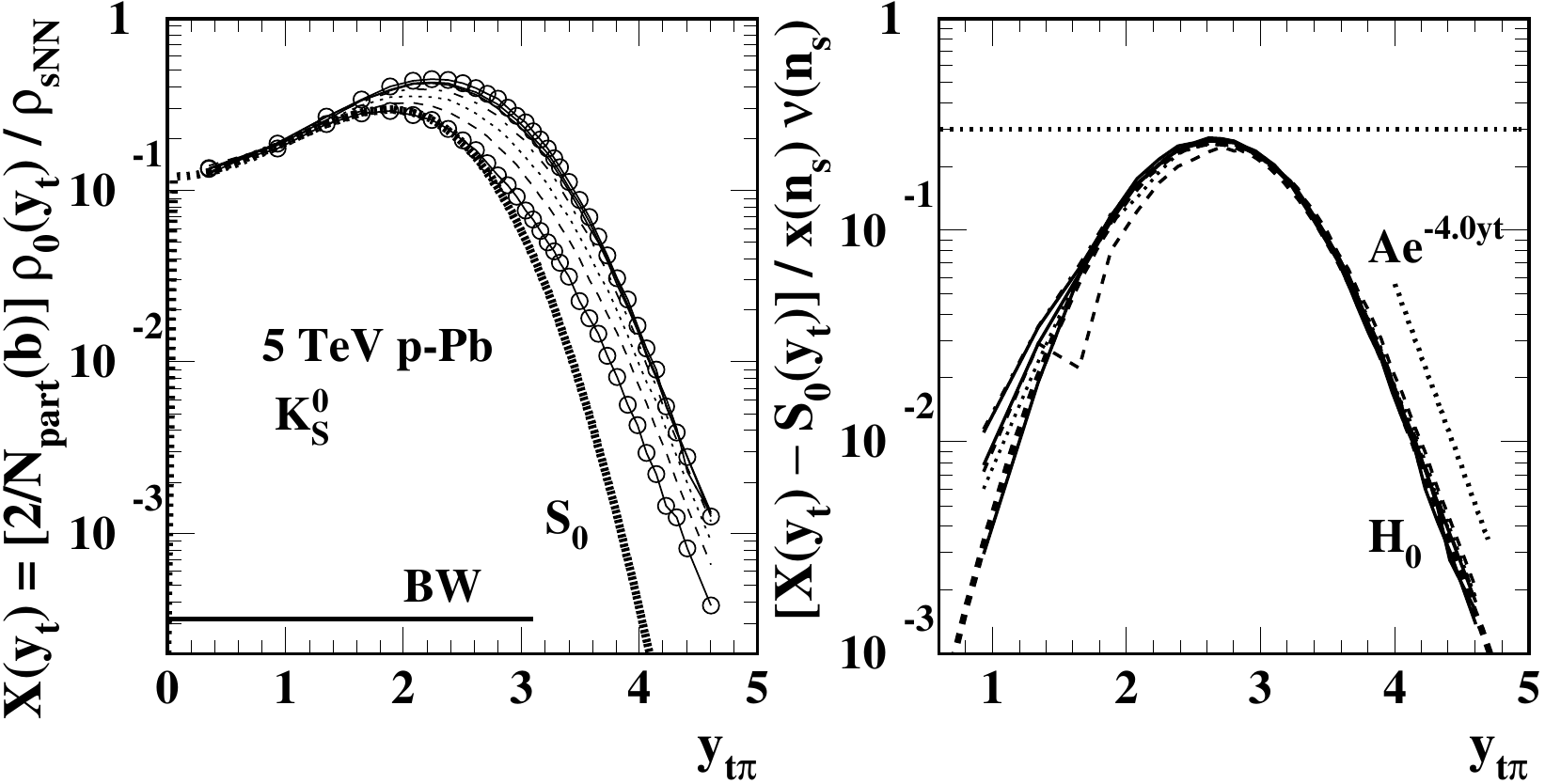}
\put(-142,105) {\bf (c)}
\put(-23,105) {\bf (d)}
	\caption{\label{kch}
 Identified-kaon spectra for 5 TeV \ppb\ collisions from Ref.~\cite{aliceppbpid}: (a), (b) charged kaons $K^\pm$; (c), (d) neutral kaons $K_S^0$. The panel descriptions  are otherwise as for pions. In panel (c) individual data points (open circles) are provided for the most-peripheral and most-central event classes.
}  
\end{figure}

Figure~\ref{kch} shows charged-kaon $K^+ + K^-$ and neutral $K^0_S$  spectra from Fig.~\ref{piddata} (c) and (d) processed in the same manner as for charged pions.
The $K^0_S$ and $K^\pm$ spectra are consistent within data uncertainties as reported in Ref.~\cite{aliceppbpid}. The TCM model functions are therefore constrained to be the same for all kaons. Whereas the $K^\pm$ data are quite limited the $K^0_S$ data subtend the spectacular interval $p_t \in [0,7]$ GeV/c.  $K^0_S$ data below $y_{t\pi} \approx 1.2$ ($\approx 0.2$ GeV/c) demonstrate that only a {\em fixed} soft component independent of \ppb\ centrality contributes in that interval, and a L\'evy distribution on $m_{tK}$ describes the data well. Actual data points (open circles) for the lowest and highest $\bar n_{ch}$ classes are shown. A usable estimate for $\hat H_0(y_t)$ obtained down to $y_t = 1.2$ ($p_t \approx 0.2$ GeV/c) confirms that the TCM hard component drops off sharply below its mode. These PID data strongly support a MB jet-spectrum lower bound near 3 GeV~\cite{fragevo}. 

\begin{figure}[h]
	\includegraphics[width=3.3in]{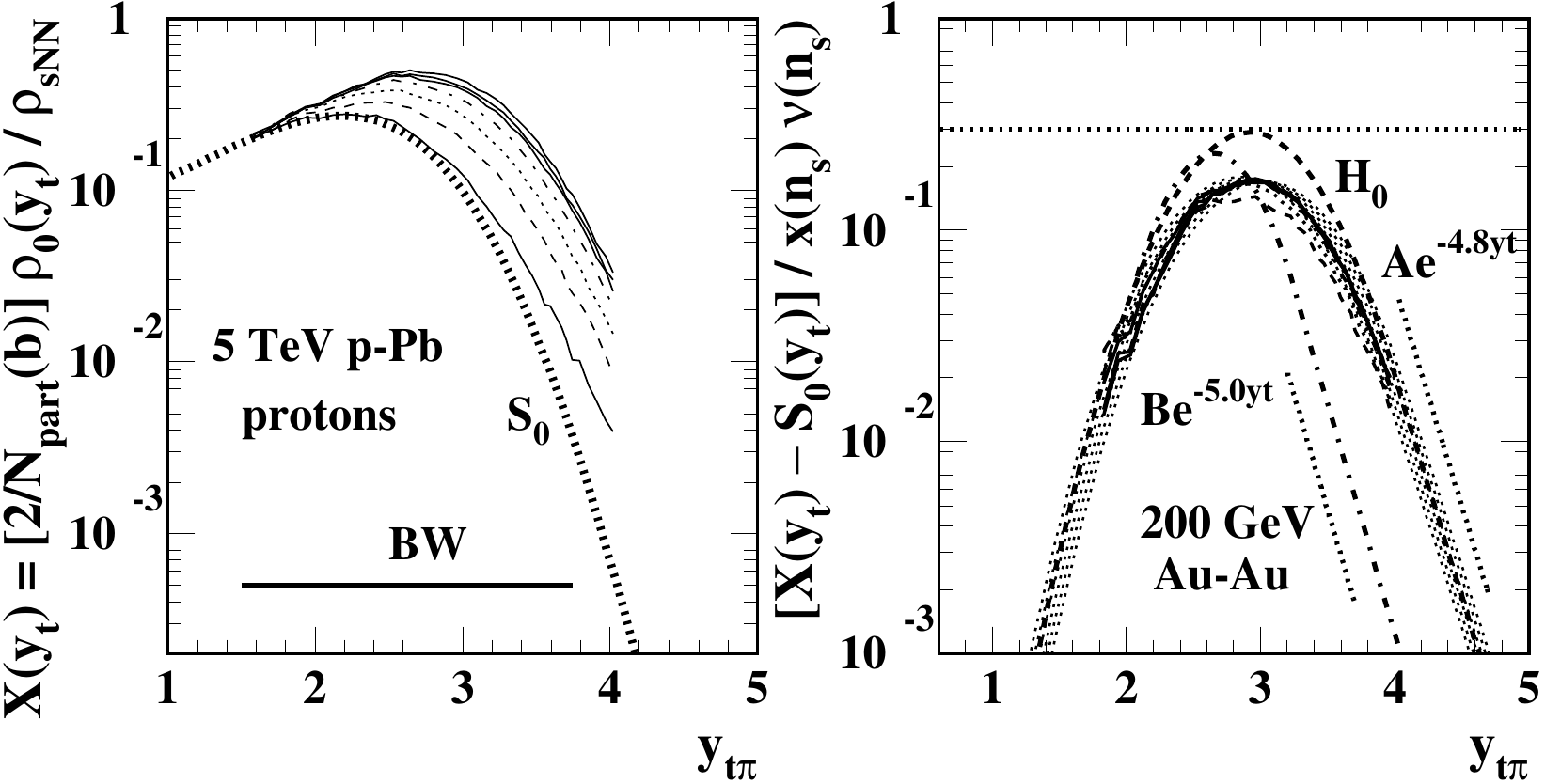}
\put(-142,105) {\bf (a)}
\put(-23,105) {\bf (b)}\\
	\includegraphics[width=3.3in]{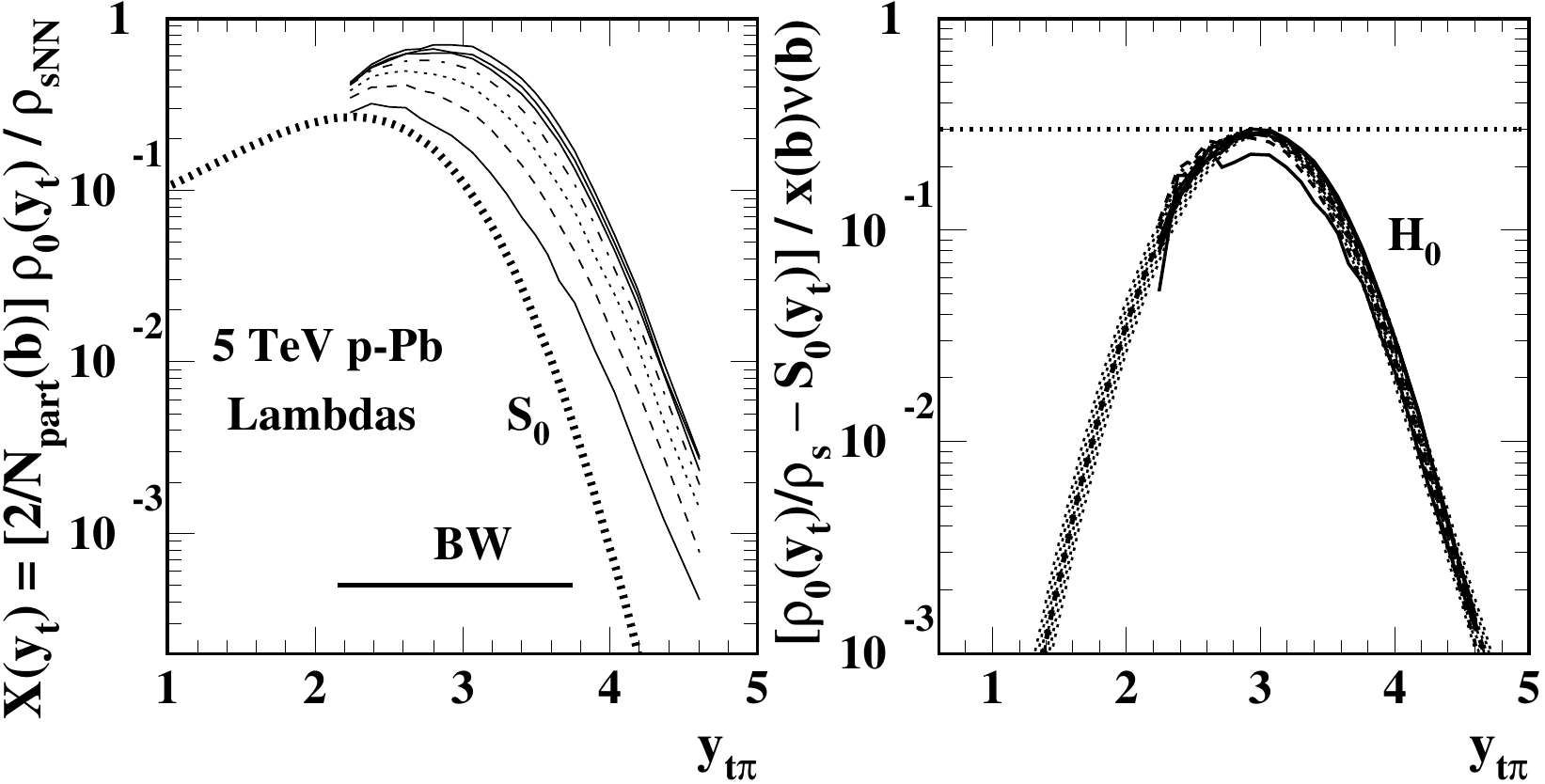}
\put(-142,105) {\bf (c)}
\put(-23,105) {\bf (d)}
	\caption{\label{protons}
(a), (b) Identified-proton $p + \bar p$ spectra for 5 TeV \ppb\ collisions from Ref.~\cite{aliceppbpid}; (c), (d) identified Lambda $\Lambda + \bar \Lambda$ spectra. The panel descriptions are otherwise as for pions. In (b) and (d) the dotted curves follow shifts on \yt\ of the hard components with increasing \nch\ or centrality. The dashed curves are defined by values in Table~\ref{pidparams}. In  (b) the dotted curves are suppressed relative to the nominal $\hat H_0(y_t)$ model (dashed) to accommodate the proton data (see text).
}   
\end{figure}

Figure~\ref{protons} shows proton $p + \bar p$ and Lambda $\Lambda + \bar \Lambda$  spectra from Fig.~\ref{piddata} (e) and (f) processed in the same manner as for charged pions.  Given the complementary \pt\ coverage of the two species the proton data were used to determine $\hat S_0(y_t)$ at lower \yt\ and the Lambda data were used to determine $\hat H_0(y_t)$ at higher \yt. 

The Lambda data are well described out to 7 GeV/c but, as noted in Sec.~\ref{alicedata} regarding Fig.~\ref{piddata} (e), the proton hard components in panel (b) appear to be strongly biased relative to the expected $\hat H_0(y_t)$. Proton soft components appear consistent with the TCM prediction [refer to data in Fig.~\ref{piddata} (e) compared to the TCM solid curves]. In order to accommodate the proton hard components $\hat H_0(y_t)$ is multiplied by  function $1 - a \exp\{- [ (y_t - \bar y_t)/\sigma_{y_{t,corr}} ]^2/2 \}$ with $\bar y_t = 2.92$, $\sigma_{y_{t,corr}} = 0.85$ and $a = 0.40$ which suppresses the hard component {\em over a limited \yt\ interval}. The result is the dotted curves in (b) corresponding to dash-dotted curves in Fig.~\ref{piddata} (e).

In each of the right panels of Fig.~\ref{protons} hard-component model $\hat H_0(y_t)$ appears as multiple curves (dotted) that accommodate systematic shifts of hard-component modes to higher \yt\ with increasing \nch\ or \ppb\ centrality. The shift is clearly apparent for baryons but no shift was required for meson data in this study. Hard-component model $\bar H_0(y_t)$ is modified by $\bar y_t \rightarrow \bar y_t + \delta_n$ where for $n \in [1,7]$ $\delta_n = (n - 4) \delta_0$ with $\delta_0 = $ 0.03 for protons and 0.015 for Lambdas. Consequences  for PID spectrum ratios are discussed in Sec.~\ref{pidratios}.

In contrast to BW fits to the same PID spectrum data as described in Ref.~\cite{aliceppbpid} the TCM spectrum descriptions above do not rely on selecting some limited \pt\ or \yt\ range based on agreement with data (see intervals labeled BW in Figs.~\ref{pions}, \ref{kch} and \ref{protons}) as discussed in Sec.~\ref{bwfits}. The TCM system is accurate, exhaustive and predictive.

\subsection{$\bf p$-$\bf Pb$ TCM PID spectrum parameters} \label{pidfracdata}

\begin{figure}[h]
	\includegraphics[width=1.64in]{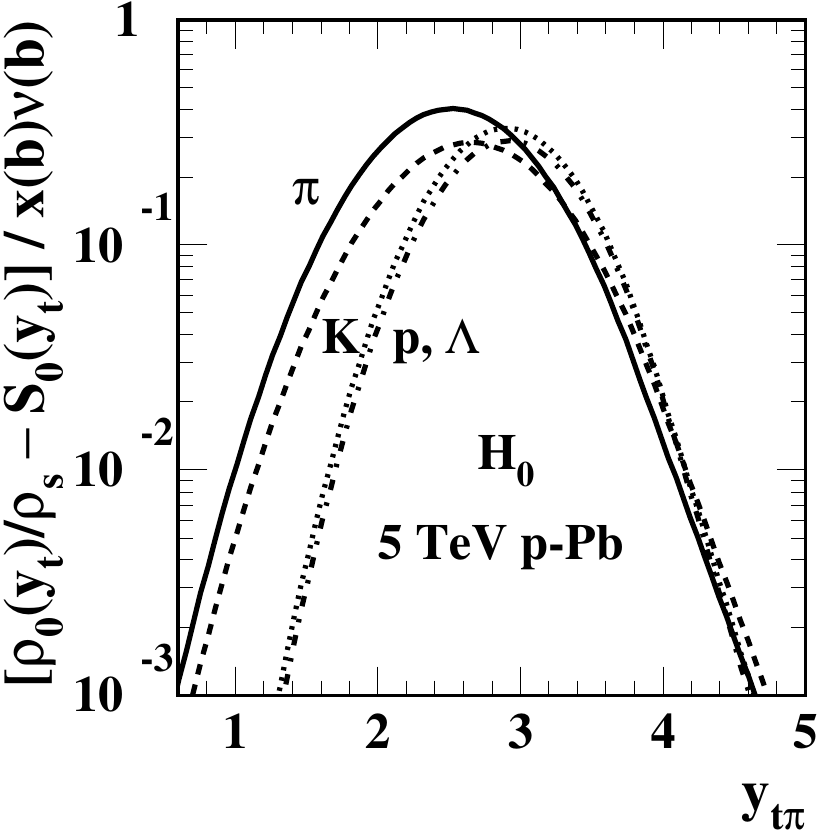}
	\includegraphics[width=1.67in]{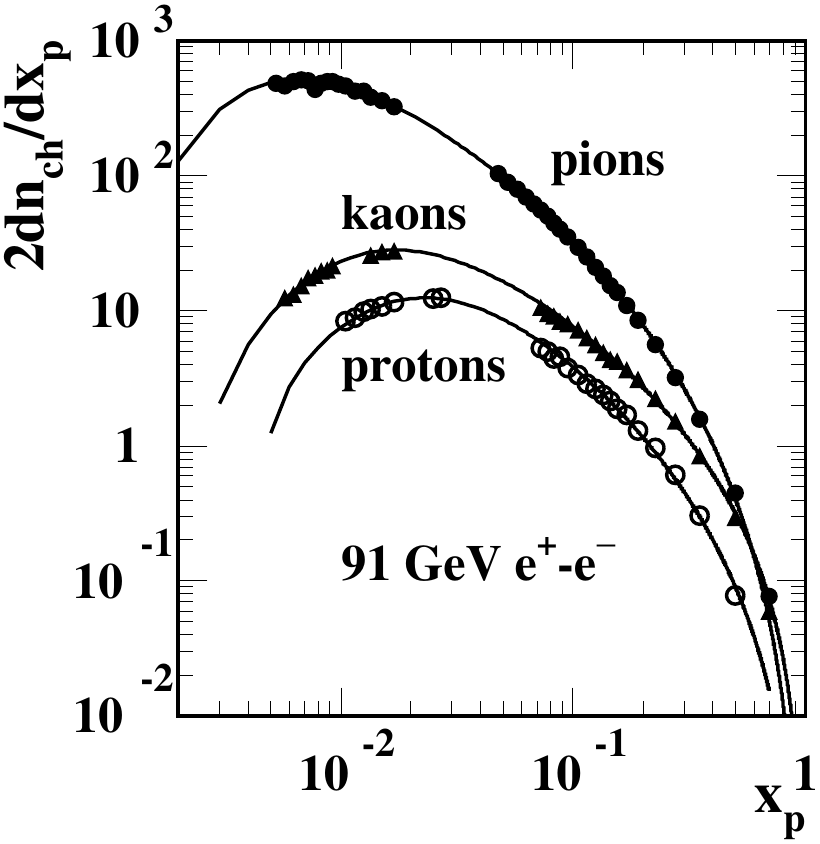}
	\caption{\label{fragdave}
		Left:	Hard-component models $\hat H_0(y_t)$ from Figs.~\ref{pions}, \ref{kch} and \ref{protons}. Models for charged and neutral kaons are assumed identical consistent with spectrum data.
		Right: \ee\ fragmentation functions for identified pions, kaons and protons from unidentified partons (mainly light quarks) from Ref.~\cite{eeprd}.
	} 
\end{figure}

Table~\ref{pidparams} shows TCM model parameters for hard component $\hat H_0(y_t)$ (first three) and soft component $\hat S_0(y_t)$ (last two). Hard-component model parameters vary slowly but significantly with hadron species. Centroids $\bar y_t$ shift to larger \yt\ with increasing hadron mass. Widths $\sigma_{y_t}$ are substantially larger for mesons than for baryons. Only $K_s^0$ and $\Lambda$ data extend to sufficiently high \pt\ to determine exponent $q$ which is substantially larger for baryons than for mesons. The combined centroid, width and exponent trends result in near coincidence among the several models for larger \yt.  Hard components from multiple hadron species all point to a common underlying parton spectrum as demonstrated in Fig.~\ref{fragdave}.

\begin{table}[h]
	\caption{TCM model parameters for unidentified hadrons $h$ from Ref.~\cite{alicetomspec} and for identified hadrons from 5 TeV \ppb\ collisions from this study: hard-component parameters $(\bar y_t,\sigma_{y_t},q)$ and soft-component parameters $(T,n)$. Numbers without uncertainties are adopted from a comparable hadron species with greater accuracy. 
	}
	\label{pidparams}
	\begin{center}
		\begin{tabular}{|c|c|c|c|c|c|} \hline
			& $\bar y_t$ & $\sigma_{y_t}$ & $q$ & $T$ (MeV) &  $n$  \\ \hline
			$ h $     &  $2.64\pm0.03$ & $0.57\pm0.03$ & $3.9\pm0.2$ & $145\pm3$ & $8.3\pm0.3$ \\ \hline
			$ \pi^\pm $     &  $2.52\pm0.03$ & $0.56\pm0.03$ & $4.0\pm1$ & $145\pm3$ & $8.5\pm0.5$ \\ \hline
			$K^\pm$    & $2.65$  & $0.58$ & $4.0$ & $200$ & $14$ \\ \hline
			$K_s^0$          &  $2.65\pm0.03$ & $0.58\pm0.02$ & $4.0\pm0.2$ & $200\pm5$ & $14\pm2$ \\ \hline
			$p$        & $2.92\pm0.02$  & $0.47$ & $4.8$ & $210\pm10$ & $14\pm4$ \\ \hline
			$\Lambda$       & $2.96\pm0.02$  & $0.47\pm0.03$ & $4.8\pm0.5$  & $210$ & $14$ \\ \hline	
		\end{tabular}
	\end{center}
\end{table}

Soft-component model parameter $T \approx 145$ MeV for pions is consistent with that for unidentified hadrons found to be universal over all A-B collision systems and collision energies~\cite{alicetomspec}. The values for higher-mass hadrons are substantially larger. L\'evy exponent $n \approx 8.5$ for pions is also consistent with that for unidentified hadrons at 5 TeV and has a  $\log(\sqrt{s}/\text{10 GeV})$ energy dependence~\cite{alicetomspec}. Soft-component exponent $n$ values for more-massive hadrons are not well-defined because the hard-component fraction is much larger than for pions. Varying $n$ then has little impact on the overall spectra.

Figure~\ref{fragdave} (left) shows hard-component model functions $\hat H_0(y_{t\pi})$ for four hadron species (pions, kaons, protons, Lambdas). The TCM spectrum hard component has previously been interpreted as a fragment distribution arising from MB dijets~\cite{fragevo}.  These similar model shapes further support that interpretation. 
With increasing hadron mass the lower-\pt\ tails and distribution modes move to the right, but the models for protons and Lambdas drop faster on the higher-\pt\ side so the high-\pt\  intercepts coincide, consistent with a common underlying parton \pt\ spectrum having a lower bound near 3 GeV

Figure~\ref{fragdave} (right) shows identified-hadron fragmentation functions (FFs) from unidentified (light) partons [Fig.~7 (left) of Ref.~\cite{eeprd}]. A shift to lower fragment momenta for pions is  expected based on those data. The main trend -- pion FFs are softer than kaon FFs are softer than proton FFs -- is consistent with the left panel. Two conclusions emerge from PID \pt\ spectra: (a)  For all \ppb\ centralities there is no apparent jet modification, no ``jet quenching.'' (b) The MB jet contribution dominates baryon production, which is far greater than  expected from the statistical model, leading to large values for baryon $z_h / z_s$.

Table~\ref{otherparams} shows PID parameters $z_0$ and $z_h / z_s$ for five hadron species that are determined from spectrum data as fixed values independent of centrality.  The choice to hold $z_0$ and $z_h / z_s$ fixed rather than $z_s$ and $z_h$ separately arises from  PID spectrum data structure as follows:
Given the TCM expression in Eq.~(\ref{pidspectcm}) the correct normalization $1/\bar \rho_{si}$ should result in data spectra coincident with $\hat S_0(y_t)$ as $y_t \rightarrow 0$ for all centralities. That condition is not achieved by fixing $z_{si}$. Empirically, the required TCM condition is met by holding $z_{hi} / z_{si}$ and $z_{0i}$ fixed as described in the previous subsection. 

\begin{table}[h]
	\caption{TCM model parameters for identified hadrons from 5 TeV \ppb\ collisions. Numbers without uncertainties are adopted from a comparable hadron species with greater accuracy. Parameters $ \bar p_{ts}$ and $\bar p_{th0}$ are determined by model functions $\hat S_0(y_t)$ and $\hat H_0(y_t)$ with parameters from Table~\ref{pidparams}.  $h$ represents results for unidentified hadrons.
		}
	\label{otherparams}
	\begin{center}
		\begin{tabular}{|c|c|c|c|c|} \hline
			&   $z_0$    &  $z_h / z_s$ &   $ \bar p_{ts}$ (GeV/c)  & $ \bar p_{th0}$ (GeV/c)  \\ \hline
			$ h$        &  $\equiv 1$  & $\equiv 1$  & $ 0.40\pm0.02$ &    $1.30\pm0.03$  \\ \hline
			$ \pi^\pm$        &   $0.70\pm0.02$  & $0.8\pm0.05$  & $0.40\pm0.02$ &    $1.15\pm0.03$  \\ \hline
			$K^\pm $   &  $ 0.125\pm0.01$   &  $2.8\pm0.2$ &  $0.60$&  $1.34$   \\ \hline
			$K_s^0$        &  $0.062\pm0.005$ &  $3.2\pm0.2$ &  $0.60\pm0.02$ &   $1.34\pm0.03$  \\ \hline
			$p $        & $ 0.07\pm0.005$    &  $7.0\pm1$ &  $0.73\pm0.02$&   $1.57\pm0.03$   \\ \hline
			$\Lambda $        &  $0.037\pm0.005$    & $7.0$ &   $0.76\pm0.02$ &    $1.65\pm0.03$ \\ \hline	
		\end{tabular}
	\end{center}
\end{table}

Separate fractions $z_s$ and $z_h$ may be derived from fixed model parameters $z_h / z_s$ and $z_0$ via the relation
\bea
z_s &=& \frac{1 + x(n_s) \nu(n_s)}{1 + (z_h / z_s) x(n_s) \nu(n_s)} z_0,
\eea
and since $z_0$ and  $z_h / z_s$ are held fixed $z_s$ and $z_h$ must then be centrality dependent per $x(n_s) \nu(n_s)$. As a consequence, for hadron species 1 and 2 (with 2 more massive)
\bea \label{zratio}
\frac{z_{x2}}{z_{x1}} &\propto& \frac{1 + (z_{h1}/z_{s1})\, x(x_s)\nu(n_s)}{ 1 +  (z_{h2}/z_{s2})\, x(x_s)\nu(n_s)},
\eea
where $x = s ~\text{or}~h$.  For increasing \ppb\ centrality  those ratios must then {\em decrease} according to parameter values in Table~\ref{otherparams}. If TCM model functions are held fixed independent of \ppb\ \nch\ or centrality then according to Eq.~(\ref{pidspectcm}) (first line) the spectrum ratio of two hadron species must decrease with increasing centrality. If that trend is not observed the assumption of fixed model functions should be questioned, as discussed in Sec.~\ref{pidratios}.

\section{$\bf p$-$\bf Pb$ PID ensemble-mean $\bf \bar p_t$} \label{pidmptt}

The TCM for PID spectra in the previous section may be tested by comparison with measured PID \mmpt\ values. A TCM for ensemble-mean \mmpt\ for unidentified hadrons is described in Sec.~\ref{ppbgeom}. In this section the \mmpt\ TCM is generalized to describe identified hadrons. Just as for PID spectra it is assumed that all hadron species share common \ppb\ geometry parameters $x(n_s)$ and $\nu(n_s)$. 

 The ensemble-mean {\em total} \pt\ for identified hadrons of species $i$ integrated over some angular acceptance $\Delta \eta$ is
\bea \label{ptintid}
\bar P_{ti} &=& \Delta \eta \int_0^\infty dp_t\, p_t^2\, \bar \rho_{0i}(p_t)
~=~ \bar P_{tsi} + \bar P_{thi}
\\ \nonumber
&=& \frac{N_{part}}{2} z_{si}  n_{sNN} \bar p_{tsNNi} + N_{bin} z_{hi} n_{hNN} \bar p_{th0i}.
\eea 
As for unidentified hadrons it is assumed that $\bar p_{tsNNi} \rightarrow \bar p_{tsi}$ is a universal quantity for each hadron species. An ensemble-mean \mmpt\ expression based on the TCM (with $n_{si} = z_{si}n_{sNN} N_{part} / 2$) then has the simple form
\bea \label{pampttcmpid}
\frac{\bar P_{ti}}{n_{si}} &=& \bar p_{tsi} + (z_{hi} / z_{si}) x(n_s)\nu(n_s) \, \bar p_{th0i}.
\eea
The corresponding TCM for conventional ratio $\bar p_{ti}$  is
\bea \label{pampttcmid}
\frac{\bar P_{ti}}{\bar n_{chi}} \hspace{-.05in} &=& \hspace{-.05in}   \bar p_{ti}   \approx \frac{\bar p_{tsi} + (z_{hi} / z_{si}) x(n_s) \nu(n_s) \, \bar p_{th0i}}{1 +  (z_{hi} / z_{si})x(n_s)\, \nu(n_s)},~~~~
\eea
assuming that the \pt\ integral extends down to $p_t = 0$ (by TCM extrapolation of $\bar \rho_{0i}$ data). The lower limit for $\bar p_{ti}$ is then $\bar p_{tsi}$. If the \pt\ acceptance has lower bound $p_{t,cut} > 0$ then $1 \rightarrow \xi_i$ above similar to Eq.~(\ref{pampttcm}).

Based on results in the previous section ratio $z_{hi} / z_{si}$ for each hadron species is held fixed independent of centrality or \nch\ using values from Table~\ref{otherparams}. That table also includes values for $\bar p_{tsi}$ and $\bar p_{th0i}$ derived from model functions $\hat S_{0i}(y_t)$ and $\hat H_{0i}(y_t)$  respectively as defined by parameters in Table~\ref{pidparams}. The centrality parameters, independent of hadron species, are taken from Table~\ref{rppbdata}.

\begin{figure}[h]
	\includegraphics[width=3.3in]{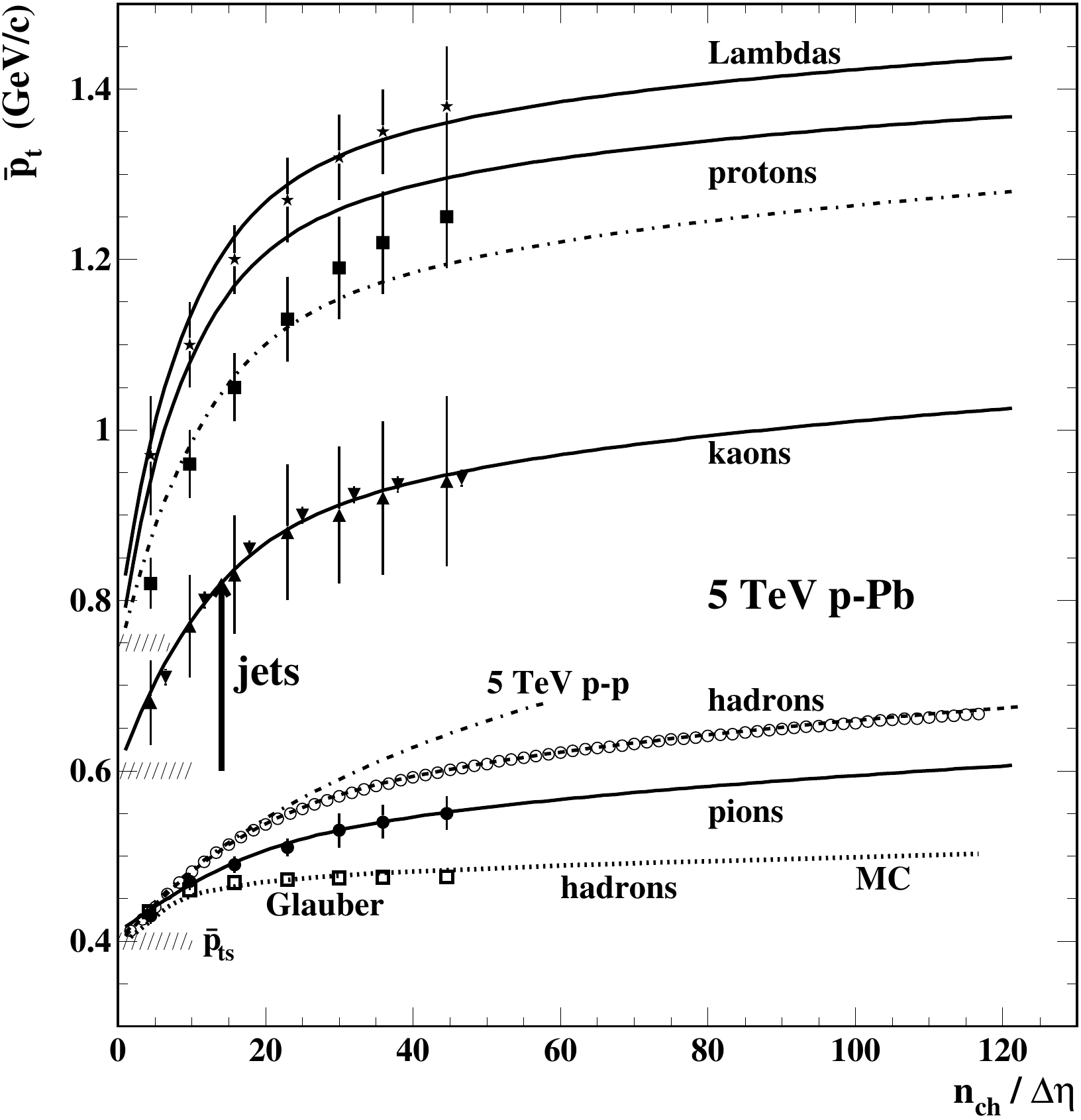}
	\caption{\label{pidmpt}
Corrected $\bar p_t$ vs \nch\ data from Ref.~\cite{aliceppbpid} (solid points) with TCM described by Eq.~(\ref{pampttcmid}) (solid curves).  $K^0_S$ data (inverted triangles) are displaced slightly to the right of the $K^\pm$ data. The proton solid curve is an expectation from TCM systematics. The dash-dotted curve represents apparent suppression of the proton hard component (see text). The open circles are unidentified-hadron data from Ref.~\cite{alicempt} with corresponding TCM (dashed) from Ref.~\cite{tommpt}. \mmpt\ values implied by a Glauber-model analysis in Ref.~\cite{aliceglauber} are represented by  open squares. The dotted curve  MC is derived from Ref.~\cite{alicempt}. The hatched bands indicate $\bar p_{ts}$ values from Table~\ref{otherparams}.
	} 
\end{figure}

Figure~\ref{pidmpt} shows $\bar p_t$ vs \nch\ data from Ref.~\cite{aliceppbpid} (solid points) compared to the TCM described by Eq.~(\ref{pampttcmid}) (solid curves). The dash-dotted curve for protons is Eq.~(\ref{pampttcmid}) with $x(n_s)$ reduced by factor 0.6 to represent the bias effect in Fig.~\ref{protons} (b). Both versions assume fixed TCM model functions. The open circles are $\bar p_t$ data for unidentified hadrons from Ref.~\cite{alicempt} with corresponding TCM from Ref.~\cite{tommpt} (dashed). The dotted curve represents a MC trend derived from Ref.~\cite{alicempt} (Fig.~3). The open squares represent a prediction in Ref.~\cite{tommpt} for unidentified hadrons based on the Glauber \ppb\ centrality analysis from Ref.~\cite{aliceglauber}.  All data and curves are corrected to full \pt\ acceptance. The same $z_h / z_s = 3.0$ value is used for both kaon species. Reported data uncertainties (bars) from Ref.~\cite{aliceppbpid} are dominated by spectrum extrapolations to $p_t = 0$. Point-to-point uncertainties are much smaller. 
Deviations of proton and Lambda data  relative to corresponding TCM curves vary smoothly from lesser to greater, consistent with  shifts on \yt\ of data and model (dotted) hard components in Fig.~\ref{protons} (b) and (d). 

Several conclusions can be drawn from the results in Fig.~\ref{pidmpt}: (a) The strong increase in \mmpt\ values with \nch\ corresponds to the quadratic relation between soft and hard TCM components in \pp\ or \nn\ collisions $\bar \rho_{hNN} \approx \alpha \bar \rho_{sNN}^2$ which in turn follows the quadratic trend for \pp\ dijet production as described in Ref.~\cite{jetspec2}. (b) The strong increase in \mmpt\ values with hadron mass correspond to the properties of parton fragmentation to jets (i.e.\ follows the spectrum hard component associated with jets). (c) Whereas PID \mmpt\ data from Ref.~\cite{aliceppbpid} (12.5 or 25 million events) extend only to $\bar \rho_0 = n_{ch} / \Delta \eta \approx 45$ (0-5\% central \ppb\ collisions reported in Ref.~\cite{aliceglauber}) the \ppb\ \mmpt\ data for unidentified hadrons extend to $\bar \rho_0 \approx 116$ according to Ref.~\cite{alicempt} from the same collaboration (106 million events). (d) \mmpt\ evolution implied by the Glauber-model analysis of Ref.~\cite{aliceglauber} (open boxes) or Monte Carlos based on a Glauber model of \pp\ collisions (dotted curve, e.g.\ PYTHIA) deviate strongly from the \ppb\ data.

\section{$\bf p$-$\bf Pb$ PID spectrum ratios} \label{pidratios}

\begin{figure}[h]
	\includegraphics[width=1.65in]{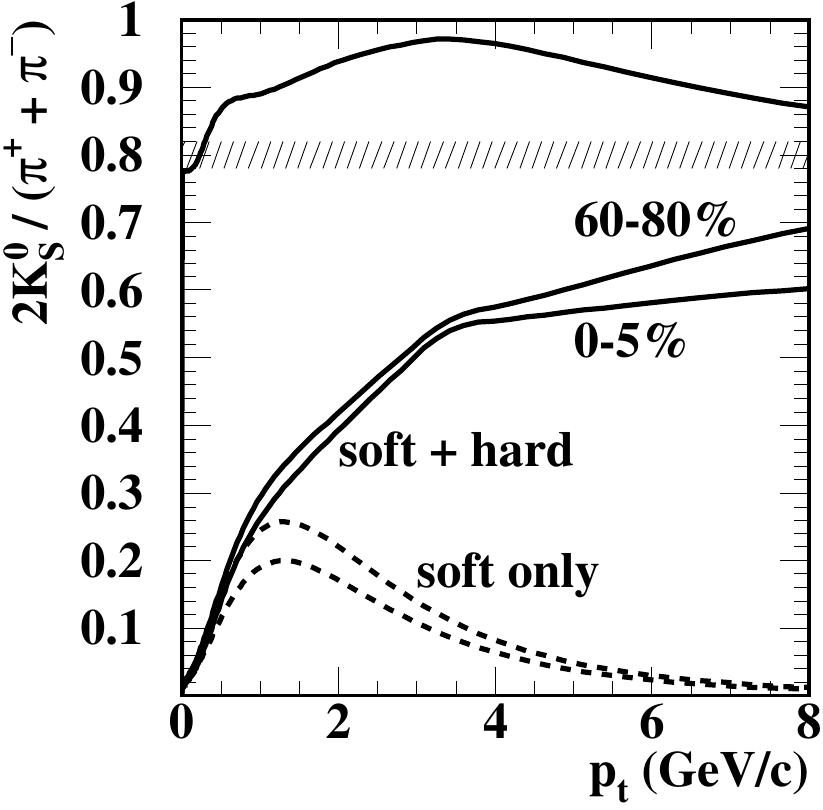}
	\includegraphics[width=1.65in]{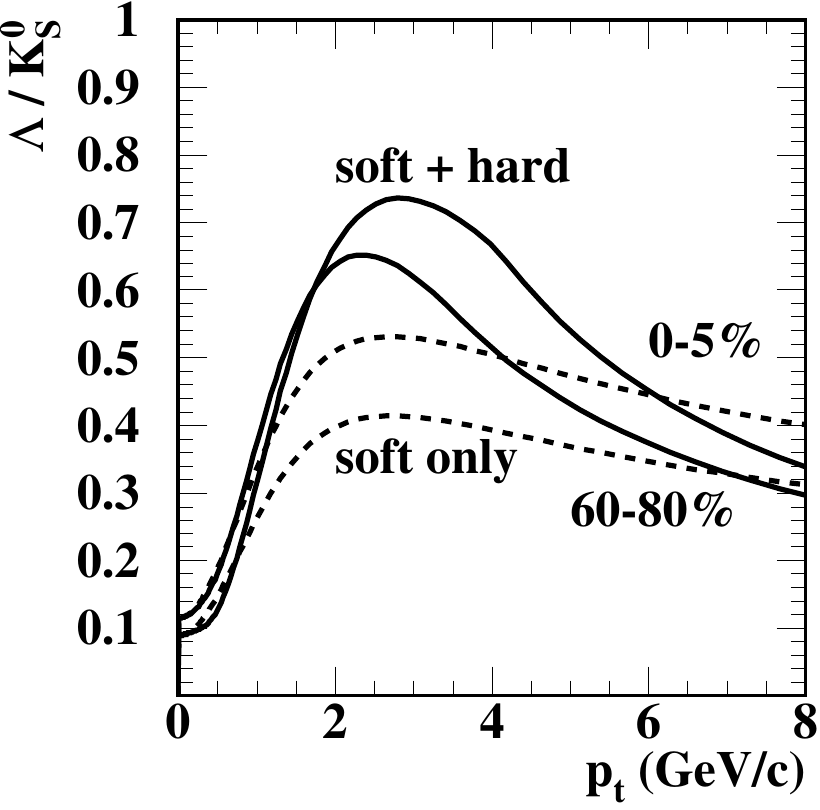}
	\put(-142,105) {\bf (a)}
	\put(-21,105) {\bf (b)}\\
	\includegraphics[width=1.64in]{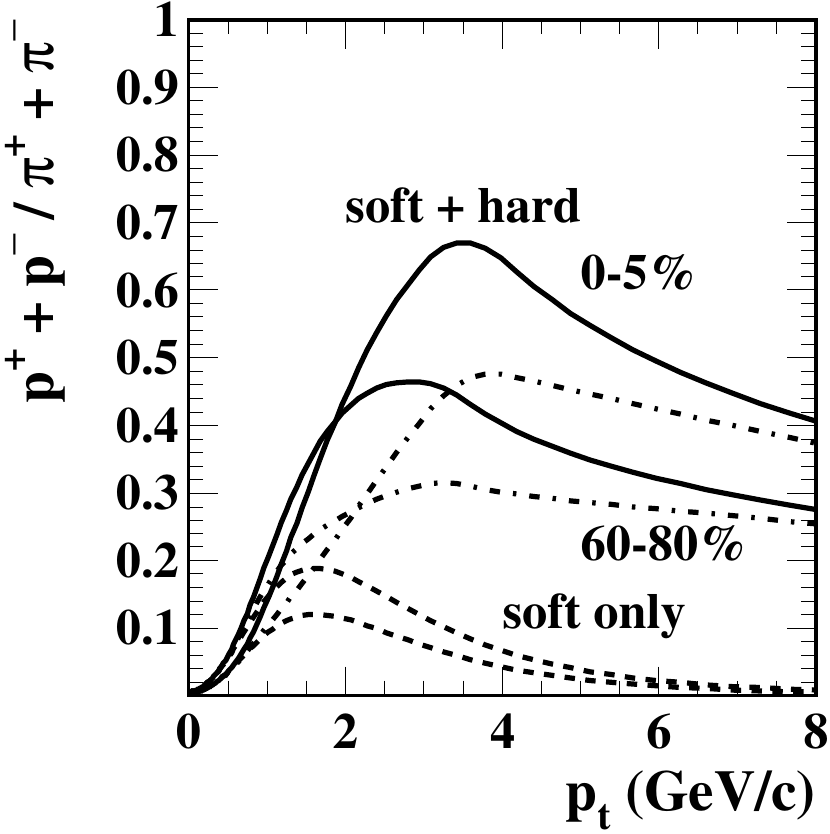}
	\includegraphics[width=1.67in]{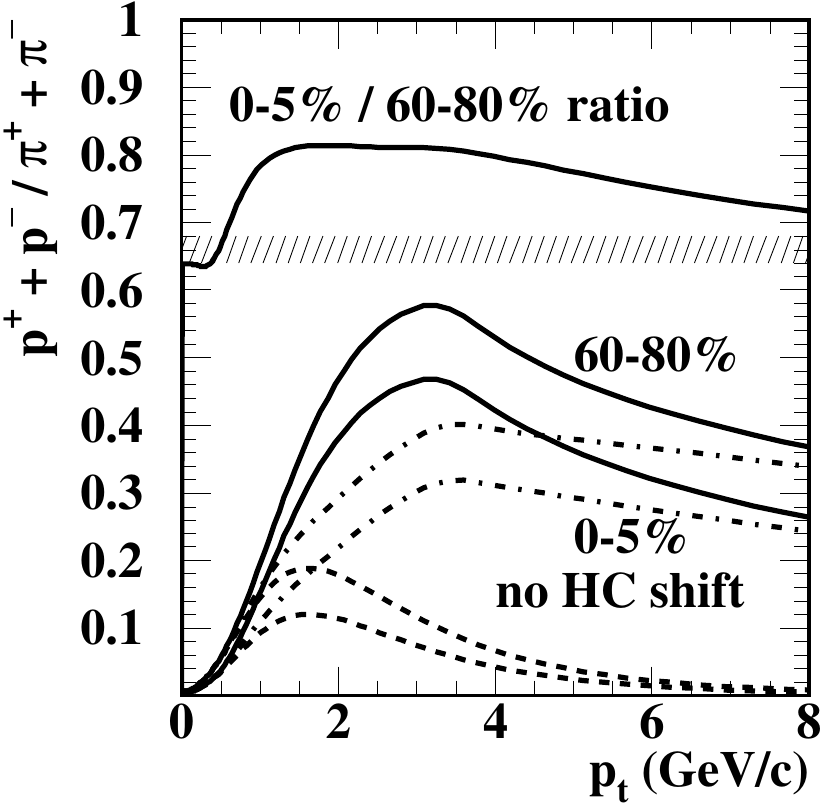}
	\put(-142,105) {\bf (c)}
	\put(-21,105) {\bf (d)}
	\caption{\label{670}
		Spectrum ratios for three combinations of identified hadrons from 5 TeV \ppb\ collisions as derived from the corresponding TCMs. Dashed curves represent soft components only. Other curves represent soft + hard. In panels (c) and (d) the solid curves correspond to expected proton spectra and the dash-dotted curves represent suppression of the hard component to accommodate spectrum data from Ref.~\cite{aliceppbpid}. In panel (d) the TCM hard-component position is fixed whereas in panel (c) the proton hard component shifts to greater \yt\ with increasing centrality following  data as shown in Fig.~\ref{protons} (b). There is no corresponding shift for pions or kaons.
	} 
\end{figure}

Figure~\ref{670} shows spectrum ratios for (a) $2 K_S^0 / (\pi^+ + \pi^-)$,  (b)$\Lambda / K_S^0$ and (c) $(p + \bar p) / (\pi^+ + \pi^-)$ for two \ppb\ centralities, 0-5\% and 60-80\% (as inferred from the Glauber analysis of Ref.~\cite{aliceglauber}). Panel (d) is discussed below. The solid curves are derived from TCM solid curves in Fig.~\ref{piddata} that describe spectrum data well (except for protons). The dash-dotted curves for protons in panel (c) are determined using the biased dash-dotted curves in Fig.~\ref{piddata} (e). The dashed curves are derived from the TCM with hard components omitted. Those three panels can be compared with data in Fig.~2 (left panels) of Ref.~\cite{aliceppbpid}. The agreement between TCM and data (referring also to proton data vs dash-dotted curves) is generally good with no TCM parameter adjustment. However, baryon data description does require relaxing the assumption that TCM model function $\hat H_0(y_t)$ is independent of \nch\ or centrality.

In Sec.~\ref{pidfracdata} the nominal TCM prediction for spectrum ratios is {uniform (on \pt) {\em decrease} with increasing \ppb\ centrality {assuming} TCM model functions independent of centrality, as illustrated in Fig.~\ref{670} (d) where the ratios are generated with  no hard-component shift. The hatched band in panel (d) is an estimate of ratio reduction determined by Eq.~(\ref{zratio}), where the ratio value 0.66 is derived from Table values 0.8 and 7 for $z_h/z_s$ and values for parameters $x$ and $\nu$ of 0.19 and 1.52 for 0-5\% and 0.10 and 1.03 for 60-80\%. The upper solid curve is the ratio of lower solid curves and is consistent with that estimate. 

However, variation of baryon/meson {\em data} ratios with increasing centrality in panels (b) and (c) changes from decreasing to increasing with increasing \pt, as noted in the top panel of Fig.~3 in Ref.~\cite{aliceppbpid}. Differential analysis of spectrum data as in Sec.~\ref{piddiff} provides deeper insight into PID spectrum evolution with \ppb\ centrality.

Spectrum-ratio centrality trends actually result from two effects working in opposition: (a) common reduction of fractions $z_s$ and $z_h$ with increasing centrality as above and (b) the effect of baryon hard components shifting to higher \pt. Spectrum ratios tend to decrease uniformly on \pt\ with increasing \nch\ or centrality because of the trend for decreasing fractions $z_s$ and $z_h$, {assuming} model functions (and their data equivalents) do not vary with centrality. The $z_s$ and $z_h$ trends are determined by spectrum data below 0.5 GeV/c where the hard-component contribution is negligible. The nearly-uniform (on \pt) decrease is illustrated in Fig.~\ref{670} (d).

In Fig.~\ref{protons} (b) and (d)  spectrum hard components for protons and Lambdas shift significantly to higher \yt\ with increasing \nch, and the TCM equivalents are shifted to accommodate the data. The difference between panels (c) and (d) in Fig.~\ref{670} is consistent with the spectrum hard component for protons shifting to substantially higher \yt\ with increasing \nch\ while no significant shift for pions is observed in Fig.~\ref{pions} (or for kaons in Fig.~\ref{kch}). The effects of such shifts for unidentified hadrons were {observed already for \pp} collisions in Refs.~\cite{alicetomspec,tommpt}. The contrast of effects in the lower panels of Fig.~\ref{670} should be larger for \pp\ collisions where all of the \nch\ variation contributes to quadratic increase of dijet production, with resulting substantial bias of \mmpt\ trends as demonstrated in Ref.~\cite{tommpt}.

Reference~\cite{aliceppbpid} interprets ratio data in the context of \pbpb\ spectrum ratios (right panels in Fig.~2 of Ref.~\cite{aliceppbpid}) as follows: Again arguing by analogy there is ``significant enhancement [of spectrum ratios] at intermediate $p_T \sim 3$ GeV/c, qualitatively reminiscent of that measured in \pbpb\ collisions. The latter [ratio trends, i.e.~``baryon / meson puzzle''] are generally discussed in terms of collective flow or quark recombination.'' But a different interpretation is indicated by TCM analysis of ratio data.

As demonstrated especially for proton/pion data in panel (c) (solid vs dashed curves) the peak structure near 3 GeV/c is dominated by spectrum hard components definitively associated with MB jet production~\cite{hardspec,fragevo,ppquad,mbdijets}.  In \ppb\ collisions the spectrum-ratio centrality trend [e.g.\ panel (c) solid or dash-dotted curves] results from a baryon hard component shifting to higher \pt\ while a meson hard component exhibits negligible shift.  In contrast, for \pbpb\ collisions the dominant variation with centrality is the {\em meson} (e.g.~pion) hard component shifting to {\em lower} \pt\ while the baryon (e.g.~proton) hard component shifts only slightly to higher \pt\ as demonstrated in Ref.~\cite{hardspec}. These TCM results then demonstrate that evolution of spectrum ratios as in Fig.~\ref{670} is dominated in any A+B collision system by MB jet production, differently for different hadron species.

\section{Systematic uncertainties} \label{sys}

Uncertainties for \ppb\ collision-geometry determination, TCM spectrum model functions and accuracy of spectrum models for identified-hadron data are discussed in the context of the TCM as a lossless data-compression strategy with a small number of degrees of freedom.

\subsection{TCM degrees of freedom} \label{tcmdof}

In contrast to a typical MC model with tens of parameters readjusted (tuned) to each individual collision system the TCM includes only a few parameters applied self-consistently to a broad array of collision systems with little or no individual adjustment. The TCM can be seen as a form of data compression: a large number of collision systems and data formats is represented by a small number of tightly-constrained parameters. The result is a simple global data model with predictive power.

For \pp\ spectra the parameters are hard/soft ratio $\alpha$ and \yt\ model parameters $(T,n;\bar y_t,\sigma_{y_t},q)$. The energy  and \nch\ systematics of \yt\ model parameters are described in Ref.~\cite{alicetomspec} covering a span from 17 GeV to 13 TeV. For unidentified hadrons $T \approx 145$ MeV is universal, $\bar y_t$ and $\sigma_{y_t}$ variations are small or negligible and $n$ and $q$ vary as $\log(\sqrt{s})$ as expected for QCD processes. The energy dependence of $\alpha$ depends on measured jet properties and is also described in Ref.~\cite{alicetomspec}. For \ppb\ collisions additional model parameters $\bar \rho_{s0}$ (\pn--\ppb\ transition point) and $m_0$ [$x(n_s)$ slope reduction factor] are introduced to accommodate \mmpt\ data~\cite{tommpt}.  For \pp\ and \pa\ collisions jet formation is assumed to be unmodified: The spectrum hard component is then approximately invariant on \nch\ in agreement with data.
Spectrum and \mmpt\ data are typically described within point-to-point uncertainties.

Within a composite TCM, A-B centrality is factorized from \nn\ (\pp) densities which are factorized from \pt\ or \yt\ dependence leading to a simple model with largely-independent degrees of freedom that can be evaluated accurately. In contrast, within a one-component model such as a QGP/flow model (all soft) or PYTHIA (all hard) multiple physical mechanisms may not be properly distinguished leading to complexity, misinterpretations and substantially increased parameter uncertainties.

For the present identified-hadron study two additional parameters are introduced, density fractions $z_s$ and $z_h$ as the combinations $z_h/z_s$ and $z_0$, and the TCM \yt\ model parameters are determined individually for each hadron species. However, the combinations $z_h/z_s$ and $z_0$ are constrained to be independent of centrality and \yt, and centrality parameters are maintained independent of hadron species, thereby achieving factorization of the TCM.

In terms of degrees of freedom the geometry parameters in Table~\ref{rppbdata} depend on only three parameters, $\alpha(\sqrt{s})$, $\bar \rho_{s0}$ and $m_0$. The last two are derived from \ppb\ \mmpt\ data in Ref.~\cite{tommpt} but the first is determined by a simple $\log(\sqrt{s})$ function derived from \pp\ data in Ref.~\cite{alicetomspec}. Parameters in Table~\ref{pidparams} include those for unidentified hadrons $h$ derived from \pp\ data as in Refs.~\cite{fragevo,ppquad} and for PID data. For the latter there are three lines (pions, kaons, baryons) of five parameters each for a total of 15. However, the hard-component parameters are closely correlated with measured PID fragmentation functions as in Fig.~\ref{fragdave}.  For the parameters in Table~\ref{otherparams} there are again three lines (pions, kaons, baryons) of two parameters each, $z_0$ and $z_h/z_s$, for a total of 6 that are newly derived from the present study and can be contrasted with the corresponding many \pt-dependent PID parameter values represented by Fig.~9 of Ref.~\cite{aliceppbpid}.  Because of correlations among PID parameters the actual number of degrees of freedom in the 5 TeV \ppb\ PID TCM is substantially less  than $15 + 6 + 2 = 23$.

\subsection{$\bf p$-Pb geometry estimation} \label{geomestimate}

As noted in Sec.~\ref{geom} accurate centrality determination is essential to establish a TCM for any A-B collision system. Entries in Table~\ref{rppbdata} reveal major discrepancies between Glauber (primed) and TCM (unprimed) centrality parameters suggesting large uncertainties in \ppb\ geometry estimation. However, available evidence indicates that the TCM version is quite accurate as argued here.

The TCM for 7 TeV \pp\ collisions provides a self-consistent description of yields, spectra and two-particle correlations over an \nch\ range corresponding to 100-fold increase in dijet production for $\bar \rho_0$ increasing to more than ten times its NSD value 6~\cite{ppprd,ppquad}. Quantitative relations among jets, 2D angular correlations and spectrum hard components have been established. A central element of the TCM is the quadratic relation $\bar \rho_h \approx \alpha \bar \rho_s^2$ between jet production per $\bar \rho_h$ and low-$x$ participant gluons per $\bar \rho_s$.

Figure~\ref{padata} (left) demonstrates close agreement between the TCM description for \ppb\ collisions (solid curve) and \mmpt\ data (open squares). As noted above, the \ppb\ \mmpt\ trend coincides with that for \pp\  collisions (dashed) up to $\bar \rho_0 \approx 20$ or four times the 5 TeV NSD value, implying that \ppb\ collisions within that interval are nearly equivalent to single peripheral \pn\ collisions (with increasing \nch) or that $N_{bin} \approx 1$. In contrast, the Glauber trend for $N_{bin}'$ in Table~\ref{rppbdata} increases in the same interval to greater than 6 and coincides with an implicit assumption that all \pn\ collisions retain the same mean \nch\ and other properties. If that were literally true the result would be the solid points in the left panel deviating greatly from \ppb\ data, with the restriction $\bar \rho_0 < 50$ for 0-5\% central collisions.

The TCM for charge densities from A-B systems is
\bea \label{nchppbx}
\frac{2}{N_{part}} \bar \rho_0 &=&  \bar \rho_{sNN}(n_s) + \nu(n_s) \bar \rho_{hNN}(n_s),
\eea
and in relation to the Glauber analysis of Ref.~\cite{aliceglauber} three cases can be considered: 
(a) $N_{part} \propto \bar \rho_0$ per Ref.~\cite{aliceglauber},
(b) $\bar \rho_{sNN}$ and $\bar \rho_{hNN}$ remain fixed but $\nu$ varies, and 
(c) the full TCM with all components is inferred from \mmpt\ data.

Figure~\ref{tcmerror} (left) shows Glauber parameter $N_{part}/2$ for 5 TeV \ppb\ collisions from Ref.~\cite{aliceglauber} (solid dots, primed values in Table~\ref{rppbdata}) and corrected Glauber values from Ref.~\cite{tomglauber} (open circles). The dashed curve is $N_{part}/2 = \bar \rho_0 / 4.5$ (the NSD value for $\bar \rho_0$ is 5.0). The solid curve is the full TCM described in Sec.~\ref{ppbgeom}. The short hatched bands indicate limits on two models based on the Glauber MC simulation (near $\bar \rho_0 = 50$) and on \mmpt\ data (near 120).

\begin{figure}[h]
	\includegraphics[width=1.65in]{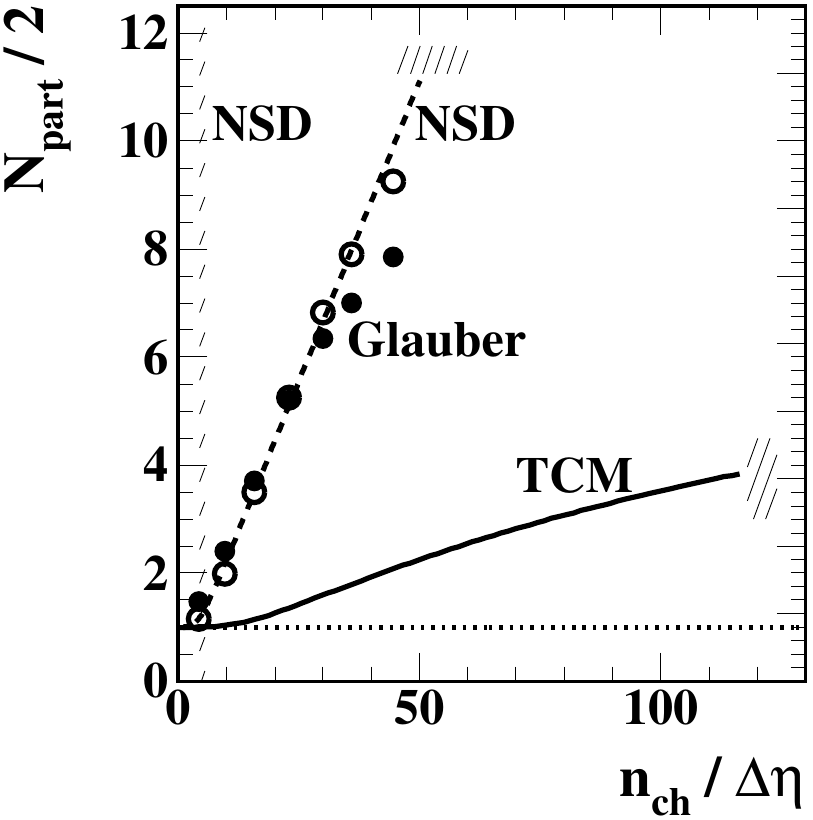}
	\includegraphics[width=1.65in]{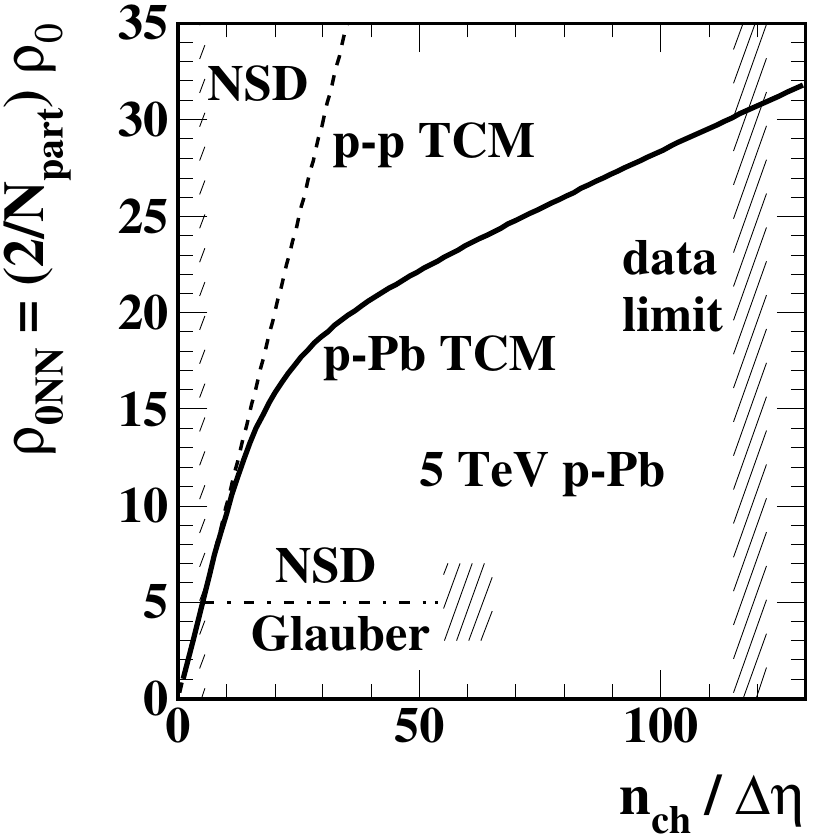}
	\caption{\label{tcmerror}
		Left:  Participant-pair number $N_{part}/2$ vs charge density $\bar \rho_0$ from a Glauber study of 5 TeV \ppb\ collisions in Ref.~\cite{aliceglauber} (points) and from a TCM that describes \mmpt\ data reported in Ref.~\cite{tomglauber} (solid). The dashed curve is $N_{part}/2 = \bar \rho_0 / 4.5$. The hatched bands denote effective limits for the two trends.
		Right:  Hadron production per participant pair: trends from the Glauber study in Ref.~\cite{aliceglauber} (dash-dotted), from the  \ppb\ \mmpt\ TCM (solid) and from the \pp\ TCM (dashed).
	} 
\end{figure}

Figure~\ref{tcmerror} (right) shows $\bar \rho_{0NN} \equiv (2/N_{part})\bar \rho_0$ (vs $\bar \rho_0$) which, for the Glauber-model analysis of Ref.~\cite{aliceglauber}, is assumed constant near the NSD value 5 as in case (a) (dash-dotted). In contrast, the \ppb\ TCM trend of case (c) follows that for \pp\ (dashed) with $N_{part}/2 \approx 1$ over a substantial interval as required by \mmpt\ data. Above a transition point ($\bar \rho_{s0} \approx 15$)  $\bar \rho_{0NN}$ continues to increase linearly but with reduced slope ($m_0 \approx 0.1$) as $N_{part}$ increases above 2. Near the limit of \mmpt\ data at $\bar \rho_0 \approx 120$ (and $N_{part} \approx 8$) $\bar \rho_{0NN} \approx 30$ is just half the value 60 reached by isolated 7 TeV \pp\ collisions for \mmpt\ data as in Ref.~\cite{alicempt} and App.~\ref{ppmptapp} but implies {\em 36-fold increase in dijet production} compared to the NSD value $\approx 5$ implicit in the Glauber analysis.

The large discrepancies between Glauber model and TCM can be interpreted as follows: The relative variations of $N_{part}$ and $\bar \rho_{0NN}$ with \nch\ condition (or \ppb\ centrality) depend on the relation between probability distributions on those parameters. The assumption $N_{part} \propto \bar \rho_0$ in Ref.~\cite{aliceglauber} (points in the left panel) implies that the distribution on $N_{part}$ (as simulated by the Glauber MC) is much broader than that on $\bar \rho_{0NN}$, so with increasing \nch\ variation of $N_{part}$ is rapid while variation of $\bar \rho_{0NN}$ is negligible. The TCM  result implies that  the distribution on $N_{part}$ must be much {\em narrower} than that on $\bar \rho_{0NN}$. 

A study in Reference~\cite{tomglauber} concludes from \mmpt\ data that the distribution on $N_{part}$ is indeed quite narrow compared to a \pp\ probability distribution on \nch, and much narrower than the Glauber MC result. For smaller \nch\ the cases (a) and (b) relating to Eq.~(\ref{nchppbx}) are clearly excluded by the coincidence of \pp\ and \ppb\ data in Fig.~\ref{padata} (left); $N_{part}$ must remain near 2 for lower \nch, contradicting the Glauber MC. Above the transition the TCM and Glauber model represent limiting cases for $N_{part}$, but \mmpt\ data also strongly favor the TCM distribution there.

A follow-up study in Ref.~\cite{tomexclude} offers an explanation: The Glauber approach with eikonal approximation estimates the number of \pn\ {\em geometric encounters} during projectile passage through a target nucleus. Each such encounter is then assumed to be an actual \pn\ collision, with multiple {\em simultaneous} collisions not only possible but likely. The result is the Glauber trends in Fig.~\ref{tcmerror}. Reference~\cite{tomexclude} suggests that simultaneous \pn\ collisions are excluded, in which case the maximum $N_{part}$ value for \ppb\ is near 8 as for the TCM trends in Fig.~\ref{tcmerror}. Exclusion of multiple simultaneous collisions is also consistent with the quadratic relation $\bar \rho_h \approx \alpha \bar \rho_s^2$ for \pp\ collisions that implies {\em full overlap for any actual \pn\ collision}.

The consequence for uncertainty estimation is that based on \mmpt\ data the actual trend for $N_{part}$  vs $\bar \rho_0$ must be close to the TCM trend in Fig.~\ref{tcmerror} (left) as a lower limit. For more-peripheral collisions the uncertainty is negligible due to the constraints of \mmpt\ data. For more-central collisions the possibility of additional mechanisms for \mmpt\ variation (jet modification, flows) could increase uncertainties, but indications from spectrum analysis (e.g.\ the present study) exclude a flow contribution. Observed spectrum hard-component trends indicate that jet production is effectively unmodified in \ppb\ collisions.

\subsection{TCM model functions}

Parameter values for PID TCM model functions on \yt\ are shown in Table~\ref{pidparams}. Coverage or acceptance on \yt\ for various hadron species varies greatly, leading to quite different uncertainties. For pions $q$ is poorly determined because the spectrum terminates at 3 GeV/c near the transition point from Gaussian to exponential tail. Gaussian parameters $\bar y_t$ and $\sigma_{y_t}$ are better determined and are consistent with the trend for unidentified hadrons from Ref.~\cite{alicetomspec}, as are soft-component parameters $(T,n)$.

The charged-kaon data are similarly restricted on \yt\ and cannot therefore compete with the neutral-kaon data with their large acceptance $p_t \in [0,7]$ GeV/c. Since spectra for the two species are reported to be statistically equivalent where they overlap (confirmed in the present study) the parameters for charged kaons $K^\pm$ are simply copied from those for neutral kaons $K_S^0$. The uncertainty for soft-component parameter $n$ is large because the hard component makes a large contribution to spectra, thus reducing sensitivity to the tail of the L\'evy distribution.

Because protons and Lambdas have similar masses their spectra are expected to be similar in form. Their \yt\ coverage is complementary in that Lambda data also extend to 7 GeV/c but proton data extend to lower \pt. Those differences are reflected in the parameter uncertainties. $(T,n)$ are determined by proton data whereas $(\bar y_t,\sigma_{y_t})$ are determined by Lambda data.  The baryon high-\yt\ tails drop much faster than those for mesons so $q$ is substantially less certain (and larger) for baryons.

In Sec.~\ref{piddataa} it is observed that the proton TCM does not describe spectrum data properly. The difference is shown in Fig.~\ref{protons} (b) and problems seem confined to the spectrum hard component. As noted in Sec.~\ref{tcmdof} the TCM is constrained such that parameters are not adjusted arbitrarily for individual collision systems or hadron species. The proton TCM provides a prediction from which the data deviate strongly. The deviation source is not evident but simple tracking inefficiency seems unlikely: The soft component is unaffected [see Fig.~\ref{piddata} (e) below 1 GeV/c], and the suppression is exactly centered on the hard-component peak [see Fig.~\ref{protons} (b)].

\subsection{Accuracy of PID spectrum parameters}

Given the PID spectrum \yt\ model parameters in Table~\ref{pidparams} those related to species abundances for soft and hard components are presented in Table~\ref{otherparams}. As noted above, those parameters are determined by spectrum properties for $y_t \rightarrow 0$ and are thus insensitive to hard-component properties. However, their values determine the relation of spectrum soft and hard components via $z_h / z_s$. The accuracy of the parameter values is then reflected in TCM descriptions of spectrum and \mmpt\ data.

Roughly speaking, $z_h / z_s$ and $z_0$ parameter values can be determined to about 10\% by comparing data to TCM soft components as in Figs.~\ref{pions} (left), \ref{kch} (c) and \ref{protons} (a). The accuracy of the resulting TCM is then demonstrated in the right panels of those figures where, except for protons, the inferred data hard components agree in amplitude with the unit-normal TCM models within the same 10\%.

Table~\ref{otherparams} also includes values for soft and hard ensemble means $\bar p_{tsi}$ and $\bar p_{th0i}$ derived from TCM model functions $\hat S_{0i}(y_t)$ and $\hat H_{0i}(y_t)$ according to parameter values in Table~\ref{pidparams}. When combined with hadron-species-independent geometry parameters $x(n_s)$ and $\nu(n_s)$ in Eq.~(\ref{pampttcmid}) predictions for ensemble-mean $\bar p_{ti}$ are produced.
	
Figure~\ref{pidmpt} shows \mmpt\ vs \nch\ data for pions, kaons, protons and Lambdas from Ref.~\cite{aliceppbpid} (solid points) vs TCM trends from Eq.~(\ref{pampttcmid}) (solid curves). The agreement for mesons is well within point-to-point uncertainties. The large error bars for charged kaons are associated with extrapolation of limited spectrum data to $y_t \rightarrow 0$ [see Fig.~\ref{kch} (a)]. But since $K^\pm$ and $K_S^0$ spectra are statistically equivalent the real extrapolation uncertainty is negligible [see Fig.~\ref{kch} (c)], and that is reflected in the close correspondence of data and TCM(s) for the two kaon species.

The situation with baryons is markedly different. Extrapolation uncertainties for protons should be much less than for Lambdas because of the lower spectrum cutoff [see Fig.~\ref{protons} (a) and (c)], but  proton hard components appear to be systematically suppressed by about 40\% {\em only near the mode} [see  Fig.~\ref{protons} (b)] as represented by the dash-dotted curve in Fig.~\ref{pidmpt}. The baryon solid curves in Fig.~\ref{pidmpt} are determined by Eq.~(\ref{pampttcmid}) with fixed hard components and no suppression, but the systematic centroid shifts in Fig.~\ref{protons} (b) and (d) lead to displacement of \mmpt\ data relative to TCM trend from lower to higher with increasing \nch. The proton discrepancy illustrates the predictive power of the TCM: it is not adjusted to accommodate individual cases. The TCM serves as a fixed reference system applicable to any A-B collision system. Anomalous behavior can then be detected and characterized accurately.

\subsection{Comparison with blast-wave model fits} \label{bwfits}

In Ref.~\cite{aliceppbpid} the BW spectrum model applied to \ppb\ collisions is described as giving ``the best description of the data over the full $p_T$ range,'' referring to individual BW fits to each hadron species and centrality class, where the ``full $p_T$ range'' depends strongly on specific hadron species (see Fig.~\ref{piddata}).  However, BW parameters $(T_{kin},\bar \beta_t)$ for each centrality class as reported in Table~5 of Ref.~\cite{aliceppbpid} are obtained from simultaneous fits to all hadron species.  

For the simultaneous fits the actual \pt\ range is determined by the low-\pt\ acceptance limit for each species as above but by ``agreement with the data at high $p_T$'' (i.e.\ the \pt\ upper limits are chosen based on fit quality -- see horizontal lines labeled BW in Figs.~\ref{pions}, \ref{kch} and \ref{protons}). But $\chi^2$ is a measure of ``agreement with the data,'' so the $\chi^2$ values in Table~5 simply reflect a process of data selection, are not indicative of the {\em likelihood} of the fit model. Note that $\chi^2$ values for seven centrality classes are all less than 1 with mean value 0.45, whereas the mean of the $\chi^2$ distribution {\em per degree of freedom} should be 1, with roughly half the values expected to be above 1 for an acceptable data description. The $\chi^2$ values in Table~5 then likely reflect a strong bias resulting from data selection via \pt\ cuts. Reference~\cite{aliceppbpid} offers the following comment on its Table~5:  ``Positive and negative variations of the parameters using the different [\pt] fit ranges...are also reported. Variations of the fit range lead to large shifts ($\sim$10\%) of the fit results (correlated across centralities)...."

It is notable that $\chi^2$ is greatest for the most-peripheral centrality class where the TCM soft component (what should be described by a hydro model if such a model were appropriate) is by far the dominant component compared to the jet contribution. For the most-central collisions (at least according to the Glauber model), where jets clearly dominate, $\chi^2$ is lowest. That combination of features suggests that the BW model is unlikely to reflect actual hadron production mechanisms in \ppb\ collisions. 

In contrast, the TCM is required to describe {\em all} available data self-consistently with minimal parametrization. Section~\ref{piddiff} illustrates the description quality. The deviations for proton data in Fig.~\ref{protons} (b) emphasize the importance of a {\em fixed} model in revealing data-model anomalies.

\section{Discussion} \label{disc}

 As summarized in Sec.~\ref{alicedata}, Ref.~\cite{aliceppbpid} interprets PID spectrum data to suggest that ``collectivity'' (e.g.\ radial flow) is manifested in \ppb\ collisions. The suggestion is based on certain terminology, preferred analysis methods and argument by analogy. If correct, \pa\ or \da\ data, initially assumed to serve as a control for QGP discovery in \aa\ collisions, would no longer serve that purpose.

\ppb\ PID spectra are said to ``flatten'' or become ``harder'' with increasing \nch\ or \ppb\ centrality, i.e.\ spectrum slopes for lower \pt\ are observed to decrease.  The slope changes exhibit ``mass ordering,'' i.e.\ the effects increase with increasing hadron mass. It is then argued by analogy that since similar effects are observed in \pbpb\ collisions, and since such effects observed in \aa\ collisions are conventionally interpreted to arise from radial flow as the ``natural explanation,'' the cause in \ppb\ collisions must also be ``collective'' radial flow. In this section such arguments are confronted with alternative evidence derived in part from the present PID \ppb\ study.

\subsection{Evidence for and against flows from $\bf p_t$ spectra} \label{noflow}

Evidence for ``collectivity'' (flows) has been reported previously for \ppb\ collisions at the LHC, including apparent indications of hydrodynamic flows~\cite{dusling,bozek,bozek2}. In \aa\ collisions the presence of a flowing dense QCD medium has been associated with ``jet quenching''~\cite{jetquenching}.  In Sec.~\ref{alicedata} BW fits to PID spectrum data as reported in Ref.~\cite{aliceppbpid} are briefly summarized. It is observed that BW parameters $\bar \beta_t$ and $T_{kin}$ have similar values for \ppb\ and \pbpb\ collisions. Those results are then interpreted as ``consistent with the presence of radial flow in \ppb\ collisions.''  It is further noted that ``a larger radial velocity in \ppb\ collisions has been suggested as a consequence of stronger radial gradients,'' albeit within a smaller  system. 

However, \ppb\ PID spectrum data as presented in the TCM formats of Sec.~\ref{piddiff} contradict such claims. Soft component $\hat S_0(y_t)$ (left panels) for a range of \ppb\ \nch\ remains consistent with isolated \pp\ collisions and with a common energy dependence extending over three orders of magnitude in $\sqrt{s}$. The inferred spectrum hard components (right panels) are approximately independent of \ppb\ \nch\ and consistent with jet properties inferred from isolated \pp\ collisions over a large energy interval~\cite{alicetomspec}. There is no significant evidence for radial flow (e.g.\ boosted \yt\ spectra) or for modification of parton fragmentation to jets (i.e.\ FF modification~\cite{fragevo}).
The argument for larger {\em gradients} in smaller systems implies that central {\em densities} must remain similar for all collision systems, but central densities must scale with participant number which should be proportional to system size.

As mentioned in Sec.~\ref{alicedata}, Ref.~\cite{aliceppbpid} offers a disclaimer about radial flow interpretations based on BW fits to spectrum data. Given results from BW fits applied to PYTHIA \pp\ spectra it is concluded that the PYTHIA CR mechanism produces results that ``...can mimic the effects of radial flow.'' The single disclaimer is juxtaposed with a number of positive statements regarding radial flow based on BW fits and on argument by analogy with \aa\ spectra, as summarized in Sec.~\ref{alicedata}, and is not mentioned in the paper summary. A more important omission is lack of acknowledgement of the contribution of minimum-bias jets to spectrum data with a peak near $p_t = 1$ GeV/c that lies well within any BW fitting interval (see left panels in Sec.~\ref{piddiff}). In the context of BW fits the large jet contribution can certainly mimic the effects of radial flow, and thus demonstrates that the BW model is inappropriate for \pt\ spectrum description.

\subsection{$\bf \bar p_t$ trends vs flow interpretations}

Figure~\ref{pidmpt} shows PID \mmpt\ data reported in Ref.~\cite{aliceppbpid} (solid points) and corresponding TCM trends (solid curves) for 5 TeV \ppb\ collisions. \mmpt\ data certainly reflect evolving \pt\ spectrum structure, but with greater statistical precision due to integration. It is  notable that the most rapid increase of \mmpt\ occurs for lowest \nch. For unidentified hadrons from 5 TeV \pp\ and \ppb\ collisions the two \mmpt\ trends are identical up to $\bar \rho_0 \approx 20$ (i.e.\  four times the NSD value 5) and it is likely that PID data for \pp\ collisions also agree with those for \ppb\ within that interval. \mmpt\ values do increase strongly with hadron mass, but for larger \nch\ the overall \mmpt\ trend is strong {\em decrease} with increasing system size from \pp\ to \ppb\ to \pbpb~\cite{tommpt}. 

Several questions arise concerning flow interpretations of \mmpt\ data: In the context of a flowing dense medium where is the transition (e.g.\ on \nch) from free-flying particles (no inertial confinement) to rescattering within a dense medium (confinement increasing with densities and system volume)?  For instance, why doesn't \mmpt\ increase {\em more} rapidly for larger \nch\ or more rapidly in \pbpb\  (than in \pp) where inertial confinement is more likely?

In a flow context it is expected that \mmpt\ increases more rapidly for more-massive hadrons because within a common velocity field particle momentum should increase with hadron mass. But that implies soft components of \pt\ spectra should be boosted to higher \yt\ proportional to hadron mass. Why is a common boost not observed for the PID spectra in Sec.~\ref{pidmpt} [$\hat S_0(y_t)$ shapes do depend on mass, but all proceed from \yt\ = 0, not a boosted value]?

TCM analysis of \pt\ spectra and \mmpt\ data provide answers. Based on differential spectrum structure as shown in Sec.~\ref{piddiff} \mmpt\ increases with \nch\ and hadron mass because of the large  MB dijet contribution to hadron production, while soft components {\em remain invariant}. More jet-related hadrons are produced but with the same hard-component \yt\ distributions, and there are copious jet-related baryons (large $z_h/z_s$ values in Table~\ref{otherparams}) as determined experimentally. There is no indication of a flow component in PID spectra.
The inverse hierarchy \pp\ $>$ \ppb\ $>$ \pbpb\ for \mmpt\ data is explained in terms of typical $\bar \rho_{sNN}$ values for the three collision systems that have the same ordering and the relation $\bar \rho_{hNN} \propto \bar \rho_{sNN}^2$ that determines MB jet production with, at least in \pp\ and \pa\ systems,  no jet modification. 

\subsection{Preferred analysis methods}

Certain analysis methods, statistical measures and plotting formats have been conventionally preferred over a range of alternatives in the analysis of high-energy particle data. \pt\ spectra are typically modeled and interpreted with the implicit assumption that they are monolithic (a single component), represented then by a single functional form. Examples include the ``power law'' model for \pp\ spectra from the S$p \bar p$S (jets plus soft process)~\cite{ua1}, Tsallis model (similar to the TCM soft component)~\cite{tsallis} and the blast-wave model for \aa\ spectra (locally-thermalized flowing medium)~\cite{blastwave}. Other choices include linear \pt\ as an independent variable rather than a logarithmic scale and the use of spectrum ratios to evaluate fit quality and relations among hadron species.

In Ref.~\cite{aliceppbpid} the BW spectrum model is applied to  individual BW fits for each hadron species and \ppb\ centrality class. The BW results are described as ``...the best descriptions of data over the full $p_T$ range'' (presumably the data ranges subtended by the various species). It is noted that for equivalent $\bar \rho_0$
values $\bar \beta_t$ for \ppb\ is ``significantly higher'' than for \pbpb\ (e.g.\ compare Figs.~11 and 14 of Ref.~\cite{tommpt}), leading to speculation that ``stronger radial gradients'' may arise in \ppb\ collisions. The same argument applied to \pp\ \mmpt\ data must conclude that such gradients are {\em highest in the smallest collision system}.

Among alternative possibilities are treatment of spectra as possibly composite, plotting spectra vs logarithmic variable \yt\ and {\em differential} study of individual spectra rather than their ratios, the goal being to extract as much information as possible from particle data. Comparison of the same data as plotted in Sec.~\ref{alicedata} vs Sec.~\ref{piddiff} illustrates the major differences in {\em visually accessible} information derived from alternative plotting formats.

The TCM results in Sec.~\ref{piddiff} provide a very different picture of mechanisms relevant to \ppb\ collisions. Instead of a monolithic model function with BW parameters varying strongly with collision conditions the TCM soft and hard model functions are approximately invariant with \ppb\ centrality, greatly simplifying the model. Because the TCM hard component is quantitatively related to measured jet properties the \mmpt\ trends attributed to radial flow in the BW context are instead related to jet physics in the TCM context. (The logarithmic \yt\ rapidity variable allows greatly improved access to low-\pt\ data where the {\em great majority} of jet fragments resides.) That jet production per final-state hadron (e.g.\ TCM ratio $\bar \rho_h / \bar \rho_s$) might be largest in the smallest collision system (hence driving up \mmpt\ values) is easily explained in terms of conventional jet physics and the quadratic relation $\bar \rho_h \propto  \bar \rho_s^2$ {\em consistent with measured jet cross sections}.

\ppb\ centrality determination based on the Glauber model~\cite{aliceglauber}, as opposed to direct inference from \mmpt\ data~\cite{tomglauber}, strongly limits information available from particle data. In Fig.~\ref{pidmpt} PID \mmpt\ data from Ref.~\cite{aliceppbpid} extend only to $\bar \rho_0 \approx 45$, assigned by the Glauber analysis to 0-5\% centrality, whereas high-statistics \mmpt\ data extend to $\bar \rho_0 \approx 116$ corresponding to {\em 4-fold increase} in dijet production. Claims for collectivity or flows in \ppb\ collisions could be much better tested with PID spectra from higher charge densities and associated jet production.

\subsection{Argument by analogy}

Reference~\cite{aliceppbpid}, arguing by analogy, suggests that  commonalities between \ppb\ spectrum and \mmpt\ data and those from \pbpb\ collisions imply the presence of collective flow in \ppb\ collisions:  ``In heavy-ion collisions, the flattening of transverse momentum distribution and its mass ordering [i.e.\ relating to variation with \nch] find their natural explanation in the collective radial expansion of the system.'' and
``The [\ppb] transverse momentum distributions show a clear evolution with multiplicity, similar to the pattern observed in high-energy pp and heavy-ion collisions, where in the latter case the effect is usually attributed to collective radial expansion.''

That a particular data characteristic might appear in \pbpb\ {\em and} \ppb\ data does not {require} similar causes, and attribution of certain trends in \aa\ collisions to flow phenomena or quark recombination may be questioned.   Spectrum and yield {\em ratios} tend to discard a substantial fraction of the information carried by particle data~\cite{mbdijets} and therefore cannot be relied on for definitive model tests. The same comments apply to fit models:  Application of a flow-related spectrum model to spectrum data does not demonstrate the presence of flow in the corresponding collision system, and differential spectrum details (e.g.\ the TCM applied to PID hadron spectra as in the present study) may exclude a flow interpretation.

\subsection{Abandoning control experiments}

In preparation for initial RHIC operations data from \dau\ collisions were promoted as a control experiment for possible discovery of QGP in more-central \auau\ collisions (e.g.\ Ref.~\cite{dautest}). Whereas a phase transition to a QGP was expected to require large energy and particle densities \dau\ collisions should represent ``cold nuclear matter'' and no transition. The apparent contrast between \auau\ and \dau\ data (e.g.\ apparent presence or absence of jet quenching) was then cited as confirming discovery~\cite{daufinalstate}.
But recent observations in \ppb~\cite{ppbridge} and even \pp~\cite{ppcms} data from the LHC of certain characteristics (e.g.\ one or more ridge structures, \pt\ spectrum ``hardening'' with increased centrality or hadron mass) conventionally associated with ``collective'' behavior (i.e.\ flows) in \aa\ collisions have led to speculation that flows and QGP might be possible even in the smallest systems, albeit corresponding to higher  energies and charge densities~\cite{dusling}.

Such reasoning betrays the function of a control experiment. Theoretical understanding of QCD and a conjectured phase transition led to a prediction: QGP may be possible in the largest \aa\ collision systems but not possible in low-density smaller systems. If phenomenon X is observed in \aa\ but not in \pa\ it is a candidate QGP indicator (e.g.\ elliptic flow, jet quenching). But if X is observed to a significant extent in both large {\em and} small collision systems it is unlikely to be associated with QGP formation and flows {\em according to the original theoretical expectation}. Concluding that QGP must be a universal manifestation even in small low-density systems betrays the role of the control experiment: In essence, theory is modified to accommodate data and is then not falsifiable.

For example, a nonjet quadrupole feature [i.e.\ $v_2 \sim \cos(2\phi)$] is accurately inferred for 2D angular correlations on $(\eta,\phi)$ from 200 GeV \pp\ collisions~\cite{ppquad}. Arguing by analogy with the $v_2$ interpretation in \aa\ collisions one might then conclude that elliptic flow must appear in the smallest collision system. But \pp\ collisions are intended to serve as a control experiment for claims of jet quenching and flows in \aa\ collisions, for instance in the form of spectrum ratio $R_{AA}$. In fact, systematics of the quadrupole component in \pp\ (e.g.\ its \nch\ dependence) make a flow interpretation unlikely there~\cite{ppquad}, casting doubt on flow interpretations of $v_2$ data in \aa\ collisions.

Similarly, interpretation (again by analogy) of certain \pt\ spectrum properties in terms of radial flow in \ppb\ collisions (as a nominal control experiment) are falsified by the PID spectrum analysis of the present study. Significant radial flow (i.e.\ corresponding to a boosted hadron source) should result in substantial suppression of spectra near zero momentum to complement enhancement (``hardening'') at higher momentum.%
\footnote{An illustration from recent hydro theory descriptions may be found in Fig.\ 16 (right) of Ref.~\cite{nature}. The hydro curve (bold solid) is strongly suppressed at lower \pt\ as the model accommodates the strong jet contribution at higher \pt.
}  
However, the spectra from Ref.~\cite{aliceppbpid} are accurately and exhaustively described by a TCM including an {\em invariant} soft component (i.e.\ consistent with a \ L\'evy distribution) and a hard component quantitatively related to jet production. That is particularly evident for the $K^0_\text{S}$ data in Fig.~\ref{kch} (c).

This \ppb\ result is consistent with previous analysis of 200 GeV \auau\ collisions~\cite{hardspec} which concluded  "Since 1/3 of the hadrons originate effectively from rapidly-moving sources (parton fragmentation), the significance of statistical-model spectrum measures (chemical potentials, decoupling temperatures, $\langle p_t \rangle$s, [$\bar \beta_t$]) attributed to an expanding bulk medium can be strongly questioned. Upon close examination of pion and proton spectra no [evidence for] identifiable radial flow is apparent."

Control experiments based on differential analysis of all available information reveal that certain data features appearing in \aa\ and even \pa\ or \da\ collisions, previously interpreted in terms of flows and QGP, are more likely manifestations of MB jets~\cite{mbdijets}.

Problems raised by claims for collectivity in small systems (i.e.\  control experiments) have been recognized~\cite{thoughts}:
``With decreasing system size, one may expect a transition from collective behavior to free streaming on scales where the mean free path...of medium [QGP, perfect liquid] constituents becomes comparable to the system size.... Experimentally, however, no indications for the existence of such an onset of collective behavior with system size have been identified so far [see (a) below]. Signatures of collectivity such as higher-cumulant flow harmonics [see (b) below] display a remarkably weak dependence on system size. Understanding the apparent absence of a minimal scale for collective behavior is a central problem of the field for which qualitatively different solutions remain to be explored in more detail....'' (a)  Experimental evidence for a ``sharp transition'' (on centrality) of jet modification within \auau\ collisions (from no jet quenching to strong jet quenching) has been established previously~\cite{anomalous}, whereas {\em no} jet modification is observed in \pp~\cite{ppquad} or \ppb\ (the present study) collisions, and (b) emphasis on what are denoted ``flow [higher] harmonics'' and their interpretation as indicators of ``collectivity'' (flows) may be strongly questioned~\cite{harmonics,harmonics2}. The present TCM study indicates no significant evidence for radial flow in \ppb\ PID hadron spectra (see Sec.~\ref{noflow}).

\subsection{Collectivity inferred from angular correlations}

While the present study focuses on \ppb\ \pt\ spectrum properties for PID hadrons, arguments for manifestations of ``collectivity'' or flows in small systems have mainly been based on one or two ``ridge'' features in 2D angular correlations~\cite{ppcms,ppbridge}. However, the \nch\ trends for those features rule out a flow interpretation as argued here.

In Ref.~\cite{ppquad} 2D angular correlations from high-statistics 200 GeV \pp\ collisions are studied. The TCM for \nch\ dependence of \pt\ spectra is first confirmed and corresponds to 100-fold increase of dijet production over  the \nch\ interval spanned by the \pp\ data. Model fits to 2D angular correlations  then reveal three components: (a) the TCM soft component manifested as a 1D peak on pseudorapidity $\eta$ centered at $\eta = 0$ (charge-neutral pairs only); (b) the TCM hard component consisting of a same-side ($\phi = 0$) 2D peak on $(\eta,\phi)$ centered also at $\eta = 0$ and an away-side ($\phi = \pi$) 1D peak on $\phi$; and (c) a quadrupole $\cos(2\phi)$ component on $\phi$ representing a third component. Component (c) should be distinguished from a contribution to the total quadrupole moment from the 2D jet peak and is thus referred to as the {\em nonjet} quadrupole.

The three correlation components have distinct dependences on charge-density soft component $\bar \rho_s$ employed as the system control parameter. As for \pt\ spectra the soft-component amplitude (a) varies as $\propto \bar \rho_s$ and the hard-component amplitude (b) varies as $\propto \bar \rho_s^2$, where ``amplitude'' here refers to the number of correlated pairs associated with a given component. Further analysis reveals that the nonjet quadrupole component (c) varies as $\propto \bar \rho_s^3$. 

Those trends are followed precisely over the full \nch\ range of data. Thus, as dijet production increases 100-fold the quadrupole amplitude {\em increases 1000-fold} and becomes clearly visible in 2D angular correlations for the larger \nch\ values. As a result of the increasing {\em two}-lobed (maxima at 0 and $\pi$) quadrupole contribution the same-side ($\phi = 0$) curvature on $\phi$ (excluding the 2D jet peak) {\em changes sign} to become the apparently-single ``ridge'' usually referred to~\cite{ppcms}, while the away-side ($\phi = \pi$) curvature {\em increases in magnitude} although that aspect is typically ignored. The full two-lobe quadrupole structure for LHC data was only recently  inferred from \ppb\ data~\cite{ppbridge}, based on visual examination of 2D angular correlations rather than precision model fits as in Ref.~\cite{ppquad}.

The observed $\propto \bar \rho_s^3$ trend for the \pp\ nonjet quadrupole leads to several important conclusions: 
(a) Jet-related and quadrupole trends are clearly distinct. 
(b) The cubic trend is followed precisely over the full \nch\ range for data down to small charge densities. A flowing dense medium (collectivity) interpretation is therefore quite unlikely. 
(c) The cubic trend can be recast as $\propto N_{part} N_{bin}$ in terms of number of participant low-$x$ gluons $N_{part} \propto \bar \rho_s$ and their binary collisions to produce dijets $N_{bin} \propto \bar \rho_s^2$. 

The $\propto N_{part} N_{bin}$ quadrupole trend for \pp\ collisions can be compared with the $\propto N_{part} N_{bin}\epsilon_{opt}^2$ trend for \auau\ collisions as determined in Ref.~\cite{nonjetquad}, where in the latter case $N_{part}$ and $N_{bin}$ refer to number of {\em nucleon} N participants and N-N binary collisions. As for \pp\ collisions the same \nch\ trend is followed precisely by the nonjet quadrupole  over the complete centrality range for \auau\ collisions. The same  trend is observed for LHC \pbpb\ collisions~\cite{nonjetquad}. There is no significant correlation with the strongly-varying modification of jet formation (jet quenching)~\cite{anomalous} attributed to energy loss in a dense medium, again making a flow interpretation unlikely.

For \pp\ collisions there is no {\em eccentricity} dependence (i.e.\ no $\epsilon_{opt}^2$ factor as appears for \aa\ collisions~\cite{nonjetquad}). That observation is consistent with a conclusion from \pt\ spectrum data that centrality plays no role in \pp\ collisions~\cite{ppprd,ppquad,pptheory,pptheory2}. Each \pp\ (or \nn) collision requires full transverse overlap~\cite{tomexclude}. It is possible that the $\propto N_{part} N_{bin} \propto \bar \rho_s^3$ trend for the \pp\ quadrupole corresponds to a {\em three-gluon} interaction resulting in QCD quadrupole radiation~\cite{gluequad}. The $\propto N_{part} N_{bin}\epsilon_{opt}^2$ trend for \aa\ collisions then suggests that the same three-gluon interaction may be responsible  as well, but that the overall \aa\ geometry must be eccentric to produce a significant {\em net} quadrupole component from the composite system. 

The present TCM study of PID \pt\ spectra from \ppb\ collisions confirms that hadron production is dominated by TCM soft and hard  components. Evidence from 2D angular correlations reveals that the nonjet quadrupole is a distinct third component and must therefore be ``carried'' by a small minority (i.e.\ few percent) of hadrons even for large \nch. That conclusion is consistent with a study of \aa\ quadrupole trends~\cite{nonjetquad} but inconsistent with a hydro/QGP narrative in which the nonjet quadrupole is carried by almost all hadrons and most hadrons emerge from a flowing dense medium. In any case, evidence from 2D angular correlations corroborates conclusions derived from \pt\ spectrum data as in the present study: collectivity (i.e.\ flows) in small collision systems is unlikely, and the same features interpreted to indicate collectivity or flows in small systems may also be misinterpreted as such in more-central \aa\ collisions.

\section{Summary}\label{summ}

Certain data features from small collision systems (\pp\ and \ppb\ collisions) at the large hadron collider (LHC), e.g.\ properties of \pt\ spectrum and 2D angular correlation data, have been interpreted recently as demonstrating the presence of ``collectivity'' (hydrodynamic flows) even for low charge densities. Such interpretations are based on arguments by analogy, that similar data features in \aa\ collisions are interpreted to indicate formation of a flowing dense medium or quark-gluon plasma (QGP). 

Such findings reverse the intended role of small collision systems as control experiments based on the assumption that for low-enough particle and energy densities free particle streaming should be inconsistent with a hydrodynamic flow description. Claimed evidence for QGP  formation in \aa\ should then not appear in a control collision system representing ``cold nuclear matter.''

The inference of collectivity in small systems implies that there is no threshold point for the onset of collectivity (and possible QGP formation). Absence of a density threshold for collectivity or flows has been recognized as a central problem for interpretation of high-energy particle data. One suggested approach is development of new theoretical mechanisms to account for the lack of such a transition, but an alternative approach is reexamination of conventional interpretations of certain data features as demonstrating collectivity, thus challenging the basic concept of collectivity or flows in any collision system.

To that end the present study is a differential analysis of  identified-hadron (PID) \pt\ spectra from 5 TeV \ppb\ collisions for seven centrality or \nch\ classes. The PID spectra are described by a two-component (soft + hard) model (TCM)  of hadron production mechanisms. The soft component is associated with longitudinal projectile-nucleon dissociation while the hard component is associated with large-angle scattering of low-$x$ partons (gluons) to form jets. \ppb\ centrality is adopted from a previous TCM study of \ppb\ ensemble-mean \mmpt\ data for unidentified hadrons. Questions to be addressed include: (a) Is spectrum evolution with \nch\ consistent with expectations for effects of radial flow (soft)? and (b) Is jet formation modified by parton interactions with a dense flowing medium (hard)? In a previous study it was observed that \mmpt\ data for peripheral \ppb\ are equivalent to those for \pp\ over a significant \nch\ interval, thus providing information on both systems from the \ppb\ data.

The main analysis results are as follows: (a) The PID spectrum TCM provides an accurate description of all spectrum and \mmpt\ data with one exception -- protons -- where the spectrum hard component is found to be suppressed by about 40\%. (b) The TCM provides access to unprecedented details of spectrum structure and systematic variation with \nch. (c) Spectrum variations with \nch\ and hadron mass that have been attributed to radial flow are actually jet manifestations represented by the spectrum hard component. (d) Approximate invariance of PID spectrum hard components with \nch\ demonstrates that the jet component of hadron production remains unmodified over a large range of hadron and jet densities.

Precise distinction between TCM soft and hard components is based on a quadratic hard-vs-soft relation that is already evident for measured eventwise-reconstructed-jet cross sections. PID spectrum soft components, including the majority of produced hadrons, retain fixed shapes independent of \nch\ or \ppb\ centrality and show no evidence for a common source boost that would indicate radial flow. The dominant contribution to the TCM hard component is from lowest-energy jets near 3 GeV which should be most sensitive to a dense flowing medium if one existed. PID spectrum data indicate that \ppb\ collisions for any \nch\ are linear superpositions of \pn\ collisions.

It could be argued that appearance of a ridge or ridges in 2D angular correlations from small systems may still indicate some form of medium formation and collectivity. Angular correlations are outside the scope of the present study, but evidence from other studies suggests  that the presence of a nonjet quadrupole in \pp\ and \pa\ collisions is the manifestation of a QCD process based on direct few-gluon interactions, not medium formation. Observation of a ridge structure in \ppb\ collisions where spectrum structure precludes a dense medium may serve to buttress such an interpretation. The role of small systems as control experiments would then be restored.

\begin{appendix}

\section{$\bf \bar p_t$ TCM for $\bf p$-$\bf p$ collisions}  \label{ppmptapp}

The present study of PID hadron spectra from 5 TeV \ppb\ collisions relies critically on \ppb\ centrality determination inferred from \mmpt\ data for unidentified hadrons as reported in Ref.~\cite{tomglauber}. This appendix briefly reviews basic  \mmpt\ TCM analysis for elementary \pp\ collisions.

Reference~\cite{alicetommpt} established that a TCM for ratio $ \bar p_t = \bar P_t / n_{ch}$ constitutes a good description of LHC data from \pp\ and \pbpb\ collisions at several energies and  provided hints as to the mechanism of \ppb\ \mmpt\ variation -- evolving from a \pp\ or \pn\ trend for smaller \nch\ to a quantitatively different but similar trend above a transition point.
Reference~\cite{alicetomspec} presented a detailed TCM analysis of \pp\ \pt\ spectra for a range of energies from 17 GeV to 13 TeV.  Soft component $\hat S_0(m_t,\sqrt{s})$ varies  weakly with energy and not at all with \nch, but hard component $\hat H_0(y_t,n_s,\sqrt{s})$ varies strongly with energy (consistent with variation of the underlying jet spectrum) and significantly with \nch\ (as established with 200 GeV spectra). 
Those new spectrum results were then incorporated within a revised \mmpt\ analysis in Ref.~\cite{tommpt}, summarized for \pp\ here and for \ppb\ in Sec.~\ref{ppb}.

Quantities $\bar p_{th}(n_s,\sqrt{s})$, $\alpha(\sqrt{s})$ and an effective detector acceptance ratio $\xi$ are used  to update results from Ref.~\cite{alicetommpt}. The TCM for charge densities averaged over some angular acceptance $\Delta \eta$ (e.g.\ 0.6 for Ref.~\cite{alicempt}) is 
\bea
\bar \rho_0 &=& \bar \rho_{s} + \bar \rho_{h}
\\ \nonumber
&=& \bar \rho_{s}[1+ x(n_s)],
\\ \nonumber
\frac{\bar \rho_0'}{ \bar \rho_s}&=& \frac{n_{ch}'}{n_s} ~=~ \xi+ x(n_s),
\eea
where $x(n_s) \equiv \bar \rho_{h} / \bar \rho_{s} \approx  \alpha \bar \rho_s$ is the ratio of hard-component to soft-component yields~\cite{ppprd} and $\alpha(\sqrt{s})$ is defined in Ref.~\cite{alicetomspec}. Primes denote uncorrected quantities. The TCM for ensemble-mean {\em total} $p_t$ integrated over some angular acceptance $\Delta \eta$ from \pp\ collisions for given $(n_{ch},\sqrt{s})$ is expressed as
\bea \label{mptsimple}
\bar P_t &=& \bar P_{ts} + \bar P_{th}
\\ \nonumber
&=& n_s \bar p_{ts} + n_h \bar p_{th}.
\eea
The conventional intensive ratio of extensive quantities
\bea \label{ppmpttcm}
\frac{\bar P_t' }{n_{ch}'} \equiv \bar p_t' &\approx & \frac{\bar p_{ts} + x(n_s) \bar p_{th}(n_s)}{\xi + x(n_s)}
\eea
(assuming $\bar P_t' \approx \bar P_t$~\cite{tommpt}) in effect partially cancels MB dijet manifestations represented by ratio $x(n_s)$.  The alternative ratio
\bea \label{niceeq}
\frac{n_{ch}'}{n_s} \bar p_t'   \approx \frac{ \bar P_t}{n_s} &= & \bar p_{ts} + x(n_s) \bar p_{th}(n_s)
\\ \nonumber
&=& \bar p_{ts} + \alpha(\sqrt{s})\, \bar \rho_s \, \bar p_{th}(n_s,\sqrt{s})
\eea
preserves the simplicity of Eq.~(\ref{mptsimple}) and provides a convenient basis for testing the TCM hypothesis precisely.

Figure~\ref{alice5a} (left) shows $\bar p_t$ data for four \pp\ collision energies from the RHIC  (solid triangles~\cite{ppprd}), the Sp\=pS  (open boxes~\cite{ua1mpt}) and the LHC (upper points~\cite{alicempt}) increasing monotonically with charge density $\bar \rho_0 = n_{ch} / \Delta \eta$. The lower points and curves correspond to full \pt\ acceptance.  For acceptance extending down to zero ($\xi = 1$), $\bar p_t' \rightarrow \bar p_t$ in Eq.~(\ref{ppmpttcm}) should vary between the universal lower limit $\bar p_{ts} \approx 0.4$ GeV/c ($n_{ch} = 0$) and  $\bar p_{th}$ ($n_{ch} \rightarrow \infty$) as limiting cases. For a lower \pt\ cut $p_{t,cut} > 0$ the lower limit is $\bar p_{ts}' = \bar p_{ts} / \xi$ (dotted lines) and the data are systematically shifted upward (upper points and curves).  The solid curves represent the \pp\ $\bar p_t$ TCM from Ref.~\cite{tommpt}. Note  that  the 7 TeV \mmpt\ data extend to $\bar \rho_0 \approx 10\, \bar \rho_{0NSD} \approx 60$ and were derived from 150 million \pp\ collision events.

  \begin{figure}[h]
   \includegraphics[width=1.65in,height=1.6in]{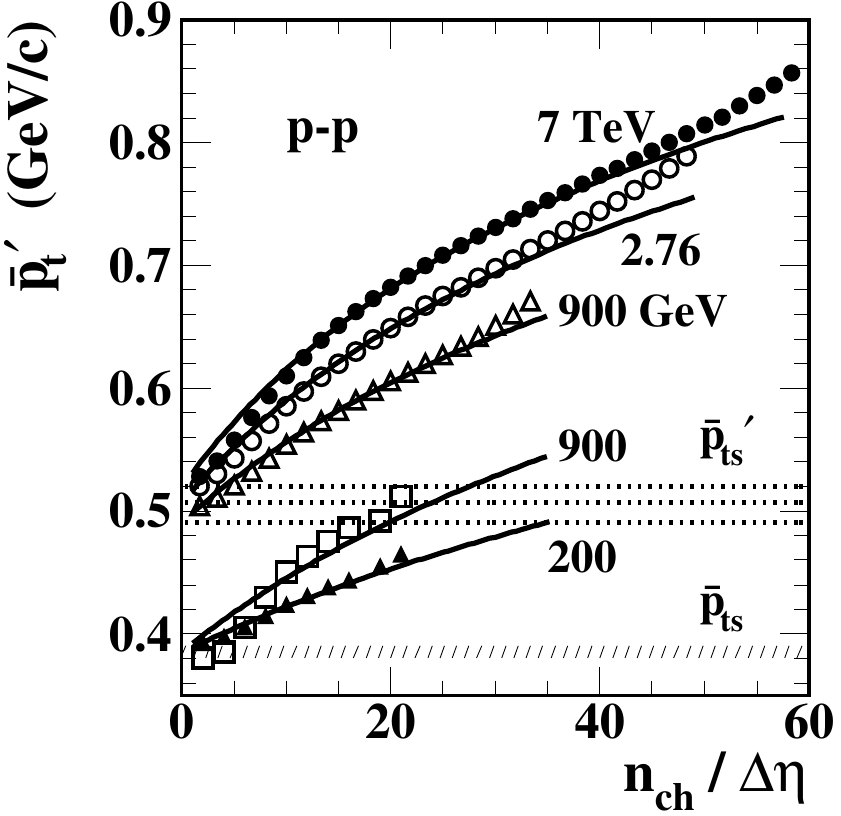}
   \includegraphics[width=1.65in,height=1.6in]{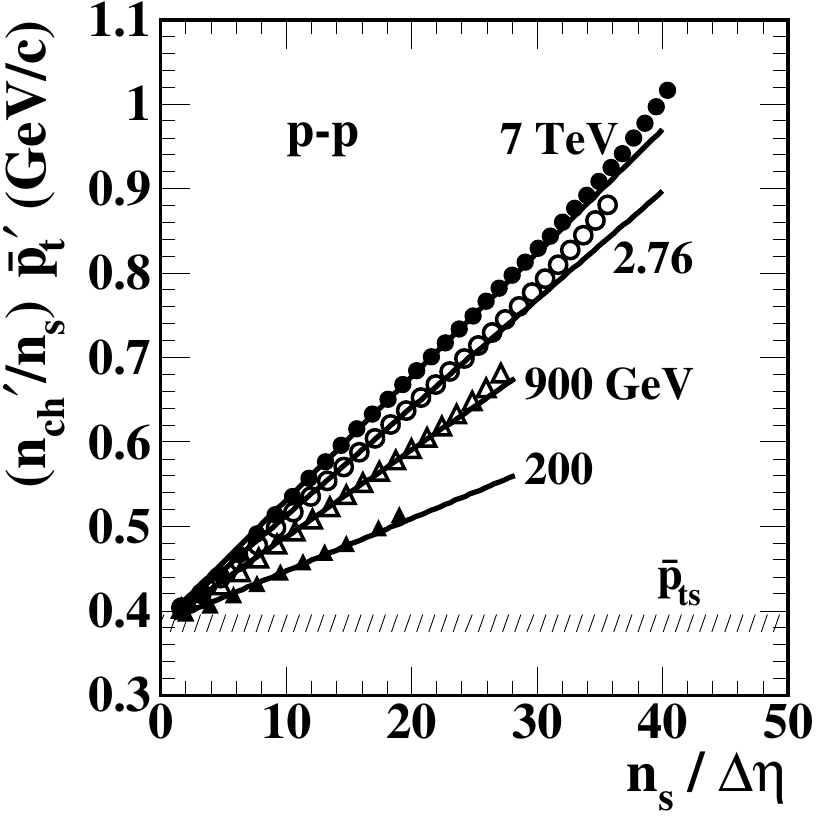}
 \caption{\label{alice5a}
 Left: \mmpt\ vs \nch\ for several collision energies. The upper group of points from Ref.~\cite{alicempt} are derived from particle data with a lower \pt\ cutoff. The lower 900 GeV data from Ref.~\cite{ua1mpt} and 200 GeV data from Ref.~\cite{ppprd} are extrapolated to zero \pt.
Right: Data from the left panel multiplied by factor ${n'_{ch}} / n_s$ that removes the jet contribution and the effect of the low-\pt\ cut on the soft component from the denominator of \mmpt. 
} 
  \end{figure}
 
Figure~\ref{alice5a} (right) shows data on the left transformed via Eq.~(\ref{niceeq}) to $(n_{ch}' / n_s) \bar p_t' \approx  \bar P_t / n_s$ (points). The TCM curves undergo the same transformation and the slopes of the resulting straight lines are $\alpha(\sqrt{s}) \bar p_{th0}(\sqrt{s})$. The data deviate significantly from the straight-line TCM because of systematic variation with \nch\ of the \pt\ spectrum hard-component shape as reported in Refs.~\cite{alicetomspec,tommpt}. However, those details are beyond the scope of this spectrum study.

The success of the \pp\ \mmpt\ TCM confirms that variation of \pp\ \mmpt\ is dominated by jet fragments from large-angle-scattered low-$x$ gluons. 
The hard yield or angular density $\bar \rho_{h} \approx \alpha(\sqrt{s})\, \bar \rho_s^2$ represents the dijet fragment density determined precisely by soft component $\bar \rho_s$. 
The \pt\ spectrum TCM hard component and underlying jet energy spectrum evolve according to the same rules~\cite{jetspec2}. 
The quadratic relation $\bar \rho_h \propto \bar \rho_s^2$ implies that \pp\ collisions are {\em noneikonal} (compared to the eikonal trend $\bar \rho_h\propto \bar \rho_s^{4/3}$). 
The quadratic trend (each participant gluon in one proton can interact with {\em any} participant gluon in the partner proton) implies that \pp\ collisions with large \nch\ are very jetty.
Reference~\cite{tommpt} demonstrates a direct connection between \mmpt\ hard component $\bar p_{th}(n_s)$, \pt\ spectrum hard component $H(p_t,n_s)$~\cite{alicetomspec} and jet spectra as in Ref.~\cite{jetspec2}. 
Thus, a variety of \pp\ data provide strong evidence that MB dijets dominate \pp\ collisions and $\bar p_t(n_{ch},\sqrt{s})$ trends.

\end{appendix}


\end{document}